\documentclass[12pt]{article}
\usepackage[left=2cm,top=2cm,right=3cm,head=1cm,foot=1cm]{geometry}
\usepackage{amsmath}
\usepackage{mathrsfs}
\usepackage{graphicx}
\usepackage{url}
\numberwithin{equation}{section}
\newcommand{\INT}{\int_{-\infty}^{\infty}\!\!}
\newcommand{\Ds}{\mathscr{D}}
\newcommand{\la}{\mathscr{L}}

\newcommand{\bra}{\langle}
\newcommand{\ket}{\rangle}
\newcommand{\half}{\tfrac{1}{2}}
\newcommand{\fourth}{\tfrac{1}{4}}
\newcommand{\al}{\alpha}
\newcommand{\G}{\Gamma}
\newcommand{\g}{\gamma}
\newcommand{\Ta}{\Theta}
\newcommand{\ta}{\theta}

\newcommand{\e}{\varepsilon}

\newcommand{\Del}{\Delta}
\newcommand{\del}{\delta}
\newcommand{\w}{\omega}
\newcommand{\z}{\zeta}

\newcommand{\s}{\sigma}
\newcommand{\tr}{\text{Tr}}
\newcommand{\ml}{\left(\begin{matrix}}
\newcommand{\mr}{\end{matrix}\right)}

\newcommand{\pa}{\partial}
\newcommand{\da}{\dagger}
\newcommand{\singleint}{\int_{t_0}^{t_f}\!\! dt\;}
\newcommand{\doubleint}{\int_{t_0}^{t_f}\!\! dt\int_{t_0}^{t_f}\!\! dt'\;}
\newcommand{\sign}{\text{sign}}
\newcommand{\dt}{\int_{t_0}^{t_f}\!\! dt}
\newcommand{\op}{\mathcal O}
\newcommand{\Scl}{S_{\text{cl}}}
\newcommand{\re}{\text{Re}}
\newcommand{\im}{\text{Im}}
\newcommand{\OUT}{\text{out}}
\newcommand{\IN}{\text{in}}
\newcommand{\minus}{\!-\!}
\newcommand{\plus}{\!+\!}
\newcommand{\Gs}{\mathscr G}
\title{}
\author{}
\begin{document}
%%%%%%%%%%%%
\title{Schwinger-Keldysh path integral for the quantum harmonic oscillator}
\date{}
\author{Yoni BenTov \\ \small Perimeter Institute for Theoretical Physics, Waterloo, Ontario N2L 2Y5}
%%%%
\maketitle
%%%%
\begin{abstract}
I review the generating function for quantum-statistical mechanics, known as the Feynman-Vernon influence functional, the decoherence functional, or the Schwinger-Keldysh path integral. I describe a probability-conserving $i\e$ prescription from a path-integral implementation of Lindblad evolution. I also explain how to generalize the formalism to accommodate out-of-time-ordered correlators (OTOCs), leading to a Larkin-Ovchinnikov path integral. My goal is to provide step-by-step calculations of path integrals associated to the harmonic oscillator.
\end{abstract}
%%%%
%%%%
\tableofcontents
%%%%
%%%%
\pagebreak
%%%%
\section{Introduction}\label{sec:intro}
%%%%
A typical review of the Schwinger-Keldysh formalism \cite{schwinger1961, keldysh1965a} pitched to high-energy theorists goes something like this \cite{mottola_large-N, calzetta1987}. Remember the Feynman path integral? It only works if the vacuum in the far future is the same as the vacuum in the far past. But when a quantum field theory is placed on a dynamical background---like an expanding universe or an evaporating black hole---there is no guarantee that the vacuum in the future will be the same as the vacuum in the past. So the Feynman path integral must be replaced by something better.
\\\\
But that makes the formalism sound like witchcraft, when it is just statistical mechanics \cite{feynman1948}. 
\\\\
The Schwinger-Keldysh formalism is about constructing generating functions for expectation values instead of transition amplitudes. It has nothing to do with dynamics, equilibrium or otherwise.\footnote{It is true that most who study nonequilibrium dynamics are primarily concerned with expectation values, but that concern is a choice \cite{srednicki_chaos}.} It is also, in its path-integral incarnation, better attributed to Feynman and Vernon \cite{feynman_vernon}, whose approach I will follow and extend.\footnote{See Kamenev and Levchenko \cite{kamenev2009} for other historical references.} 
\\\\
To encapsulate my presentation within a consistent, unambiguous framework, I will start all the way from the path-integral decomposition itself, presenting the Schwinger-Keldysh formalism more or less in the style of Sections~6~and~7 of the textbook by Srednicki \cite{srednicki2007}. With out-of-time-ordered correlators (OTOCs) \cite{larkin1968} in the collective consciousness \cite{shenker_stanford, kitaev_soft_mode}, I will further construct a generating function for 4-point expectation values, which I will call the Larkin-Ovchinnikov path integral.\footnote{For an operator treatment with examples, see Aleiner et al.~\cite{aleiner2016}; for further physical interpretation, see Stanford \cite{stanford_manybody}. For another path-integral treatment, see Haehl et al.~\cite{haehl2019}}
\\\\
In this entire article I will consider only a single example: The harmonic oscillator. Again, and again, and again. I will review its Feynman path integral, then calculate its Schwinger-Keldysh and Larkin-Ovchinnikov path integrals for various choices of density matrix. 
\\\\
Building on all of that, I will review the path-integral implementation \cite{strunz1997, buchhold2016} of Lindblad evolution for open systems \cite{lindblad1976}; I will wrap up by formulating a probability-conserving version of the $i\e$ prescription from $S$-matrix theory. Although I will consider only single-particle quantum mechanics, I remind you that quantum mechanics is quantum field theory in $0+1$ dimensions. Everything in this paper should be understood with that mindset. %To generalize to higher dimensions, just append an additional label for Fourier modes---the resulting phenomenology will depend on the number of spatial dimensions, but the formalism itself will not. 
\\\\
Whether your affiliation is particle physics, cosmology, open systems, biology, or finance, I invite you to Bring Your Own Motivation. This is a paper about path integrals. I wish only to present a series of correct mathematical steps. 
\vfill
\pagebreak
%%%%
\section{Generating functions in quantum mechanics}\label{sec:generating functions}
%\\\\
Let $\hat q$ be a quantum-mechanical coordinate operator, and let $\hat H$ be the time-independent Hamiltonian.\footnote{I will generalize to Lindblad evolution in Sec.~\ref{sec:iepsilon}.} The Heisenberg-picture quantum field operator $\hat q(t)$ corresponding to the time-independent operator $\hat q$ is defined as the solution of the Heisenberg equation:
\begin{equation}\label{eq:heisenberg eq}
	\pa_t \hat q(t) = i[\hat H, \hat q(t)] \implies \hat q(t) = e^{\,i\hat H t} \hat q\, e^{-i\hat H t}\;.
\end{equation}
Let $|\psi\ket$ and $|\psi'\ket$ be arbitrary quantum states. The purpose of this paper is to study generating functions for quantities of the form
\begin{equation}\label{eq:objects}
	\bra \psi'| e^{-i\hat H t'}\;\hat q(t_n)\, ...\, \hat q(t_1)\; e^{\,i\hat H t}|\psi\ket\;.
\end{equation}
 When $|\psi'\ket \neq |\psi\ket$ and $t' \neq t$, those quantities are called transition amplitudes; when $|\psi'\ket = |\psi\ket$ and $t' = t$, they are called expectation values.\footnote{See p.~217 of Feynman \& Hibbs \cite{feynmanhibbs}. I am not sure whether the standard usage of ``transition amplitude'' includes the case of expectation value, but I treat the two as exclusive.} In both cases I will also call them correlation functions or correlators. 
\\\\
The generating functions will be expressed as path integrals. Whether those path integrals obviate any need to mention the operator expressions in Eqs.~(\ref{eq:heisenberg eq}) and~(\ref{eq:objects}) in the first place is not my concern.\footnote{See the introduction to Chapter 9 of Weinberg, Vol.~I \cite{weinberg1} for comments about unitarity whose significance I still do not fully appreciate.} 
\\\\
In Eq.~(\ref{eq:objects}), no relation among the times $t_1,...,t_n$ is assumed---if the operators are to be ordered in some way, I will include the appropriate ordering symbol. One theme of this review is the path-integral description of arbitrarily-ordered products of fields, not just of time-ordered products. 
\\\\
For transition amplitudes, it should always be understood that $t < t_1,...,t_n < t'$. For expectation values, there are two options: Either $t' = t < t_1,...,t_n$ or $t_1, ..., t_n < t' = t$. In this paper I will consider only the former. 
\subsection{Heisenberg picture and operator insertions}
Eq.~(\ref{eq:objects}) is a lot to unpack, and ultimately it will be defined by the examples that follow. But as a preliminary matter, I will set $n = 1$ and write down a transition amplitude and an expectation value. 
\subsubsection{Transition amplitude}
For a transition amplitude, write $t = t_0$, $t' = t_f$, $|\psi\ket = |\psi_0\ket$, $|\psi'\ket = |\psi_f\ket$, and insert $\hat 1 = \INT dq_0|q_0\ket\bra q_0|$ to the left of $|\psi_0\ket$, and $\hat 1 = \INT dq_f|q_f\ket \bra q_f|$ to the right of $\bra \psi_f|$. The result is
\begin{equation}
	\INT \!\!dq_f\INT \!\!dq_0\;\bra \psi_f|q_f\ket \bra q_0|\psi_0\ket\; \bra q_f|e^{-i\hat H t_f}\, \hat q(t_1)\, e^{\,i\hat H t_0}|q_0\ket\;.
\end{equation}
The primordial task in this paper will be to review the generating function for the object
\begin{equation}\label{eq:1pt}
	\bra q_f|e^{-i\hat H t_f}\;\hat q(t_1) \;e^{\,i\hat H t_0} |q_0\ket\;,
\end{equation}
which is the 1-point transition amplitude in the coordinate basis. I call that the primordial task because it will lead to the Feynman path integral over a finite interval, out of which all of the other path integrals are built. 
\\\\
To make sense of the time dependence in Eq.~(\ref{eq:1pt}), you may prefer to rewrite it as
\begin{equation}\label{eq:1pt amplitude with time dependence outside operator}
	\bra q_f|e^{-i(t_f-t_1)\hat H}\;\hat q\;e^{-i(t_1-t_0)\hat H}|q_0\ket\;. 
\end{equation}
With the evolution operator $\hat U(t) \equiv e^{-i t\hat H}$, you could interpret Eq.~(\ref{eq:1pt amplitude with time dependence outside operator}) in the Schrodinger picture as an off-diagonal matrix element of $\hat q$ between the time-dependent states $\hat U(t_1\!-\!t_0)|q_0\ket$ and $\hat U(t_f\!-\!t_1)^\da |q_f\ket$. 
\\\\
But I will work in the Heisenberg picture: Operators are functions of time, and states describing the system are not.\footnote{As Srednicki points out \cite{srednicki2007}, the object $|q,t\ket \equiv e^{\,i\hat H t}|q\ket$ is an instantaneous eigenstate of the quantum field $\hat q(t)$; that is, $\hat q(t)|q,t\ket = q |q,t\ket$. In that notation, Eq.~(\ref{eq:1pt}) would read $\bra q_f,t_f|\hat q(t_1)|q_0,t_0\ket$. But when attempting to weave a continuous thread between scattering theory and open systems, I found it distracting to countenance the faintest whiff of a time-dependent state. I have concluded that the notational convenience is usually not worth it, but I dabble (see Sec.~\ref{sec:decay}).\label{ft:time-dependence}} The source-free Feynman path integral is the evolution operator in the coordinate basis [see Eq.~(\ref{eq:evolution operator in coordinate basis})], and Eq.~(\ref{eq:1pt}) is what the Feynman path integral with sources generates [see Eq.~(\ref{eq:1pt amplitude})]. The operator $\hat q(t)$ corresponds to the path-integration variable $q(t)$, and, as far as I am concerned, the evolution operators $\hat U(t_f) = e^{-it_f\hat H}$ and $\hat U(t_0)^\da = e^{\,it_0\hat H}$ are there because the mathematics of path integrals over finite intervals says so. 
%%%
\subsubsection{Expectation value}
%%%
Now for an expectation value. Return to Eq.~(\ref{eq:objects}) with $n = 1$, set $t = t' = t_0$, $|\psi\ket = |\psi'\ket = |\psi_0\ket$, and insert $\hat 1 = \INT dq_0|q_0\ket\bra q_0|$ to the left of $|\psi_0\ket$, and $\hat 1 = \INT dq_0'|q_0'\ket \bra q_0'|$ to the right of $\bra\psi_0|$. The result is
\begin{align}\label{eq:expectation with n=1}
	&\INT \!\!dq_0'\INT \!\!dq_0\;  \bra q_0|\psi_0\ket \bra \psi_0|q_0'\ket\; \bra q_0'|e^{-i\hat H t_0}\;\hat q(t_1)\;e^{\,i\hat H t_0}|q_0\ket\;.
\end{align}
Next, insert $\hat 1 = e^{\,i\hat H t_f}e^{-i\hat H t_f} = \INT dq_f\;e^{\,i\hat H t_f}|q_f\ket \bra q_f|e^{-i\hat H t_f}$ to the left of $\hat q(t_1)$ to get
\begin{equation}\label{eq:expectation with n=1 v2}
	\INT \!\!dq_f\INT \!\!dq_0'\INT \!\!dq_0\;  \bra q_0|\psi_0\ket \bra \psi_0|q_0'\ket\; \bra q_0'|e^{-i\hat H t_0}\; e^{\,i\hat H t_f} | q_f\ket \bra q_f|e^{-i\hat H t_f}q(t_1)\;e^{\,i\hat H t_0}|q_0\ket\;.
\end{equation}
See how Eq.~(\ref{eq:1pt}) emerges? That is how transition amplitudes can be used to build expectation values. And that is the key to unlock path integrals that generate arbitrarily-ordered products of fields. 
\\\\
Note that $\hat\rho \equiv |\psi_0\ket\bra \psi_0|$ is the density matrix for the system in state $|\psi_0\ket$. In terms of a density matrix, the expectation-value version of Eq.~(\ref{eq:objects}) would read
\begin{equation}\label{eq:unordered n-pt expectation}
	\tr\left[ \hat q(t_n)\,...\, \hat q(t_1)\;e^{\,i\hat H t_0}\hat\rho\;e^{-i\hat H t_0}\right]\;,\;\; t_1,...,t_n > t_0\;.
\end{equation} 
I will continue to call that an expectation value regardless of whether the state is pure. 
\\\\
Once again, I will comment on the time dependence of the operators. Given Eq.~(\ref{eq:heisenberg eq}), you could, if you wish, insert a factor of $\hat 1 = e^{\,i\hat H t_0}e^{-i\hat H t_0}$ to the left of each $\hat q(t_i)$, for $i = 1,...,n-1$, and rewrite Eq.~(\ref{eq:unordered n-pt expectation}) as $\tr\left[\hat q(t_n\!-\!t_0)\,...\,\hat q(t_1\!-\!t_0)\,\hat\rho\right]$. But the mathematics of path integrals guides against doing so: The path-integration variable $q(t)$ depends only on the parameter $t$, not on both $t$ and its lower bound $t_0$. 
\\\\
Alternatively, e.g., for $n = 1$, you could once again define $\hat U(t) = e^{-i\hat H t}$ and rewrite the 1-point expectation value as
\begin{equation}
	\tr\left[\hat q(t_1)\;e^{\,i\hat H t_0}\hat\rho\; e^{-i\hat H t_0}\right] = \tr\left[ \hat q\;\hat U(t_1\!-\!t_0)\hat\rho\, \hat U(t_1\!-\!t_0)^\da\right]\;.
\end{equation}
That could be interpreted in the Schrodinger picture as the expectation value of $\hat q$ in the time-dependent state described by a density matrix $\hat U(t_1\!-\!t_0)\hat\rho\, \hat U(t_1\!-\!t_0)^\da$. But as I said before, I will insist on the Heisenberg picture: The density matrix for the system describes the system's state, and the system's state does not evolve. 
\\\\
Now that the formulas have marinated a bit, I will derive their path-integral expressions. 
%%%%
\subsection{Transition amplitudes}
%%%%
I will begin with the 1-point transition amplitude\footnote{Instructors usually begin with the $0$-point amplitude, then explore time ordering from higher-point functions. Instead I begin with the 1-point amplitude, because its path-integral decomposition is only slightly more complicated and yet demonstrates the inherent time ordering of operator insertions. Two calculations for the price of one.} in Eq.~(\ref{eq:1pt}), repeated below for convenience: 
\begin{equation}
	\bra q_f|e^{-i\hat H t_f}\;\hat q(t_1) \;e^{\,i\hat H t_0} |q_0\ket\;.
\end{equation}
\subsubsection{Path-integral decomposition}\label{sec:path-integral decomposition}
\begin{figure*}
	\centering
	\frame{\begin{tabular}{lcc}
			\\
			\includegraphics[width=5cm]{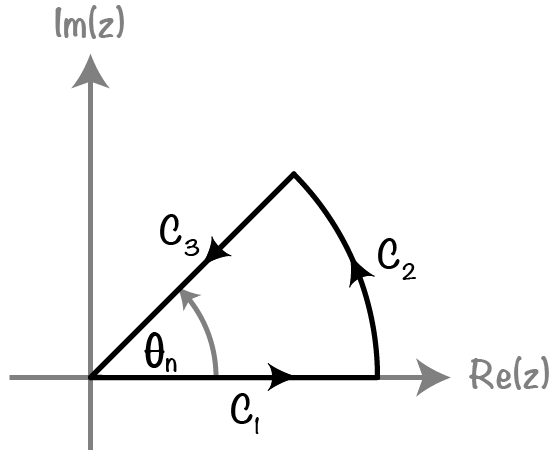}\\
	\end{tabular}}
	\caption{Rotate the complex Gaussian integral into a real one.}
	\label{fig:contour}
\end{figure*}

Split the time interval into an arbitrarily large number $N$ of arbitrarily small steps $\e$:
\begin{equation}
	T \equiv t_f \!-\! t_0 \equiv N \e\;.
\end{equation}
Let $n$ be the number of steps required to get from $t_0$ to $t_1$:
\begin{equation}\label{eq:t1-t0=ne}
	t_1\!-\!t_0 \equiv n \e\;.
\end{equation}
Although I will keep $t_f$ and $t_0$ finite, the goal is to calculate correlation functions of fields at \textit{any} intermediate times, allowing in principle the limits $t_0 \to -\infty$ and $t_f \to +\infty$. So $n$ and $N$$-$$n$ should also be taken large. 
\\\\
Insert a copy of $\hat 1 = \INT dq~ |q\ket \bra q|$ between each instance of $e^{\,i\hat H\e}$:
\begin{align}
	&\bra q_f|e^{-i\hat H t_f} \hat q(t_1)\, e^{i\hat H t_0}|q_0\ket = \bra q_f|e^{-i(t_f-t_1)\hat H} \hat q\, e^{-i(t_1-t_0)\hat H}|q_0\ket = \bra q_f| \underbrace{e^{-i\e \hat H} ...\, e^{-i\e \hat H}}_{N-n \text{ copies}} \hat q \underbrace{e^{-i\e \hat H}...\, e^{-i\e \hat H}}_{n\text{ copies}}|q_0\ket \nonumber\\
	&= \INT \!\! dq_{N-1}\, ... \INT\!\!  dq_1\; \bra q_f|e^{-i\e \hat H}|q_{N-1}\ket \, ...\, \bra q_{n+2}| e^{-i\e \hat H} |q_{n+1}\ket \;\times \nonumber\\
	&\qquad\qquad \qquad\qquad\qquad\qquad\qquad\bra q_{n+1}|e^{-i\e \hat H} \underbrace{\hat q| q_n \ket}_{q_n|q_n\ket}  \bra q_n|e^{-i\e \hat H} |q_{n-1}\ket ... \bra q_1|e^{-i\e \hat H} |q_0\ket \nonumber\\
	&= \INT \!\! dq_{N-1} ... \INT\!\!  dq_1 \;q_n\; \prod_{j\,=\,0}^{N-1} \bra q_{j+1} |e^{-i\e \hat H}|q_j\ket\;,\;\; q_N \equiv q_f\;.
\end{align}
Now specialize to Hamiltonians of the form
\begin{equation}
	\hat H = \half \hat p^2 + V(\hat q)\;,\;\; [\hat q,\hat p] = i\hat 1\;.
\end{equation}
Even though I will not end up doing anything with it in this paper, I want to keep track of where I choose to evaluate the field in each infinitesimal interval.\footnote{See the Appendices of Matacz \cite{matacz2000} for an example in which that matters. A last-minute reference sweep revealed that he too wrote about Feynman-Vernon path integrals for oscillators \cite{matacz_oscillators}.} So I will express the incremental evolution operator as\footnote{To go further for the special case $V(\hat q)$~\!\!$=$\!\!~$\half m^2 \hat q^2$, write $e^{-i\half \hat p^2}$~\!\!\!\!$=$\!\!~$e^{-i\fourth \hat p^2}e^{-i\fourth \hat p^2}$, and use the Baker-Campbell-Hausdorff formula multiple times, as follows. \\ \indent First note that $[\hat q^2,\hat p^2]$~$=$~$2i\{\hat q,\hat p\}$. Then use $e^A e^B$~$=$~$e^{A+B+\half [A,B]+...}$ with $A$~$=$~$-i\al \e V(\hat q)$ and $B$~$=$~$-i\e\fourth \hat p^2$ to get $e^{-i\al \e V(\hat q)}e^{-i\e\fourth \hat p^2}$~$=$~$e^{-i\e\left(\fourth\hat p^2 + \half\al m^2 \hat q^2\right)-i\al \e^2 \tfrac{1}{8}m^2\{\hat q,\hat p\} + O(\e^3)}$~$\equiv$~$e^X$. \\ \indent Similarly, use $e^B e^{A'}$~$=$~$e^{B+A'+\half[B,A']+...}$ with $B$~$=$~$-i\e\fourth \hat p^2$ (again), and $A'$~$=$~$-i(1$$-$$\al)\e V(\hat q)$ to get $e^{-i\e\fourth \hat p^2}e^{-i(1-\al)\e V(\hat q)}$~$=$~$e^{-i\e\left(\fourth\hat p^2 + (1-\al)\half m^2 \hat q^2\right) + i(1-\al)\e^2 \tfrac{1}{8}m^2 \{\hat q,\hat p\}+O(\e^3)}$~$\equiv$~$e^Y$. \\ \indent Finally, use $e^X e^Y$~$=$~$e^{X+Y+\half [X,Y] + ... }$ to get $e^{-i\al \e V(\hat q)}e^{-i\e\half\hat p^2} e^{-i(1-\al)\e V(\hat q)}$~$=$~$e^{-i\e\hat H - i\left(\al-\half\right)\e^2\half m^2\{\hat q,\hat p\} + O(\e^3)}$. For the special case $\al = \half$---that is, for midpoint regularization---the decomposition is valid to $O(\e^3)$. \\ \indent As a mildly interesting aside, I point out that $e^{-i\e\hat H}$~$\approx$~$\left(\hat 1-i\e\half \hat H\right)\left(\hat 1+i\e \half \hat H\right)^{-1}$ is also unitary and accurate to $O(\e)$ \cite{discreteQM}. But interpreting the resulting path integral and its discretization error in terms of an effective action and the renormalization group seems for the birds.}
\begin{equation}
	e^{-i\e\hat H} = e^{-i\al \e V(\hat q)} e^{-i\e\half \hat p^2} e^{-i(1-\al)\e V(\hat q)}e^{\,O(\e^2)}\;,\;\; \al \in [0, 1]\;.
\end{equation}
Let $|q\ket$ be an eigenstate of $\hat q$, and let $|p\ket$ be an eigenstate of $\hat p$:
\begin{equation}
	\hat q|q\ket = q|q\ket\;,\;\; \hat p|p\ket = p|p\ket\;,\;\; \bra q|p\ket = e^{\,ipq}\;.
\end{equation}
The 0-point transition amplitude over interval $\e$ is:
\begin{align}
	C_{j+1,j} &\equiv \bra q_{j+1}|e^{-i\e\hat H}|q_j\ket \approx \bra q_{j+1}|e^{-i\al\e V(\hat q)} e^{-i\e\half \hat p^2} e^{-i(1-\al)\e V(\hat q)}|q_j\ket \nonumber\\
	&= \INT \frac{dp_j}{2\pi} \;e^{-i\e\half p_j^2}\, \bra q_{j+1}|e^{-i\al\e V(\hat q)}|p_j\ket \bra p_j|e^{-i(1-\al)\e V(\hat q)}|q_j\ket \nonumber\\
	&= \INT \frac{dp_j}{2\pi}\;e^{-i\e \half p_j^2}\; e^{-i\al \e V(q_{j+1})} \bra q_{j+1}|p_j\ket\; e^{-i(1-\al)\e V(q_j)} \bra p_j|q_j\ket  \nonumber\\
	&= \frac{1}{2\pi} e^{-i\e\left[\al V(q_{j+1}) + (1-\al)V(q_j)\right]}\INT\!\! dp_j\;e^{-i\e \half p_j^2}\, e^{i\left(q_{j+1}-q_j\right)p_j} \nonumber\\
	%&= \frac{1}{2\pi} e^{-i\e\left[\al V(q_{j+1}) + (1-\al)V(q_j)\right]}\INT\!\! dp_j\;e^{-i\e\half\left[p_j^2-2\left(\frac{q_{j+1}-q_j}{\e}\right)p_j \right]} \nonumber\\
	&= \frac{1}{2\pi}\, e^{-i\e\left[\al V(q_{j+1}) + (1-\al)V(q_j)\right]}\INT\!\! dp_j\;e^{-i\e\half\left\{ \left[p_j-\left(\frac{q_{j+1}-q_j}{\e}\right)\right]^2-\left(\frac{q_{j+1}-q_j}{\e}\right)^2 \right\}} \nonumber\\
	&= \frac{1}{2\pi}\,e^{\;i\e\left\{ \half \left(\frac{q_{j+1}-q_j}{\e}\right)^2 - \left[\al V(q_{j+1}) + (1-\al) V(q_j)\right] \right\} } \INT \!\! dp\;e^{-i\e\frac{1}{2}p^2}\;. \label{eq:steps for C_j+1,j}
\end{align}
Notice that I did not yet do the integral: Shifting by a constant in the complex plane does not merit review, but rotating to get the phase right does. 
\\\\
Consider the contour integral
\begin{equation}\label{eq:I_C}
	I_C(a,n) \equiv \int_C dz\;e^{\,ia z^n}\;,
\end{equation}
with $a$ real and positive, $n$ an integer greater than or equal to 1, and $C = C_1 + C_2 + C_3$ the contour depicted in Fig.~\ref{fig:contour}. 
(The case in Eq.~(\ref{eq:steps for C_j+1,j}) is with $a < 0$, but it is cleaner to make the point with $a > 0$.) The obvious part is that the integral converges with increasing radius only in the upper half-plane; the less obvious part is that one should only integrate up to $1/n^{\text{th}}$ of the upper-right quadrant: 
\begin{equation}
	0 < \ta < \ta_n \equiv \frac{\pi}{2n}\;.
\end{equation}
The integration variable is parametrized along each segment of the contour as follows:
\begin{align}
	&C_1:\qquad z\,=\,x\;,\;\;x\;\text{ from }\;0 \text{ to } R\;; \nonumber\\
	&C_2:\qquad z\,=\,R\, e^{\,i\ta}\;,\;\; \ta\;\text{ from }\;0 \text{ to } \ta_n\;; \nonumber\\
	&C_3:\qquad z\,=\,r\,e^{\,i\ta_n}\;,\;\; r\;\text{ from }\;R \text{ to } 0\;.
\end{align}
The integral in Eq.~(\ref{eq:I_C}) encloses no poles and therefore evaluates to zero:
\begin{equation}
	I_C(a,n) = \int_{0}^R\!\! dx\;e^{\,ia x^n}+ \int_{0}^{\ta_n}\!\! R\,e^{\,i\ta}i\,d\ta\;e^{ia(Re^{i\ta})^n} + \int_{R}^{0}\!\! dr\;e^{\,i\ta_n}\;e^{\,ia(re^{i\ta_n})^n}\;=\;0.
\end{equation} 
The task is to relate the complex integral to a real integral, so I want the imaginary part of
\begin{equation}
	e^{\,ia (re^{\,i\ta_n})^n} = e^{iar^n \cos(n\ta_n)}e^{-ar^n\sin(n\ta_n)}
\end{equation}
to be zero:
\begin{equation}
	\cos(n\ta_n) \equiv 0 \implies \ta_n = \frac{\pi}{2n}(2k+1)\;,
\end{equation}
with $k$ any integer. But I must also maintain convergence of the real part:
\begin{equation}
	\sin(n\ta_n) > 0 \implies 0 < \ta_n < \frac{\pi}{n}\;.
\end{equation}
That fixes $k = 0$ and hence $\ta_n = \frac{\pi}{2n}$. Taking $R \to \infty$ therefore produces the desired relation:
\begin{equation}
	\int_0^\infty \!\! dx\;e^{\,ia x^n} = e^{\,i\ta_n}\int_0^\infty \!\! dr\;e^{-ar^n}\;.
\end{equation}
Now I can repeat the analogous steps for the negative domain of $x$, take $n = 2$, and combine the results to recover the usual, sloppily asserted result for the complex Gaussian integral:\footnote{I will explain why I fixated on this, other than for basic dignity. For real-valued Gaussian integrals, the standard method is to square the integral, switch to polar coordinates, then square-root the result. For the complex Gaussian integral, $I_C(a,1) = 0$ would then produce $\frac{i}{a}$, leading to the correct answer. But naively considering $I_C(a,2) = 0$ with a contour $C$ over the whole upper half-plane would erroneously suggest an overall phase of $e^{\,i\pi/2} = i$ instead of $e^{\,i\pi/4} = \sqrt{i}$.}
\begin{equation}
	\int_{-\infty}^\infty \!\!dx\;e^{\,ia x^2} = e^{\,i\pi/4}\int_{-\infty}^{\infty}\!\!dr\;e^{-ar^2} = \sqrt{\frac{i\pi}{a}}\;.
\end{equation}
With the analog of that for negative $a$, I can write down the infinitesimal matrix element:
\begin{equation}\label{eq:C_j+1,j}
	C_{j+1,j} = \frac{1}{\sqrt{2\pi i \e}}\,e^{\;i\e\left\{ \half \left(\frac{q_{j+1}-q_j}{\e}\right)^2 - \left[\al V(q_{j+1}) + (1-\al) V(q_j)\right] \right\} }\;.
\end{equation}
%\\\\
I will now choose $\al = 0$ (endpoint regularization) and proceed as usual. Defining $\dot q_j \equiv (q_{j+1}-q_j)/\e$ and
\begin{equation}
	\la(q) \equiv \half \dot q^2 - V(q)\;,
\end{equation}
I arrive at the path-integral representation of the 1-point transition amplitude: 
\begin{align}\label{eq:path integral for 1pt amplitude}
	\bra q_f|e^{-i\hat H t_f} \hat q(t_1)\, e^{\,i\hat H t_0} |q_0\ket &= (2\pi i\e)^{-N/2} \INT dq_{N-1} ... \INT dq_1 \;q_n\;e^{\,i\e \sum_{j\,=\,0}^{N-1} \la (q_j)} \nonumber\\
	&\equiv \int_{q(t_0)\,=\,q_0}^{q(t_f)\,=\,q_f}\!\!\!\!\Ds q(\cdot)\;\;q(t_1)\;\;e^{\,i \int_{t_0}^{t_f}dt\;\la (q(t))}\;.
\end{align}
\subsubsection{Feynman path integral}
It is standard practice in probability and statistics to calculate moments of a distribution by introducing an auxiliary variable and taking derivatives with respect to it \cite{generatingfunctionology}. Scrutinizing Eq.~(\ref{eq:path integral for 1pt amplitude}) from that point of view suggests introducing an auxiliary field $J(t)$ and defining the following generating function:
\begin{equation} \label{eq:feynman}
	Z(q_0,t_0;q_f, t_f|J(\cdot)) \equiv \int_{q(t_0)\,=\,q_0}^{q(t_f)\,=\,q_f}\!\!\!\!\Ds q(\cdot)\;e^{\,i \int_{t_0}^{t_f}dt\;\left[\la(q(t)) \,+\,J(t)\, q(t)\right]}\;.
\end{equation}
For notational convience, I will usually suppress the time arguments on the left-hand side.\footnote{Further comments about notation. First, I prefer to put $q_0$ on the left and $q_f$ on the right, whereas most people seem to prefer the opposite. Second, I will usually just write $J$ instead of $J(\cdot)$ in $Z(q_0,q_f|J(\cdot))$. Third, as a related matter, I do not use bracket notation for functionals---$Z(q_0,q_f|J)$ is a functional of $J$ but a function of $q_0$ and $q_f$, so I would have to write something like $Z(q_0,q_f)[J(\cdot)]$, which would look ridiculous. I also say ``generating function'' instead of ``generating functional,'' because it sounds better and there is zero risk of confusion.} In terms of Eq.~(\ref{eq:feynman}), the 1-point amplitude is
\begin{equation} \label{eq:1pt amplitude}
		\bra q_f|e^{-i\hat H t_f} \hat q(t_1)\, e^{\,i\hat H t_0} |q_0\ket = \left. -i\frac{\del}{\del J(t_1)} Z(q_0,q_f|J)\, \right|_{J\,=\,0}\;.
\end{equation}
The derivation leading to Eq.~(\ref{eq:path integral for 1pt amplitude}) also makes clear that, for $n > 1$, $Z(q_0,q_f|J)$ generates amplitudes of only time-ordered products of fields:
\begin{equation} \label{eq: npt amplitude}
	\left. (-i)^n \frac{\del^n}{\del J(t_1) ... \del J(t_n)} Z(q_0,q_f|J)\, \right|_{J\,=\,0} = \bra q_f|e^{-i\hat H t_f}\; \mathcal T\! \left( \hat q(t_1)\; ...\; \hat q(t_n) \right) e^{\,i\hat H t_0}|q_0\ket\;,
\end{equation}
with $\mathcal T(...)$ the time-ordering symbol.\footnote{For two operators $\hat A(t)$ and $\hat B(t)$, the time-ordering symbol is $\mathcal T\!\left(\hat A(t)\hat B(t')\right) \equiv \Ta(t\minus t') \hat A(t)\hat B(t')$ $+$ $\Ta(t'\minus t)\hat B(t') \hat A(t)$. It is just mathematical shorthand whose general definition for $t_1,...,t_n$ is too annoying to write.\label{ft:time-ordering symbol}} 
\\\\
I will refer to the amplitude-generating function in Eq.~(\ref{eq:feynman}) as the Feynman path integral, even though the history is more complicated, and even though Feynman wrote down many path integrals. 
\subsubsection{Feynman path integral and wavefunction}\label{sec:path integral and wavefunction}
Often lost in the cauldron of graduate school is the relation between the Feynman path integral and the wavefunction. Before continuing, I just want to point out something simple. 
\\\\
The evolution operator $\hat U(t\!-\!t_0) = e^{-it\hat H}e^{\,it_0\hat H}$ has coordinate-basis matrix elements
\begin{equation}\label{eq:evolution operator in coordinate basis}
	\bra q|\hat U(t\!-\!t_0)|q_0\ket = Z(q_0,t_0;q,t|0) = \int_{q(t_0)\,=\,q_0}^{q(t)\,=\,q}\!\!\!\!\Ds q(\cdot)\;e^{\,i\int_{t_0}^{t}dt'\;\la(q(t'))}\;.
\end{equation}
The source-free Feynman path integral is the \textit{evolution operator} in the coordinate basis. 
\\\\
Meanwhile, a system in a state $|\psi\ket$ has wavefunction $\psi(q) \equiv \bra q|\psi\ket$. That is a fixed function for all time (again, Heisenberg picture). I could, for fun, combine that fixed function with Eq.~(\ref{eq:evolution operator in coordinate basis}) and define
\begin{equation}
	\psi(q,t) \equiv \INT dq_0\;\bra q|\hat U(t\!-\!t_0)|q_0\ket \bra q_0|\psi\ket = \INT dq_0\;\psi(q_0)\;\int_{q(t_0)\,=\,q_0}^{q(t)\,=\,q}\!\!\!\!\Ds q(\cdot)\;e^{\,i\int_{t_0}^{t}dt'\;\la(q(t'))}\;.
\end{equation}
That is the Schrodinger-picture wavefunction for the system. So the Feynman path integral is \textit{related to} the Schrodinger-picture wavefunction, but it \textit{is} the evolution operator. 
%%%%
\subsection{Expectation values}
Instead of the left-hand side of Eq.~(\ref{eq:1pt amplitude}), consider Eq.~(\ref{eq:unordered n-pt expectation}) with $n = 1$:
\begin{equation} \label{eq:1pt expectation}
	%\bra q_0|e^{-i\hat H t_0} \hat q(t_1)\, e^{\,i\hat H t_0} | q_0\ket\;.
	\tr\left(\hat q(t_1)\,e^{\,i\hat H t_0}\hat\rho\,e^{-i\hat H t_0}\right)\;.
\end{equation}
Or consider the special case $\hat\rho = |q_0\ket \bra q_0|$, a pure state of definite coordinate value, in which case Eq.~(\ref{eq:1pt expectation}) would become $\bra q_0|e^{-i\hat H t_0} \hat q(t_1)\,e^{\,i\hat H t_0}|q_0\ket$. That cannot be generated from Eq.~(\ref{eq:feynman}), since setting $t_f = t_0$ there would just leave you with 1.\footnote{Skipping ahead to Eqs.~(\ref{eq:operator form of feynman path integral}) and~(\ref{eq:operator form of schwinger-keldysh}), I should point out that $\bra q_0|e^{-i\hat H t_0}\,\mathcal T\!\left(e^{-i\singleint J(t)\hat q(t)}\right)e^{\,i\hat H t_0}|q_0\ket$ could, technically, generate Eq.~(\ref{eq:1pt expectation}) with $\hat\rho = |q_0\ket\bra q_0|$. But good luck formulating effective field theory based on that, except as the $J' = 0$ limit of Eq.~(\ref{eq:keldysh}).} No problem, nothing mysterious: Just find another generating function. 
\subsubsection{Schwinger-Keldysh path integral}
I have already shown in Eq.~(\ref{eq:expectation with n=1 v2}) that the expectation value in Eq.~(\ref{eq:1pt expectation}) is built from Eq.~(\ref{eq:1pt}) and the conjugate of the zero-point amplitude---the new ingredient is Eq.~(\ref{eq:1pt amplitude}), the Feynman path integral. Inserting that into Eq.~(\ref{eq:expectation with n=1 v2}) leads to a new generating function: 
\begin{align}
	&\tr\left(\hat q(t_1)\, e^{\,i\hat H t_0} \hat \rho\, e^{-i\hat H t_0}\right) \nonumber\\
	&= \INT\!\! dq_f \INT\!\! dq_0 \INT\!\! dq'_0 \;\bra q_f| e^{-i\hat H t_f} \hat q(t_1)\, e^{\,i\hat H t_0} |q_0\ket \underbrace{\bra q_0|\hat \rho | q'_0\ket}_{\equiv\; \rho(q_0,q_0')} \bra q'_0| e^{-i\hat H t_0} e^{\,i\hat H t_f} |q_f\ket \nonumber \\
	&= \INT \!\!dq_f \INT \!\!dq_0 \INT \!\!dq'_0\; \left. (-i)\frac{\del}{\del J(t_1)} Z(q_0,q_f|J) \right|_{J\,=\,0} \rho(q_0,q'_0) \left. Z(q'_0,q_f|J')^* \phantom{\frac{\del}{\del}}\!\!\!\!\right|_{J'\,=\,0} \nonumber\\
	&=\left. (-i)\frac{\del}{\del J(t_1)}\INT \!\!dq_f \INT \!\!dq_0 \INT \!\!dq'_0\;\rho(q_0,q'_0)\;Z(q_0,q_f|J)\;Z(q'_0,q_f|J')^* \right|_{J\,=\,J'\,=\,0}\;.
\end{align}
If I define
\begin{equation}\label{eq:keldysh}
	Z(J,J') \equiv \INT dq_f \INT dq_0 \INT dq'_0\;\rho(q_0,q'_0)\;Z(q_0,q_f|J)\;Z(q'_0,q_f|J')^*\;,
\end{equation}
then I will find
\begin{equation}
	\tr\left(\hat q(t_1)\;e^{\,i\hat H t_0} \hat\rho\, e^{-i\hat H t_0} \right) = \left.-i\frac{\del}{\del J(t_1)} Z(J,J') \right|_{J\,=\,J'\,=\,0}\;.
\end{equation}
%\\\\
The expectation-value-generating function $Z(J,J')$ in Eq.~(\ref{eq:keldysh}) is the Schwinger-Keldysh path integral. No highfalutin appeal to nonequilibrium dynamics, no spells. 
\subsubsection{Two-point correlators}\label{sec:2pt correlators}
Take two derivatives of the generating function $Z(q_0,q_f|J)$ in Eq.~(\ref{eq:feynman}) to get a time-ordered transition amplitude: 
\begin{equation}\label{eq:time-ordered 2pt amplitude}
	\bra q_f|e^{-i\hat H t_f}\,\mathcal T\!\left(\hat q(t_1)\hat q(t_2)\right)e^{\,i\hat H t_0}|q_0\ket = \left.(-i^2) \frac{\del^2}{\del J(t_1)\del J(t_2)} Z(q_0,q_f|J) \right|_{J\,=\,0}\;.
\end{equation}
Take two derivatives of the generating function $Z(J,J')$ in Eq.~(\ref{eq:keldysh}) to get a collection of expectation values: 
\begin{align}
	&\tr\left[ \mathcal T\!\left( \hat q(t_1)\hat q(t_2)\right) e^{\,i\hat H t_0} \hat\rho\, e^{-i\hat H t_0}\right] = \left.(-i)^2 \frac{\del^2}{\del J(t_1) \del J(t_2)} Z(J,J') \right|_{J\,=\,J'\,=\,0}\;, \nonumber\\
	&\tr\left[ \mathcal T^*\! \left(\hat q(t_1)\hat q(t_2)\right) e^{\,i\hat H t_0} \hat\rho\, e^{-i\hat H t_0} \right] = \left.(+i)^2 \frac{\del^2}{\del J'(t_1) \del J'(t_2)} Z(J,J') \right|_{J\,=\,J'\,=\,0}\;, \label{eq:time-ordered 2pt expectation values}\\
	&\tr\left(\hat q(t_2)\hat q(t_1)\, e^{\,i\hat H t_0} \hat\rho\, e^{-i\hat H t_0} \right) = \left.\frac{\del^2}{\del J(t_1) \del J'(t_2)} Z(J,J') \right|_{J\,=\,J'\,=\,0}\;, \nonumber\\
	&\tr\left(\hat q(t_1)\hat q(t_2) \,e^{\,i\hat H t_0} \hat\rho\, e^{-i\hat H t_0} \right) = \left. \frac{\del^2}{\del J'(t_1) \del J(t_2)} Z(J,J') \right|_{J\,=\,J'\,=\,0}\;. \label{eq:unordered 2pt expectation values}
\end{align}
In Eq.~(\ref{eq:time-ordered 2pt expectation values}), $\mathcal T^*$ is the antitime-ordering symbol.\footnote{For two operators $\hat A(t)$ and $\hat B(t)$, the antitime-ordering symbol is $\mathcal T^*\!\left(\hat A(t)\hat B(t)\right) \equiv \Ta(t\minus t')\hat B(t')\hat A(t)$ $+$ $\Ta(t'\minus t) \hat A(t)\hat B(t')$. Compare to Footnote~\ref{ft:time-ordering symbol}. By the way, I struggle with whether to write ``anti-time-ordering'' or ``antitime-ordering''---does the ``anti'' modify ``time-ordering,'' or is the operator ordered in ``antitime''?} The expectation values in Eq.~(\ref{eq:time-ordered 2pt expectation values}) are called the Feynman and Dyson correlators, and the expectation values in Eq.~(\ref{eq:unordered 2pt expectation values}) are called the Wightman correlators. 
\subsubsection{The influence phase}
Logarithms of generating functions are given special names: The log of the Feynman path integral is called the effective action, and the log of the thermal partition function is called the free energy. The log of the Schwinger-Keldysh path integral is called the \textit{influence phase}, 
\begin{equation}
	\Phi(J,J') \equiv -i\ln Z(J,J')\;.
\end{equation} 
Feynman and Vernon \cite{feynman_vernon} called it that\footnote{Dueling precedents seem to have led me to the sociologically befuddled terminology of ``Schwinger-Keldysh influence phase,'' but what can I do. If it were up to me I would name the whole formalism after Feynman and Vernon, since Keldysh did not use path integrals, and Schwinger's paper is unreadable.\label{ft:influence phase}} because they interpreted $J$ as the coordinate of an additional oscillator, whereas in this paper I treat $J$ only as an auxiliary variable for probing the field.
\\\\
The influence phase has two important general properties. To derive the first property, insert the operator form of the amplitude-generating function, 
\begin{equation}\label{eq:operator form of feynman path integral}
	Z(q_0,q_f|J) \equiv \int_{q(t_0)\,=\,q_0}^{q(t_f)\,=\,q_f}\!\!\!\! \Ds q(\cdot)\;e^{\,i \int_{t_0}^{t_f} \left[ \la(q(t)) + J(t) q(t)\right]} = \bra q_f | e^{-it_f\hat H} \,\mathcal T\!\left(e^{\,i\int_{t_0}^{t_f} dt\, J(t)\hat q(t)} \right) e^{\,it_0\hat H} | q_0\ket \;,
\end{equation}
into the definition of the expectation-value-generating function in Eq.~(\ref{eq:keldysh}). If the density matrix is normalized, then setting $J'=J$ will reduce the generating function to 1:
\begin{equation}\label{eq:Z(J,J)=1}
	Z(J,J) = \tr\,\hat\rho = 1\;.
\end{equation}
So any influence phase will satisfy
\begin{equation}\label{eq:influence phase property 1}
	\Phi(J,J) = 0\;.
\end{equation}
To derive the second property, return to Eq.~(\ref{eq:keldysh}) itself, regardless of how you choose to express $Z(q_0,q_f|J)$, and exchange the sources:
\begin{align}
	Z(J',J) &= \INT dq_f \INT dq_0\INT dq_0'\;\rho(q_0,q_0')\;Z(q_0,q_f|J')\,Z(q_0',q_f|J)^* \nonumber\\
	&= \INT dq_f \INT dq_0'\INT dq_0\;\rho(q_0',q_0)\;Z(q_0',q_f|J')\,Z(q_0,q_f|J)^* \nonumber\\
	&= \left( \INT dq_f\INT dq_0' \INT dq_0\;\rho(q_0',q_0)^*\;Z(q_0',q_f|J')^*\,Z(q_0,q_f|J)\right)^*\;.
\end{align}
Since the density matrix is hermitian---i.e., $\rho(q_0',q_0)^* = \rho(q_0,q_0')$---the quantity in parentheses is the original generating function:
\begin{equation}
	Z(J',J) = Z(J,J')^*\;.
\end{equation}
So any influence phase will also satisfy
\begin{equation}\label{eq:influence phase property 2}
	\Phi(J',J) = -\Phi(J,J')^*\;.
\end{equation}
I will apply those properties in due time. But now I want to express the correlation functions from Eqs.~(\ref{eq:time-ordered 2pt expectation values}) and~(\ref{eq:unordered 2pt expectation values}) in terms of the influence phase. Taking one derivative with respect to $J(t)$ and another with respect to $J'(t')$ would produce
\begin{equation}
	\frac{\del^2 Z(J,J')}{\del J(t) \del J'(t')} = e^{\,i\Phi(J,J')}\left(i \frac{\del^2\Phi(J,J')}{\del J(t)\del J'(t')} - \frac{\del\Phi(J,J')}{\del J(t)}\frac{\del\Phi(J,J')}{\del J'(t')}\right)\;.
\end{equation}
All influence phases I calculate in this paper will satisfy
\begin{equation}
	\left.\frac{\del\Phi(J,J')}{\del J(t)}\right|_{J\,=\,J'\,=\,0} = \left.\frac{\del\Phi(J,J')}{\del J'(t)}\right|_{J\,=\,J'\,=\,0} = 0\;,
\end{equation}
in which case
\begin{equation}
	\left.\frac{\del^2 Z(J,J')}{\del J(t) \del J'(t')}\right|_{J\,=\,J'\,=\,0} = \left. i \frac{\del^2\Phi(J,J')}{\del J(t)\del J'(t')}\right|_{J\,=\,J'\,=\,0}\;.
\end{equation}
That is the key formula, leading to the desired expressions for the 2-point correlators:
\begin{align}
	&\tr\left[ \mathcal T\!\left( \hat q(t_1)\hat q(t_2)\right) e^{\,i\hat H t_0} \hat\rho\, e^{-i\hat H t_0}\right] = \left.-i \frac{\del^2 \Phi(J,J')}{\del J(t_1) \del J(t_2)} \right|_{J\,=\,J'\,=\,0}\;, \nonumber\\
	&\tr\left[ \mathcal T^*\! \left(\hat q(t_1)\hat q(t_2)\right) e^{\,i\hat H t_0} \hat\rho\, e^{-i\hat H t_0} \right] = \left.-i \frac{\del^2 \Phi(J,J')}{\del J'(t_1) \del J'(t_2)} \right|_{J\,=\,J'\,=\,0}\;, \label{eq:time-ordered 2pt expectation values in terms of influence phase}\\
	&\tr\left(\hat q(t_2)\hat q(t_1)\, e^{\,i\hat H t_0} \hat\rho\, e^{-i\hat H t_0} \right) = \left. i\frac{\del^2 \Phi(J,J')}{\del J(t_1) \del J'(t_2)} \right|_{J\,=\,J'\,=\,0}\;, \nonumber\\
	&\tr\left(\hat q(t_1)\hat q(t_2)\, e^{\,i\hat H t_0} \hat\rho\, e^{-i\hat H t_0} \right) = \left. i\frac{\del^2 \Phi(J,J')}{\del J'(t_1) \del J(t_2)} \right|_{J\,=\,J'\,=\,0}\;. \label{eq:unordered 2pt expectation values in terms of influence phase}
\end{align}
\subsubsection{Aside: Wightman transition amplitudes}
Neurosis beckons a brief rewinding to Sec.~\ref{sec:2pt correlators}: What if I wanted to generate a 2-point transition amplitude 
\begin{equation}
	\bra q_f|e^{-i\hat H t_f}\, \hat q(t_2)\,\hat q(t_1)\, e^{\,i\hat H t_0}|q_0\ket\; ?
\end{equation}
%\\\\
I would need to go forward in time with one source, then again go forward in time with a different source, with both sources defined over the whole range between $t_0$ and $t_f$.\footnote{I thank Paolo Glorioso for helping me think through this.} To do that, it seems I must introduce an intermediate branch that goes backward: 
\begin{align}
	&\bra q_f|e^{-i\hat H t_f} \hat q(t_2)\hat q(t_1) e^{\,i\hat H t_0}|q_0\ket = \bra q_f|e^{-i\hat H t_f} \hat q(t_2) e^{\,i\hat H t_0} e^{-i\hat H t_0}e^{\,i\hat H t_f}e^{-i\hat H t_f}\hat q(t_1) e^{\,i\hat H t_0} |q_0\ket \nonumber\\
	&= \INT dq'_0 \INT dq'_f\; \bra q_f|e^{-i\hat H t_f}\hat q(t_2) e^{\,i\hat H t_0}|q'_0\ket \bra q'_0| e^{-i\hat H t_0} e^{\,i\hat H t_f} |q'_f\ket \bra q'_f| e^{-i\hat H t_f} \hat q(t_1) e^{\,i\hat H t_0} |q_0\ket \nonumber\\
	&= \INT dq'_0 \INT dq'_f\left[ \left. (-i) \frac{\del}{\del J''(t_2)} Z(q_0',q_f|J'') \right|_{J''\,=\,0} \right] \times \nonumber\\
	&\qquad \qquad\qquad\left[ \left. Z(q'_0,q'_f|J')^*\right|_{J'\,=\,0}\right]\left[ \left.(-i)\frac{\del}{\del J(t_1)} Z(q_0,q'_f|J)\right|_{J\,=\,0} \right] \nonumber\\
	&= \left. (-i)^2 \frac{\del^2}{\del J(t_1) \del J''(t_2)} Z(q_0,q_f|J,J',J'') \right|_{J\,=\,J'\,=\,J''\,=\,0}\;,
\end{align}
with
\begin{equation}\label{eq:who knows}
	Z(q_0,q_f|J,J',J'') \equiv \INT dq'_0 \INT dq'_f\;Z(q_0,q_f'|J)\,Z(q'_0,q'_f|J')^*\, Z(q'_0,q_f|J'')\;.
\end{equation}
That is a specimen never before witnessed by these two eyes. Is it useful?
\subsection{Out-of-time-ordered correlators}
Bridging the gap between superconductors \cite{larkin1968} and black holes \cite{dray_thooft, thooft_string_theory}, Kitaev\footnote{Motivated by the work of Shenker and Stanford \cite{shenker_stanford}.} emphasized arbitrarily-ordered 4-point correlation functions, known as ``out-of-time-ordered correlators'' or ``OTOCs,'' in establishing equivalences between effective field theories \cite{kitaev_soft_mode}. 
\\\\
The Schwinger-Keldysh path integral can generate arbitrarily-ordered 2-point correlators, as in Eq.~(\ref{eq:unordered 2pt expectation values}). But taking more than two derivatives would time-order the corresponding operators, so $Z(J,J')$ could never generate a correlation function like\footnote{There is some debate about which type of arbitrarily-ordered 4-point function is important \cite{which_otoc}, rooted in poorly chosen words by Maldacena et al.~\cite{chaos_bound} It does not matter whether the partitioning of fractional powers of the density matrix is a UV or IR effect, and I am not interested in traditional definitions of chaos---I am interested in moments of the system in the ensemble described by $\rho$. Who cares about the Loschmidt echo.}
\begin{equation}\label{eq:OTOC}
	\tr\left(\hat q(t_4)\hat q(t_3)\hat q(t_2)\hat q(t_1)\, e^{\,i\hat H t_0} \hat \rho\,e^{-i\hat H t_0}\right)\;.
\end{equation}
The mentality by now should be clear: If you want a correlation function that your existing generating functions cannot generate, then find another generating function. 
\subsubsection{Larkin-Ovchinnikov path integral}
To find a path integral that could generate Eq.~(\ref{eq:OTOC}), I will follow reasoning analogous to that which led to Eq.~(\ref{eq:who knows}): 
\begin{align}
	&\tr\left(\hat q(t_4)\hat q(t_3)\hat q(t_2)\hat q(t_1)\, e^{\,i\hat H t_0} \hat \rho\,e^{-i\hat H t_0}\right) \nonumber\\
	&= \tr\left( e^{-i\hat H t_0}\hat q(t_4) e^{\,i\hat H t_f}e^{-i\hat H t_f}\hat q(t_3) e^{\,i\hat H t_0} e^{-i\hat H t_0}\hat q(t_2) e^{\,i\hat H t_f} e^{-i\hat H t_f}\hat q(t_1) e^{\,i\hat H t_0} \hat\rho \right) \nonumber\\
	&\nonumber\\
	&= \INT dq_0\,dq'_0\,dq''_0\,dq_f\,dq'_f\;\bra q''_0|e^{-i\hat H t_0}\hat q(t_4) e^{\,i\hat H t_f} |q'_f\ket \bra q'_f| e^{-i\hat H t_f}\hat q(t_3) e^{\,i\hat H t_0} |q'_0\ket \times \nonumber\\
	&\qquad \bra q'_0| e^{-i\hat H t_0}\hat q(t_2) e^{\,i\hat H t_f} |q_f\ket \bra q_f|e^{-i\hat H t_f}\hat q(t_1) e^{\,i\hat H t_0} |q_0\ket \bra q_0|\hat\rho|q''_0\ket \nonumber\\
	&\nonumber\\
	&= \INT dq_0\,dq'_0\,dq''_0\,dq_f\,dq'_f\left[\left. (+i)\frac{\del }{\del J'''(t_4)}Z(q''_0,q'_f|J''')^* \right|_{J'''\,=\,0}\right]\left[\left.(-i)\frac{\del}{\del J''(t_3)}Z(q'_0,q'_f|J'') \right|_{J''\,=\,0}\right] \times \nonumber\\
	&\qquad \left[\left.(+i)\frac{\del}{\del J'(t_2)} Z(q'_0,q_f|J')^* \right|_{J'\,=\,0} \right] \left[\left.(-i)\frac{\del}{\del J(t_1)}Z(q_0,q_f|J) \right|_{J\,=\,0} \right] \rho(q_0,q''_0) \nonumber\\
	&\nonumber\\
	&= \left.\frac{\del^4}{\del J(t_1)\del J'(t_2)\del J''(t_3)\del J'''(t_4)} Z(J,J',J'',J''')\right|_{J\,=\,J'\,=\,J''\,=\,J'''\,=\,0}\;,
\end{align}
with
\begin{align}\label{eq:LO}
	&Z(J,J',J'',J''') \equiv \nonumber\\
	&\qquad \INT dq_0 \, dq'_0 \, dq''_0 \, dq_f \, dq'_f\;\rho(q_0,q''_0)\;Z(q_0,q_f|J)\, Z(q'_0,q_f|J')^*\,Z(q'_0,q'_f|J'')\, Z(q''_0,q'_f|J''')^*\;.
\end{align}
I will call that the Larkin-Ovchinnikov path integral, because it generates Larkin-Ovchinnikov correlation functions \cite{larkin1968}.\footnote{I am continuing the trend of naming path integrals after famous Russians who did not use path integrals. See Footnote~\ref{ft:influence phase}.}
\subsubsection{Generalized influence phase}
Adopting the terminology of Feynman and Vernon, I will also refer to
\begin{equation}\label{eq:generalized influence phase}
	\Phi(J,J',J'',J''') \equiv -i\ln Z(J,J',J'',J''')
\end{equation}
as an influence phase, modified by ``generalized'' or ``Larkin-Ovchinnikov'' depending on context and whimsy. 
\\\\
I want to record for later use some expressions for Eq.~(\ref{eq:OTOC}) in terms of derivatives of $\Phi(J,J',J'',J''')$ instead of those of $Z(J,J',J'',J''')$, analogous to Eqs.~(\ref{eq:time-ordered 2pt expectation values in terms of influence phase}) and~(\ref{eq:unordered 2pt expectation values in terms of influence phase}). First, there is the following general expression:
\begin{align}
	&\tr\left(\hat q(t_4)\hat q(t_3)\hat q(t_2)\hat q(t_1)\;e^{\,i t_0\hat H}\hat\rho\,e^{-it_0\hat H}\right) = \left.\frac{\del^4 Z(J,J',J'',J''')}{\del J(t_1)\del J'(t_2)\del J''(t_3)\del J'''(t_4)} \right|_{J\,=\,J'\,=\,J''\,=\,J'''\,=\,0} \nonumber\\
	&\nonumber\\
	&= i\frac{\del^4\Phi}{\del J(t_1)\del J'(t_2)\del J''(t_3)\del J'''(t_4)} + i^2\left(\frac{\del\Phi}{\del J(t_1)}\frac{\del^3\Phi}{\del J'(t_2)\del J''(t_3)\del J'''(t_4)} +\frac{\del\Phi}{\del J'(t_2)}\frac{\del^3\Phi}{\del J(t_1)\del J''(t_3)\del J'''(t_4)} \right. \nonumber\\
	&\left. + \frac{\del\Phi}{\del J''(t_3)}\frac{\del^3\Phi}{\del J(t_1)\del J'(t_2) \del J'''(t_4)} + \frac{\del\Phi}{\del J'''(t_4)}\frac{\del^3\Phi}{\del J(t_1)\del J'(t_2) \del J''(t_3)}\right) + i^2\left(\frac{\del^2\Phi}{\del J(t_1)\del J'(t_2)}\frac{\del^2\Phi}{\del J''(t_3)\del J'''(t_4)} \right. \nonumber\\
	&\left.+\frac{\del^2\Phi}{\del J'(t_2)\del J''(t_3)}\frac{\del^2\Phi}{\del J(t_1)\del J'''(t_4)} + \frac{\del^2\Phi}{\del J(t_1)\del J''(t_3)}\frac{\del^2\Phi}{\del J'(t_2)\del J'''(t_4)}\right) + i^3\left( \frac{\del\Phi}{\del J(t_1)}\frac{\del\Phi}{\del J'(t_2)}\frac{\del^2\Phi}{\del J''(t_3)\del J'''(t_4)}\right. \nonumber\\
	& + \frac{\del\Phi}{\del J'(t_2)}\frac{\del\Phi}{\del J''(t_3)}\frac{\del^2\Phi}{\del J(t_1)\del J'''(t_4)} + \frac{\del\Phi}{\del J''(t_3)}\frac{\del\Phi}{\del J'''(t_4)}\frac{\del^2\Phi}{\del J(t_1)\del J'(t_2)} + \frac{\del\Phi}{\del J(t_1)}\frac{\del\Phi}{\del J'''(t_4)}\frac{\del^2\Phi}{\del J'(t_2)\del J''(t_3)} \nonumber\\
	&\left. + \frac{\del\Phi}{\del J(t_1)}\frac{\del\Phi}{\del J''(t_3)}\frac{\del^2\Phi}{\del J'(t_2)\del J'''(t_4)} + \frac{\del\Phi}{\del J'(t_2)}\frac{\del\Phi}{\del J'''(t_4)} \frac{\del^2\Phi}{\del J(t_1)\del J''(t_3)}\right) + i^4 \frac{\del\Phi}{\del J(t_1)}\frac{\del\Phi}{\del J'(t_2)}\frac{\del\Phi}{\del J''(t_3)}\frac{\del\Phi}{\del J'''(t_4)}\;,
%	&= i\frac{\del^4\Phi}{\del J(t_1)\del J'(t_2)\del J''(t_3)\del J'''(t_4)} + i^2\left( \frac{\del^2\Phi}{\del J'(t_2)\del J''(t_3)} \frac{\del^2\Phi}{\del J(t_1)\del J'''(t_4)} + \frac{\del^2\Phi}{\del J(t_1)\del J''(t_3)} \frac{\del^2\Phi}{\del J'(t_2)\del J'''(t_4)}\right) \nonumber\\
%	&+i^2\left( \frac{\del^3\Phi}{\del J(t_1)\del J'(t_2)\del J''(t_3)} \frac{\del\Phi}{\del J'''(t_4)} + \frac{\del\Phi}{\del J''(t_3)}\frac{\del^3\Phi}{\del J(t_1)\del J'(t_2)\del J'''(t_4)}\right) \nonumber\\
%	&+i^3 \left( \frac{\del^2\Phi}{\del J(t_1)\del J'(t_2)} \frac{\del\Phi}{\del J''(t_3)}\frac{\del\Phi}{\del J'''(t_4)} + \frac{\del\Phi}{\del J'(t_2)}\frac{\del^2\Phi}{\del J(t_1)\del J''(t_3)}\frac{\del\Phi}{\del J'''(t_4)} + \frac{\del\Phi}{\del J'(t_2)}\frac{\del\Phi}{\del J''(t_3)}\frac{\del^2\Phi}{\del J(t_1) \del J'''(t_4)}\right) \nonumber\\
%	&+ \left. i^4 \frac{\del \Phi}{\del J(t_1)}\frac{\del \Phi}{\del J'(t_2)}\frac{\del \Phi}{\del J''(t_3)}\frac{\del \Phi}{\del J'''(t_4)} \right|_{J\,=\,J'\,=\,J''\,=\,J'''\,=\,0}\;.
\end{align}
which just expresses the combinatorics of exponentials. (All expressions are evaluated for $J\,=\,J'\,=\,J''\,=\,J'''\,=\,0$, and I have used $Z(0,0,0,0) = 1$.)  
\\\\
For an action invariant under $q(t) \to -q(t)$, the odd derivatives will be zero. For an action that is quadratic in $q(t)$, all derivative powers higher than two will be zero. So for all calculations in this paper,\footnote{In Sec.~\ref{sec:first excited state} I will calculate the Schwinger-Keldysh path integral for a particular non-quadratic action. But I will not calculate the Larkin-Ovchinnikov path integral for that case.} the above formula will reduce to
\begin{align}
	&\tr\left(\hat q(t_4)\hat q(t_3)\hat q(t_2)\hat q(t_1)\;e^{\,i t_0\hat H}\hat\rho\,e^{-it_0\hat H}\right) =\nonumber\\
		& -\left(\frac{\del^2\Phi}{\del J(t_1)\del J'(t_2)}\frac{\del^2\Phi}{\del J''(t_3)\del J'''(t_4)}+\frac{\del^2\Phi}{\del J'(t_2) \del J''(t_3)} \frac{\del^2\Phi}{\del J(t_1) \del J'''(t_4)} + \frac{\del^2\Phi}{\del J(t_1) \del J''(t_3)} \frac{\del^2\Phi}{\del J'(t_2) \del J'''(t_4)}\right)\;. \label{eq:otoc in terms of generalized influence phase}
\end{align}
\subsubsection{Generalized influence phase for quadratic actions}
On the right-hand side of Eq.~(\ref{eq:otoc in terms of generalized influence phase}) I mean the Larkin-Ovchinnikov influence phase, $\Phi(J,J',J'',J''')$, not the Schwinger-Keldysh influence phase, $\Phi(J,J')$. But in the special case of quadratic actions, the two are related. %In the spirit of Feynman and Vernon, I want to establish some general properties of $\Phi(J,J',J'',J''')$ for quadratic actions and relate it to $\Phi(J,J')$.
\\\\
To establish that relation, I will recall some observations of Feynman and Vernon. Start with the general form\footnote{I assume translational invariance for now to simplify the notation.}
\begin{align}
	\Phi(J,J') &= \half \doubleint\Big[G_F(t\!-\!t')\,J(t)J(t') - G_D(t\!-\!t')\,J'(t)J'(t') \Big. \nonumber\\
	&\qquad \qquad \Big.- G_<(t\!-\!t')\,J(t)J'(t')-G_>(t\!-\!t')\,J'(t)J(t')\Big]\;.
\end{align}
Imposing $\Phi(J,J) = 0$ [Eq.~(\ref{eq:influence phase property 1})] gives the first constraint:
\begin{equation}\label{eq:first constraint}
	G_F(t)-G_D(t)-G_<(t)-G_>(t) = 0\;.
\end{equation} 
Imposing $\Phi(J,J')^* = -\Phi(J',J)$ [Eq.~(\ref{eq:influence phase property 2})] gives the second constraint:
\begin{equation}\label{eq:second constraint}
	G_D(t) = G_F(t)^*\;,\;\; G_<(t) = -G_>(t)^*\;.
\end{equation}
Swapping $t$ and $t'$ in the integrand and recognizing that $\Phi(J,J')$ must remain invariant gives the third constraint:
\begin{equation}\label{eq:third constraint}
	G_F(t) = G_F(-t)\;,\;\; G_D(t) = G_D(-t)\;,\;\; G_<(t) = G_>(-t)\;.
\end{equation}
All right, back to $\Phi(J,J',J'',J''')$. 
\\\\
To see what happens when sources are set equal in pairs, it is again useful to invoke the operator form of the amplitude-generating function from Eq.~(\ref{eq:operator form of feynman path integral}). With that, I can express the Schwinger-Keldysh and Larkin-Ovchinnikov generating functions as:
\begin{align}
	&Z(J,J') = \tr\left[ \mathcal T^*\left( e^{-i\int_{t_0}^{t_f} dt\,J'(t)\, \hat q(t)} \right) \mathcal T \left( e^{\,i\int_{t_0}^{t_f} dt\,J(t)\,\hat q(t)} \right) e^{\,i t_0\hat H}\hat\rho\,e^{-it_0\hat H}\right]\;,\;\; \text{ and } \label{eq:operator form of schwinger-keldysh}\\
	&\nonumber\\
	&Z(J,J',J'',J''') = \tr\left[ \mathcal T^*\left( e^{-i\int_{t_0}^{t_f} dt\,J'''(t)\, \hat q(t)} \right) \mathcal T \left( e^{\,i\int_{t_0}^{t_f} dt\,J''(t)\,\hat q(t)} \right) \times \right. \nonumber\\
	&\qquad\qquad\qquad\qquad\qquad\qquad\left.\mathcal T^*\left( e^{-i\int_{t_0}^{t_f} dt\,J'(t)\, \hat q(t)} \right) \mathcal T \left( e^{\,i\int_{t_0}^{t_f} dt\,J(t)\,\hat q(t)} \right) e^{\,i t_0\hat H} \hat\rho\,e^{-it_0\hat H}\right]\;. \label{eq:operator form of larkin-ovchinnikov}
\end{align}
There are three pairs that will reduce $Z(J,J',J'',J''')$ to the Schwinger-Keldysh generating function:
\begin{equation}\label{eq:reduce LK}
	Z(J,J',J'',J'') = Z(J,J')\;,\;\; Z(J,J,J'',J''') = Z(J'',J''')\;,\;\; Z(J,J',J',J''') = Z(J,J''')\;.
\end{equation} 
But I still need to constrain the $J'J''$, $JJ''$, and $J'J'''$ terms in the influence phase. First I will isolate the dependence on $J$ and $J'$. Setting $J''' = J$ would not make the $J$-dependent factors cancel; setting $J = J''' = 0$ instead would leave the remaining time-ordered and antitime-ordered operators in the wrong order:
\begin{equation}
	Z(0,J',J'',0) = \tr\left[\mathcal T \left( e^{\,i\int_{t_0}^{t_f} dt\,J''(t)\,\hat q(t)} \right) \mathcal T^*\left( e^{-i\int_{t_0}^{t_f} dt\,J'(t)\, \hat q(t)} \right) e^{\,i t_0\hat H} \hat\rho\,e^{-it_0\hat H} \right]\;.
\end{equation}
I do not know how to turn that into a Schwinger-Keldysh function, but I do know that it generates the same two point functions:
\begin{align}
	&\left.\frac{\del^2 Z(0,J',J'',0)}{\del J'(t_1) \del J'(t_2)} \right|_{J'\,=\,J''\,=\,0} = -\tr\left[\mathcal T^*\!\left(\hat q(t_1) \hat q(t_2)\right)\, e^{\,i t_0\hat H} \hat\rho\,e^{-it_0\hat H}\right] = \left.-\frac{\del^2 Z(J,J')}{\del J'(t_1) \del J'(t_2)} \right|_{J\,=\,J'\,=\,0}\;,\nonumber\\
	&\left.\frac{\del^2 Z(0,J',J'',0)}{\del J''(t_1) \del J''(t_2)} \right|_{J'\,=\,J''\,=\,0} = -\tr\left[\mathcal T\left(\hat q(t_1) \hat q(t_2)\right)\, e^{\,i t_0\hat H} \hat\rho\,e^{-it_0\hat H}\right] = \left.-\frac{\del^2 Z(J,J')}{\del J(t_1)\del J(t_2)} \right|_{J\,=\,J'\,=\,0}\;,\nonumber\\
	&\left.\frac{\del^2 Z(0,J',J'',0)}{\del J''(t_2) \del J'(t_1)}\right|_{J'\,=\,J''\,=\,0} = \tr\left[\hat q(t_2)\hat q(t_1) \,e^{\,i t_0\hat H}\hat\rho\,e^{-it_0\hat H} \right] = \left.\frac{\del^2 Z(J,J')}{\del J(t_1) \del J'(t_2)} \right|_{J\,=\,J'\,=\,0}\;. \label{eq:reduce LK 2}
\end{align}
All I need is that third one. For the $JJ''$ and $J'J'''$ terms, I instead consider
\begin{align}
	&Z(J,0,J'',0) = \tr\left[ \mathcal T\left( e^{\,i\int_{t_0}^{t_f} \! dt\, J''(t)\, \hat q(t)} \right) \mathcal T\left( e^{\,i\int_{t_0}^{t_f} \! dt\, J(t)\,\hat q(t)} \right) e^{\,i t_0\hat H} \hat\rho\,e^{-it_0\hat H} \right]\;, \text{ and } \nonumber\\
	&Z(0,J',0,J''') = \tr\left[\mathcal T^*\!\left( e^{-i\int_{t_0}^{t_f}\! dt\, J'''(t)\,\hat q(t)} \right) \mathcal T^*\!\left( e^{-i\int_{t_0}^{t_f}\! dt\, J'(t)\,\hat q(t)} \right) e^{\,i t_0\hat H} \hat\rho\,e^{-it_0\hat H}\right]\;.
\end{align}
The off-diagonal correlation functions are
\begin{align}
	&\left.\frac{\del^2 Z(J,0,J'',0)}{\del J(t_1) \del J''(t_2)} \right|_{J\,=\,J''\,=\,0} = \left.\frac{\del^2 Z(0,J',0,J''')}{\del J'(t_1) \del J'''(t_2)} \right|_{J'\,=\,J'''\,=\,0} \nonumber\\
	&\qquad= -\tr\left[\hat q(t_2)\hat q(t_1)\, e^{\,i t_0\hat H} \hat\rho\,e^{-it_0\hat H}\right] = \left.-\frac{\del^2 Z(J,J')}{\del J(t_1) \del J'(t_2)} \right|_{J\,=\,J'\,=\,0}. \label{eq:reduce LK 3}
\end{align}
Notice the extra minus sign. Eqs.~(\ref{eq:reduce LK}),~(\ref{eq:reduce LK 2}), and~(\ref{eq:reduce LK 3}) tell me everything I need to know. The Larkin-Ovchinnikov phase for quadratic actions is
\begin{align}
	&\Phi(J,J',J'',J''') = \half \doubleint \big\{ \big. \nonumber\\
	&\;\;+ G_F(t\!-\!t')\left[ J(t) J(t') + J''(t)J''(t') \right] - G_D(t\!-\!t')\left[ J'(t)J'(t') + J'''(t)J'''(t') \right] \nonumber\\
	&\;\;-G_<(t\!-\!t')\left[ J(t)J'(t') + J''(t)J'''(t') + J(t) J'''(t') + J'(t) J''(t') - J(t) J''(t') - J'(t) J'''(t')\right] \nonumber\\
	&\;\;-G_>(t\!-\!t')\left[ J'(t)J(t') + J'''(t)J''(t') + J'''(t)J(t') + J''(t)J'(t') - J''(t) J(t') - J'''(t) J'(t')\right]\nonumber\\
	&\big. \big\}\;. \label{eq:general form of LK phase}
\end{align}
Inserting that into Eq.~(\ref{eq:otoc in terms of generalized influence phase}), I obtain
%\\\\
\begin{align}\label{eq:otoc in terms of G_<}
	&\tr\left(\hat q(t_4)\hat q(t_3)\hat q(t_2)\hat q(t_1)\, e^{\,it_0\hat H}\hat\rho\, e^{-it_0\hat H}\right) \nonumber\\
	&\qquad = -\left[G_<(t_1\minus t_2) G_<(t_3\minus t_4) + G_<(t_2\minus t_3)G_<(t_1\minus t_4) + G_<(t_1\minus t_3) G_<(t_2\minus t_4)\right]\;.
\end{align}
That is the form of the relation that I will end up using, but for posterity I could recall that $G_<(t\!-\!t') = -i\tr\left(\hat q(t')\hat q(t)\, e^{\,it_0\hat H}\hat\rho\, e^{-it_0\hat H}\right)$, in which case
\begin{align}\label{eq:otoc in terms of 2pt}
\tr\left(\hat q(t_4)\hat q(t_3)\hat q(t_2)\hat q(t_1)\, e^{\,it_0\hat H}\hat\rho\, e^{-it_0\hat H}\right) &= \tr\left(\hat q(t_2)\hat q(t_1)\, e^{\,it_0\hat H}\hat\rho\, e^{-it_0\hat H}\right) \tr\left(\hat q(t_4)\hat q(t_3)\,e^{\,it_0\hat H}\hat\rho\, e^{-it_0\hat H}\right)\nonumber\\
& + \tr\left(\hat q(t_3)\hat q(t_2)\, e^{\,it_0\hat H}\hat\rho\, e^{-it_0\hat H}\right) \tr\left(\hat q(t_4)\hat q(t_1)\,e^{\,it_0\hat H}\hat\rho\, e^{-it_0\hat H}\right) \nonumber\\
&+ \tr\left(\hat q(t_3)\hat q(t_1)\,e^{\,it_0\hat H}\hat\rho\, e^{-it_0\hat H}\right)\tr\left(\hat q(t_4)\hat q(t_2)\,e^{\,it_0\hat H}\hat\rho\, e^{-it_0\hat H}\right)\;.
\end{align}
Defining, just for the moment, the notation $\ll \hat\op \gg\, \equiv \tr\left(\hat\op\; e^{\,it_0\hat H}\hat\rho\, e^{-it_0\hat H}\right)$, and relabeling the times as $(t_1,t_2,t_3,t_4) \equiv (t_4',t_3',t_2',t_1')$, I can express Eq.~(\ref{eq:otoc in terms of 2pt}) as
\begin{align}
	&\ll \hat q(t_1') \hat q(t_2') \hat q(t_3') \hat q(t_4') \gg\;\; = \;\; \ll \hat q(t_3')\hat q(t_4') \gg \, \ll \hat q(t_1')\hat q(t_2') \gg\nonumber\\
	&\qquad + \; \ll \hat q(t_2')\hat q(t_3')\gg \, \ll \hat q(t_1')\hat q(t_4')\gg\; + \; \ll \hat q(t_2')\hat q(t_4') \gg \, \ll \hat q(t_1')\hat q(t_3') \gg\;.
\end{align}
Given that, I have no idea what Larkin and Ovchinnikov intended to express in Eq.~(23) of their paper \cite{larkin1968}. 
\subsection{Vacuum expectation values}\label{sec:wrong iepsilon}
The Schwinger-Keldysh formalism is about a choice of generating function, not about nonequilibrium dynamics. To belabor the point, I will entertain the phenomenological $i\e$ prescription from $S$-matrix theory, which does not conserve probability. (I will explore an improved, probability-conserving, but ultimately still phenomenological version in Sec.~\ref{sec:iepsilon}.)
%\\\\
\subsubsection{From Feynman path integral}
Start with Eq.~(\ref{eq:time-ordered 2pt amplitude}) and insert a complete set of energy eigenstates $|n\ket$, $\hat H|n\ket = E_n|n\ket$; assume a unique ground state with energy $E_0$ and a nonzero energy gap $\Del_n \equiv E_n-E_0$. The two-point amplitude is:
\begin{align}
	&\bra q_f|e^{-i\hat H t_f} \mathcal T\left(\hat q(t_1) \hat q(t_2)\right) e^{\,i\hat H t_0} |q_0\ket = \sum_{n,m\,=\,0}^\infty e^{-iE_n t_f}e^{+iE_m t_0} \bra q_f|n\ket \bra n| \mathcal T\!\left(\hat q(t_1)\hat q(t_2)\right) |m\ket \bra m|q_0\ket \nonumber\\
	&\qquad = e^{-iE_0T} \left[\phantom{\sum_{\{\text{all beside } n\,=\,m\,=\,0\}}} \!\!\!\!\!\!\!\!\!\!\!\!\!\!\!\!\!\!\!\!\!\!\!\!\!\!\!\!\!\!\!\!\!\!\!\!\!\!\!\bra q_f|0\ket \bra q_0|0\ket^* \bra 0|\mathcal T\!\left(\hat q(t_1)\hat q(t_2)\right) | 0\ket \right. \nonumber\\
	&\qquad \left.\qquad +\!\!\!\!\!\!\!\!\!\!\!\!\!\! \sum_{\{\text{all beside } n\,=\,m\,=\,0\}}\!\!\!\!\!\!\!\!\!\!\!\! e^{-i\Del_n t_f} e^{+i\Del_m t_0} \bra q_f|n\ket \bra n| \mathcal T\!\left(\hat q(t_1)\hat q(t_2)\right) |m\ket \bra m|q_0\ket\; \right]\;.
\end{align} 
As explained by Srednicki \cite{srednicki2007} and countless others, the standard prescription amounts to replacing $\Del_n$ by $(1$$-$$i\e)\Del_n$, with $\e$ an arbitrarily small positive real number. Sending $t_f \to +\infty$ and $t_0 \to -\infty$ with $\e$ held fixed would then suppress all excited-state contributions relative to the ground-state contribution. As long as the ground-state wavefunction $\psi_0(q) \equiv \bra q|0\ket$ is nonzero at $q = q_0$ and $q = q_f$, then I can divide by $\bra q_f|0\ket \bra q_0|0\ket^*$ when calculating transition amplitudes. Therefore,
\begin{equation}
	\bra 0| \mathcal T\!\left(\hat q(t_1)\hat q(t_2) \right)|0\ket = \lim_{\e\to 0}\left.(-i^2)\frac{\del^2}{\del J(t_1)\del J(t_2)} Z_\e(q_0,q_f|J)\right|_{J\,=\,0}\;,
\end{equation}
with 
\begin{align}\label{eq:feynman Z with wrong iepsilon}
	Z_\e(q_0,q_f|J) \propto \int_{q(-\infty)\,=\,q_0}^{q(\infty)\,=\,q_f} \!\!\!\!\Ds q(\cdot) \;e^{\,i\int_{-\infty}^{\infty} dt\left[ \la_\e(q(t)) + J(t)q(t) \right]}\;,\;\;\la_\e(q) = (1\!+\!i\e)\half \dot q^2 - (1\!-\!i\e) \half m^2 q^2\;.
\end{align} 
Now I will adapt that reasoning to the Schwinger-Keldysh path integral.
\subsubsection{From Schwinger-Keldysh path integral}
The arbitrarily-ordered 2-point correlator is:
\begin{align}
	&\tr\left(\hat q(t_2)\hat q(t_1) \;e^{\,i\hat H t_0} \hat\rho\, e^{-i\hat H t_0}\right) = \tr\left(e^{-i\hat H t_0} \hat q(t_2)\hat q(t_1) e^{\,i\hat H t_0} \hat\rho\right) \nonumber\\
	&= \sum_{n,n'\,=\,0}^{\infty} \bra n'| e^{-i\hat H t_0} \hat q(t_2) \hat q(t_1) e^{\,i\hat H t_0}|n\ket \bra n| \hat\rho |n'\ket \nonumber\\
	&= e^{\,i E_0 t_0} e^{-i E_0 t_0}\bra 0|\hat q(t_2)\hat q(t_1)|0\ket \rho_{00} + \sum_{n\,=\,1}^\infty e^{\,i E_n t_0}\bra 0| \hat q(t_2)\hat q(t_1) |n\ket \rho_{n0} \nonumber\\
	&\qquad + \sum_{n'\,=\,1}^{\infty} e^{-i E_{n'} t_0} \bra n'| \hat q(t_2)\hat q(t_1)|0\ket \rho_{0n'}+ \sum_{n,n'\,=\,1}^\infty e^{\,i E_n t_0} e^{-i E_{n'} t_0} \bra n'| \hat q(t_2)\hat q(t_1) |n\ket \rho_{nn'} \nonumber\\
	&= \bra 0|\hat q(t_2)\hat q(t_1)|0\ket \rho_{00}+ e^{\,i E_0 t_0} \sum_{n\,=\,1}^\infty e^{\,i\Del_n t_0} \bra 0|\hat q(t_2)\hat q(t_1)|n\ket \rho_{n0} \nonumber\\
	&\qquad + e^{-i E_0 t_0}\sum_{n'\,=\,1}^\infty e^{-i\Del_{n'}t_0} \bra n'|\hat q(t_2)\hat q(t_1)|0\ket \rho_{0n'}  + \sum_{n,n'\,=\,1}^\infty e^{\,i \Del_n t_0} e^{-i \Del_{n'} t_0} \bra n'| \hat q(t_2)\hat q(t_1) |n\ket \rho_{nn'}\;.
\end{align}
Adapting the $i\e$ prescription to this case would amount to replacing $\Del_n \to (1$$-$$i\e) \Del_n$ on the forward branch and $\Del_{n'} \to (1$$+$$i\e) \Del_{n'}$ on the backward branch. Then $t_{0} \to -\infty$ would once again suppress all excited-state contributions, and as long as $\rho_{00} \neq 0$ I can divide by $\rho_{00}$. Therefore, 
\begin{align}
	\bra 0|\hat q(t_2)\hat q(t_1)|0\ket &= \left.\frac{\del^2}{\del J(t_1) \del J'(t_2)} Z_\e(J,J') \right|_{J\,=\,J'\,=\,0}\;,
\end{align}
with
\begin{align}\label{eq:keldysh Z with wrong iepsilon}
	Z_\e(J,J') &\propto \INT dq_f \INT dq_0 \INT dq'_0 \;\rho(q_0,q'_0)\; \times \nonumber\\
	&\int_{q(-\infty) \,=\, q_0}^{q(\infty)\,=\,q_f}\!\!\!\!\!\!\!\!\!\! \Ds q(\cdot) \int_{q'(-\infty)\, =\, q'_0}^{q'(\infty)\,=\,q_f}\!\!\!\!\!\!\!\!\!\! \Ds q'(\cdot)\;e^{\,i \int_{-\infty}^{\infty} dt\left[ \la_\e(q(t)) - \la_{-\e}(q'(t)) + J(t)q(t)-J'(t)q'(t)\right]}\;.
\end{align}
That provides a path-integral description of Wightman correlators.\footnote{For the amplitude-generating function, the usual argument goes a step further: I could change $q_0$ and $q_f$ to whatever I want as long as $\psi_0(q_0)$ and $\psi_0(q_f)$ remain nonzero, in which case I could once again divide by $\psi_0(q_f)\psi_0(q_0)^*$. The boundary conditions on the field do not matter. Presumably there is an analogous argument for $Z_\e (J,J')$, where as long as the density matrix has nonzero overlap with the ground state, the initial values of the field should not matter; the final values of the field also should not matter, provided that they agree (i.e., $q(\infty) = q'(\infty)$). Maybe there is a kind of stationary-state argument to be made from Eq.~(\ref{eq:intermediate form of Z from alternative order of integration}) or Eq.~(\ref{eq:unnormalized generating function for iepsilon}).} 
\\\\
I do not know whether the effective field theory of Wightman correlators is important in equilibrium, but the above would be how to formulate it. Out of equilibrium, however, I doubt that this setup makes sense \cite{Bonga2015, Glorioso2018}.
\vspace{10pt}
\pagebreak
\section{Harmonic oscillator}\label{sec:oscillator}
The formalism established, it is time for the First of Examples, Quadratic Action of the Free Field, Pinnacle of Tractability, and Mother of Hubris:
\begin{equation}
	V(q) = \half m^2 q^2\;.
\end{equation}
\subsection{Feynman path integral}
The first task is to calculate the Feynman path integral with sources and fixed boundary conditions:
\begin{align}
	Z(q_0,q_f|J) = \int_{q(t_0)\,=\,q_0}^{q(t_f)\,=\,q_f} \!\! \! \! \Ds q(\cdot)\;e^{\,i\int_{t_0}^{t_f}\!dt\,\left(\half \dot q^2-\half m^2 q^2 + Jq\right)}\;.
\end{align} 
If you think that deriving the amplitude-generating function for the harmonic oscillator is beneath the scope of a contemporary research article, then stop reading and calculate it. Start to finish, including the normalization. Now. 
\\\\
There are at least two methods I know of to calculate $Z(q_0,q_f|J)$. The first is to return to the time-sliced definition from Sec.~\ref{sec:path-integral decomposition} and evaluate the iterated integrals, which would be analogous to calculating integrals like $\int_{a}^{b}dx\,f(x)$ by pre-Riemannian brute force \cite{euler_summation}. I tried but could not complete that calculation (hara-kiri); I review the calculation for $m = 0$ and $J = 0$ in Appendix~\ref{sec:free particle from time slicing}. 
\\\\
The more erudite method is to deal with continuum field theory directly, applying the method of stationary phase. That is the approach taken by most modern books, but be warned that, in general, even (0+1)-dimensional quantum field theory needs regularization, a 1-loop determinant, and 2-loop counterterms \cite{steiner_proceedings, Grosche1987, bastianelli1992, pvn_book}.
\\\\
I will proceed with the continuum method. Split up the integration variable into a classical path satisfying the boundary conditions, plus fluctuations: 
\begin{equation}\label{eq:split field into classical plus quantum}
	q(t) = \bar q(t) + Q(t)\;,
\end{equation}
with
\begin{equation}
	\ddot{\bar q}(t) + m^2\bar q(t) = J(t)\;,\;\; \bar q(t_0) = q_0\;,\;\; \bar q(t_f) = q_f\;, \label{eq:field eq for oscillator}
\end{equation}
and
\begin{equation}
	Q(t_0) = Q(t_f) = 0\;,
\end{equation}
with $Q(t)$ otherwise unconstrained. 
\subsection{Classical solution}\label{sec:classical solution}
I will review how to solve Eq.~(\ref{eq:field eq for oscillator}). Yes, really. 
%\\\\
\subsubsection{Green's function}\label{sec:Green's function}
First consider the homogeneous equation for a function $G(t)$ with modified initial conditions, as follows:\footnote{This is what Haberman calls an ``alternative'' formulation of the Green's function \cite{haberman}. If I remember correctly (the book is quarantined in my office), he further imposes causality on the solution of Eq.~(\ref{eq:green's function definition}), whereas I prefer to impose causality only on the solution of Eq.~(\ref{eq:field eq for oscillator}).}
\begin{equation}\label{eq:green's function definition}
	\ddot G(t) + m^2 G(t) = 0\;,\;\; G(0) = 0\;,\;\; \dot G(0) = 1\;.
\end{equation}
Integrate that equation against a Gaussian profile in time:
\begin{equation}
	\INT dt\;e^{-\frac{1}{2\s^2}t^2 + i\w t}\left[\ddot G(t) + m^2 G(t)\right] = 0\;.
\end{equation}
Integrating by parts twice produces the following equation:
\begin{align}\label{eq:integrate by parts}
	&\left.e^{-\frac{1}{2\s^2}t^2+i\w t}\left[\dot G(t)-i\w G(t)\right] \right|_{t\,=\,-\infty}^{\infty} + \frac{1}{\s^2}\INT\! dt\;e^{-\frac{1}{2\s^2}t^2+i\w t}\; t\left[\dot G(t)-i\w G(t)\right] \nonumber\\
	&\qquad +\left(-\w^2+m^2\right)\INT dt\;e^{-\frac{1}{2\s^2}t^2+i\w t} G(t) = 0\;.
\end{align}
Since $e^{-\frac{1}{2\s^2}t^2}$ is even in $t$, the boundary terms are zero. The integral in the first line of Eq.~(\ref{eq:integrate by parts}) is some number. Therefore, doing all of that and then taking $\s \to \infty$ leaves the Fourier-transformed equation:
\begin{equation}
	\left(-\w^2+m^2\right) \tilde G(\w) = 0\;,\;\; \tilde G(\w) \equiv \INT dt\;e^{\,i\w t}\,G(t)\;.
\end{equation}
Either $\tilde G(\w) = 0$ or $\w = \pm m$, so $G(t) = c_+\,e^{\,im t} + c_-\,e^{-imt}$ for some constants $c_\pm$. Because the problem is manifestly real, I will instead work with the usual trigonometric functions:
\begin{equation}\label{eq:general homogeneous solution}
	G(t) = A\,\cos(mt) + B\,\sin(mt)\;.
\end{equation}
Since $G(0) = A$, the requirement $G(0) = 0$ means $A = 0$ and $G(t) = B\,\sin(mt)$. Then $\dot G(t) = Bm\,\cos(mt)$, in which case $\dot G(0) = Bm \equiv 1 \implies B = 1/m$. Therefore, the Green's function is
\begin{equation}\label{eq:green's function}
	G(t) = \frac{1}{m}\sin(mt)\;.
\end{equation}
I reviewed that for two reasons. First, to avoid lazily tossing boundary terms without explaining why.\footnote{Sure, in this case you could just start from Eq.~(\ref{eq:general homogeneous solution})---I know what an ansatz is. But as a student I never found that satisfying, and in any case I am front-loading this for Sec.~\ref{sec:sourced damped oscillator}.} Second, to emphasize that, in this formulation, the Green's function is as written in Eq.~(\ref{eq:green's function})---it is not the operator-inverse of $\pa_t^2+m^2$, and it is odd in time. 
\vfill
\pagebreak
\subsubsection{Homogeneous solution}\label{sec:homogeneous solution}
The homogenous part, $q_h(t)$, of the classical field is defined as the general solution of
\begin{equation}
	\ddot q_h(t) + m^2 q_h(t) = 0\;.
\end{equation}
Boundary conditions are imposed on the \textit{total} classical field, not just on the homogeneous part; and they will be imposed at times $t_0$ and $t_f$, not at zero. So at this stage all I can do is repeat the form of Eq.~(\ref{eq:general homogeneous solution}), but shifted by $t_0$:
\begin{equation}\label{eq:homogeneous solution}
	q_h(t) = A\,\dot G(t\!-\!t_0) + B\,G(t\!-\!t_0)\;,
\end{equation}
for some constants $A$ and $B$ to be determined shortly.
\subsubsection{Particular solution}\label{sec:particular solution}
The particular part, $q_p(t)$, of the classical field is defined as the general solution of
\begin{equation}
	\ddot q_p(t) + m^2 q_p(t) = J(t)\;,
\end{equation}
modulo homogeneous solutions, \textit{and subject to causality}. In terms of the Green's function in Eq.~(\ref{eq:green's function}), that means
\begin{equation}\label{eq:particular solution}
	q_p(t) = \int_{t_0}^{t_f}\!\!dt'\;\Ta(t\minus t')\;G(t\minus t')\;J(t')\;.
\end{equation}
I reviewed this reasoning to make Eq.~(\ref{eq:particular solution}) nonnegotiable. 
\subsubsection{Complete solution}\label{sec:complete solution}
The classical field is the sum of its homogeneous and particular parts, subject to boundary conditions:
\begin{equation}
	\bar q(t) = q_p(t) + q_h(t)\;,\;\; \bar q(t_0)\,=\,q_0\;,\;\;\bar q(t_f)\,=\,q_f\;.
\end{equation}
In Eq.~(\ref{eq:particular solution}) the limits are such that $t'$ is between $t_0$ and $t_f$, in which case $\Ta(t_0\minus t') = 0$. Since $G(0) = 0$ and $\dot G(0) = 1$, the initial condition implies
\begin{equation}
	\bar q(t_0) = A\times 1 + B\times 0 + 0 = A \equiv q_0\;.
\end{equation} 
Similarly, $\Ta(t_f\minus t') = 1$. So the final condition will fix $B$ in terms of $q_f$ and $J(t)$:
\begin{align}
	&\bar q(t_f) = q_0\,\dot G(t_f\minus t_0) + B\,G(t_f\minus t_0) + \int_{t_0}^{t_f}\!\! dt'\;G(t_f\minus t')\,J(t') \equiv q_f \nonumber\\
	&\implies B = \frac{1}{G(t_f\minus t_0)}\left[q_f-q_0\,\dot G(t_f\minus t_0) - \int_{t_0}^{t_f}\!\! dt'\;G(t_f\minus t')\,J(t')\right]\;.
\end{align}
Defining
\begin{equation}
	T \equiv t_f\minus t_0\;,
\end{equation}
and recognizing the trigonometric identity
\begin{equation}
	G(T)\,\dot G(t\minus t_0) - \dot G(T)\, G(t\minus t_0) = G(t_f\minus t)\;,
\end{equation}
I arrive at the complete solution for the classical field:
\begin{align}\label{eq:classical solution for oscillator}
	\bar q(t) &= \frac{1}{G(T)}\Bigg\{ q_f\,G(t\minus t_0) + q_0\,G(t_f\minus t) \Bigg. \nonumber\\
	&\qquad \qquad \qquad \Bigg.+ \int_{t_0}^{t_f}\!\! dt'\left[ G(T)\,\Ta(t\minus t')\,G(t\minus t') - G(t\minus t_0)\,G(t_f\minus t')\right]J(t')\Bigg\}\;.
\end{align}
\subsection{Action on classical solution}
Next I will need to evaluate the action
\begin{equation}\label{eq:action for harmonic oscillator with source}
	S(q) = \singleint\left[\half \dot q(t)^2 - \half m^2 q(t)^2 + J(t)q(t)\right]
\end{equation}
on the classical solution in Eq.~(\ref{eq:classical solution for oscillator}). Since $\dot q^2 = \frac{d}{dt}\left(q\dot q\right)-q\ddot q$, and since the classical solution $\bar q$ was defined to satisfy $\ddot{\bar q}+m^2\bar q = J$, I find
\begin{align}\label{eq: Scl definition}
	S_{\text{cl}}(q_0,q_f|J) &\equiv S(\bar q) = \left.\half \bar q \dot{\bar q}\right|_{t\,=\,t_0}^{t_f} + \half \singleint J(t)\bar q(t)\;.
\end{align}
Again, that is the action on classical solutions, not the classical action. Eq.~(\ref{eq:action for harmonic oscillator with source}) is the classical action, which is a function of the quantum field. Confusing, I know. 
\\\\
The derivative of Eq.~(\ref{eq:classical solution for oscillator}) is (remember that $G(0) = 0$):
\begin{align}
	\dot{\bar q}(t) &= \frac{1}{G(T)}\Bigg\{ q_f\dot G(t\minus t_0)-q_0\dot G(t_f\minus t) \Bigg. \nonumber\\
	&\qquad \qquad \qquad \Bigg. + \int_{t_0}^{t_f}\!\! dt'\left[G(T)\Ta(t\minus t')\dot G(t\minus t') - \dot G(t\minus t_0)G(t_f\minus t')\right]J(t') \Bigg\}\;.
\end{align}
At the initial time (remember that $\dot G(0) = 1$), I find
\begin{align}
	\dot{\bar q}(t_0) &= \frac{1}{G(T)}\left\{ q_f-q_0\dot G(T)-\int_{t_0}^{t_f}\!\! dt'\, G(t_f\minus t')J(t') \right\}.
\end{align}
At the final time, after using $\dot G(T) \dot G(t_f-t) - \dot G(T)G(t_f-t) = G(t-t_0)$, I find
\begin{align}
	\dot{\bar q}(t_f) &= \frac{1}{G(T)}\left\{ q_f\dot G(T)-q_0 + \int_{t_0}^{t_f}\!\!dt'\,G(t'\minus t_0)J(t') \right\}\;.
\end{align}
Therefore, 
\begin{align}
	\left. \bar q \dot{\bar q}\right|_{t\,=\,t_0}^{t_f} &= \frac{q_f}{G(T)}\left\{ q_f\dot G(T)-q_0 + \int_{t_0}^{t_f}\!\!dt'\,G(t'\minus t_0) J(t') \right\} \nonumber\\
	&-\frac{q_0}{G(T)}\left\{ q_f-q_0\dot G(T)-\int_{t_0}^{t_f}\!\! dt' \, G(t_f\minus t')J(t') \right\} \nonumber\\
	&\nonumber\\
	&=\frac{1}{G(T)}\left\{ \dot G(T)\left(q_f^2+q_0^2\right) -2q_fq_0 + \int_{t_0}^{t_f}\!\! dt\left[q_f G(t'\minus t_0) + q_0\,G(t_f\minus t') \right]J(t')\right\}\;.
\end{align}
Meanwhile, 
\begin{align}
	\singleint J(t)\bar q(t) &= \frac{1}{G(T)}\singleint\left[q_f G(t\minus t_0)+q_0 G(t_f\minus t)\right]J(t) \nonumber\\
	&+\frac{1}{G(T)}\doubleint \left[ G(T)\,\Ta(t\minus t')\,G(t\minus t') - G(t\minus t_0)\,G(t_f\minus t')\right]J(t)J(t')\;.
\end{align}
Therefore:
\begin{align}
	\Scl(q_0,q_f|J) &= \frac{1}{2G(T)}\left\{ \dot G(T)\left(q_f^2+q_0^2\right) - 2q_fq_0 + \singleint\left[q_f G(t\minus t_0) + q_0 G(t_f\minus t)\right]J(t)\right. \nonumber\\
	&\left. +\doubleint \left[ G(T)\,\Ta(t\minus t')\,G(t\minus t') - G(t\minus t_0)\,G(t_f\minus t')\right]J(t)J(t') \right\}.
\end{align}
\subsection{Path integral over fluctuations}
With the splitting of the field into a classical solution plus fluctuations [Eq.~(\ref{eq:split field into classical plus quantum})], the action becomes
\begin{equation}
	\int_{t_0}^{t_f}\!\! dt\left[\half \dot q(t)^2\!-\!\half m^2 q(t)^2 \!+\! J(t)q(t)\right] = \Scl(q_0,q_f|J) + \int_{t_0}^{t_f}\!\! \left(\half \dot Q^2 - \half m^2 Q^2\right)\;.
\end{equation}
The square has been completed; the fluctuations are source-free. 
\\\\
At this stage, the generating function for the harmonic oscillator has the form
\begin{equation}\label{eq:unnormalized form of oscillator generating function}
	Z(q_0,t_0;q_f,t_f|J) = C(T)\,e^{\,i S_{\text{cl}}(q_0,t_0;q_f,t_f|J)}\;,
\end{equation}
with
\begin{equation} \label{eq:path integral over fluctuations}
	C(T) \equiv Z(0,t_0;0,t_f|0) = \int_{Q(t_0)\,=\,0}^{Q(t_f)\,=\,0} \!\! \!\! \Ds Q(\cdot)\;e^{\,i\int_{t_0}^{t_f} \!dt\,\left(\half \dot Q^2 - \half m^2 Q^2\right)}\;.
\end{equation}
Because the time dependence will now be important, I have specified it explicitly.
\\\\
First, notice that $C(T)$ does not depend on the external source $J$, so it will be sufficient to consider
\begin{equation}
	Z(q_0,t_0;q_f,t_f|0) = C(T)\,e^{\,i S_{\text{cl}}(q_0,t_0;q_f,t_f|0)}\;,\;\; S_{\text{cl}}(q_0,t_0;q_f,t_f|0) = \frac{\dot G(T)(q_f^2+q_0^2)-2q_fq_0}{2G(T)}\;.
\end{equation}
Second, recognize that $Z(q_0,t_f;q_f,t_f|0)$ is just the transition amplitude $\bra q_f|e^{-i\hat HT}|q_0\ket$, which satisfies a composition law:
\begin{align}
	Z(q_0,t_0;q_2,t_2|0) &= \bra q_2|e^{-it_2\hat H} e^{\,it_0\hat H}|q_0\ket = \bra q_2|e^{-it_2\hat H} e^{\,it_1\hat H} e^{-it_1\hat H} e^{\,it_0\hat H}|q_0\ket  \nonumber\\
	&= \INT dq_1\;\bra q_2|e^{-it_2\hat H}e^{\,it_1\hat H}|q_1\ket \bra q_1|e^{-it_1\hat H}e^{\,it_0\hat H}|q_0\ket \nonumber\\
	&= \INT dq_1\; Z(q_1,t_1;q_2,t_2|0)\,Z(q_0,t_0;q_1,t_1|0)\;. \label{eq:composition law}
\end{align}
Because that composition law is nonlinear, imposing it on Eq.~(\ref{eq:unnormalized form of oscillator generating function}) will fix $C(T)$.
\subsubsection{Sum of actions on classical solutions}
The first thing to show is that the sum of two $S_{\text{cl}}$s will collect into an $S_{\text{cl}}$ plus a quadratic term to be integrated over.
\begin{align}
	&S_{\text{cl}}(q_0,t_0;q_1,t_1|0) + S_{\text{cl}}(q_1,t_1;q_2,t_2|0) = \frac{\dot G(t_1\!-\!t_0)(q_1^2+q_0^2)-2q_1q_0}{2G(t_1\!-\!t_0)} + \frac{\dot G(t_2\!-\!t_1)(q_2^2+q_1^2)-2q_2q_1}{2G(t_2\!-\!t_1)}
	&\nonumber\\
	&= \frac{G(t_2\minus t_1)\left[ \dot G(t_1\minus t_0)(q_1^2+q_0^2)-2q_1q_0\right] + G(t_1\minus t_0)\left[\dot G(t_2\minus t_1)(q_2^2+q_1^2)-2q_2q_1\right]}{2G(t_1\!-\!t_0)G(t_2\!-\!t_1)}\;. \label{eq:sum of actions 1}
\end{align} 
First observe that
\begin{equation}
	G(t_2\minus t_1)\dot G(t_1\minus t_0) + G(t_1\minus t_0)\dot G(t_2\minus t_1) = G(t_2\minus t_0)\;.
\end{equation}
Thank you, trigonometry. So the numerator in Eq.~(\ref{eq:sum of actions 1}) is
\begin{align}
	N &\equiv G(t_2\minus t_1)\left[ \dot G(t_1\minus t_0)(q_1^2+q_0^2)-2q_1q_0\right] + G(t_1\minus t_0)\left[\dot G(t_2\minus t_1)(q_2^2+q_1^2)-2q_2q_1\right] \nonumber\\
	&\nonumber\\
	&= G(t_2\minus t_1)\dot G(t_1\minus t_0)q_0^2 + G(t_1\minus t_0)\dot G(t_2\minus t_1) q_2^2  \nonumber\\
	&+ G(t_2\minus t_0) q_1^2-2\left[G(t_2\minus t_1)q_0 + G(t_1\minus t_0)q_2\right] q_1 \nonumber\\
	&\nonumber\\
	&= G(t_2\minus t_1)\dot G(t_1\minus t_0)q_0^2 + G(t_1\minus t_0)\dot G(t_2\minus t_1) q_2^2\nonumber\\
	&+G(t_2\minus t_0)\!\!\!\!\!\!\!\!\!\!\!\!\underbrace{\left\{q_1^2-2\left[\frac{G(t_2\minus t_1)q_0 + G(t_1\minus t_0)q_2}{G(t_2\minus t_0)}\right]q_1\right\}}_{\begin{matrix}\left(q_1-\left[\frac{G(t_2- t_1)q_0 + G(t_1- t_0)q_2}{G(t_2- t_0)}\right]\right)^2 - \left[\frac{G(t_2- t_1)q_0 + G(t_1- t_0)q_2}{G(t_2- t_0)}\right]^2 \end{matrix}} \nonumber\\
	&\nonumber\\
	&= G(t_2\minus t_1)\dot G(t_1\minus t_0)q_0^2 + G(t_1\minus t_0)\dot G(t_2\minus t_1) q_2^2 \nonumber\\
	&+ G(t_2\minus t_0)\left(q_1-...\right)^2-\frac{[G(t_2\minus t_1)q_0 + G(t_1\minus t_0)q_2]^2}{G(t_2\minus t_0)}\;.
\end{align}
The part of $N$ that can be moved outside the integral over $q_1$ is therefore:
\begin{align}
	&N - G(t_2\minus t_0)(q_1-...)^2 \nonumber\\
	&= \tfrac{\left[G(t_2\minus t_0)\dot G(t_1\minus t_0) - G(t_2\minus t_1)\right] G(t_2\minus t_1)q_0^2 + \left[G(t_2\minus t_0)\dot G(t_2\minus t_1) -G(t_1\minus t_0)\right]G(t_1\minus t_0)q_2^2 - 2G(t_2\minus t_1)G(t_1\minus t_0)q_0q_2}{G(t_2\minus t_0)}\;.
\end{align}
With more trigonomagic, namely
\begin{align}
	&G(t_2\minus t_0)\dot G(t_1\minus t_0) - G(t_2\minus t_1) = \dot G(t_2\minus t_0)G(t_1\minus t_0)\;,\;\;\text{ and } \nonumber\\
	&G(t_2\minus t_0)\dot G(t_2\minus t_1) - G(t_1\minus t_0) = \dot G(t_2\minus t_0) G(t_2\minus t_1)\;,
\end{align}
that will simplify further into
\begin{align}
	&N-G(t_2\minus t_0)(q_1-...)^2 \nonumber\\
	&= \frac{G(t_2\minus t_1)G(t_1\minus t_0) \dot G(t_2\minus t_0)q_0^2 + G(t_2\minus t_1) G(t_1\minus t_0) \dot G(t_2\minus t_0)q_2^2 - 2G(t_2\minus t_1)G(t_1\minus t_0)q_0q_2}{G(t_2\minus t_0)} \nonumber\\
	&= G(t_2\minus t_1)G(t_1\minus t_0)\left( \frac{\dot G(t_2\minus t_0)\left(q_0^2+q_2^2\right) - 2q_0 q_2}{G(t_2\minus t_0)}\right) \nonumber\\
	&= 2G(t_2\minus t_1)G(t_1\minus t_0) S_{\text{cl}}(q_0,t_0;q_2,t_2|0)\;.
\end{align}
A Christmas miracle. (Or just basic quantum mechanics.) With that, I can compose two amplitudes:
\begin{align}
	&\INT dq_1\;Z(q_0,t_0;q_1,t_1|0)\,Z(q_1,t_1;q_2,t_2|0) \nonumber\\
	&\qquad = C(t_1\minus t_0) C(t_2\minus t_1) e^{\,i S_{\text{cl}}(q_0,t_0;q_2,t_2|0)} \INT dq_1\;e^{\,\frac{iG(t_2-t_0)}{2G(t_2-t_1)G(t_1-t_0)}\left(q_1^2 \;-\;...\right)} \nonumber\\
	&\qquad = C(t_1\minus t_0) C(t_2\minus t_1) e^{\,i S_{\text{cl}}(q_0,t_0;q_2,t_2|0)} \sqrt{\frac{2\pi i\,G(t_2\minus t_1)G(t_1\minus t_0)}{G(t_2\minus t_0)}} \nonumber\\
	&\qquad  \equiv Z(q_0,t_0;q_2,t_2|0) = C(t_2\minus t_0)\,e^{\,i S_{\text{cl}}(q_0,t_0;q_2,t_2|0)} \nonumber\\
	&\nonumber\\
	&\implies C(t_2\minus t_0) \sqrt{2\pi i\,G(t_2\minus t_0)} = C(t_1\minus t_0) \sqrt{2\pi i\,G(t_1\minus t_0)}\;C(t_2\minus t_1)\sqrt{2\pi i\,G(t_2\minus t_1)}\;.
\end{align}
The solution is
\begin{equation}
	C(t) = \frac{1}{\sqrt{2\pi i\,G(t)}}\;. \label{eq:normalization for harmonic oscillator}
\end{equation}
\subsection{Result}
The generating function for the harmonic oscillator is
\begin{equation} \label{eq:generating function for harmonic oscillator}
	Z(q_0,t_0;q_f,t_f|J) = \frac{1}{\sqrt{2\pi i\,G(T)}}\;e^{\,iS_{\text{cl}}(q_0,t_0;q_f,t_f|J)}\;,\;\; G(t) = \frac{1}{m}\sin(m t)\;,\;\;T = t_f-t_0\;,
\end{equation}
with
\begin{align}
	&S_{\text{cl}}(q_0,t_0;q_f,t_f|J) = \frac{1}{2G(T)}\!\left[\dot G(T)\left(q_f^2\!+\!q_0^2\right) - 2q_f q_0\right] + \frac{1}{G(T)}\!\dt \left[ G(t\!-\!t_0)\,q_f + G(t_f\!-\!t)\,q_0\right] J(t) \nonumber\\
	&\qquad -\frac{1}{2G(T)}\dt \dt'\left[\Ta(t\!-\!t')\,G(t_f\!-\!t)\,G(t'\!-\!t_0) + \Ta(t'\!-\!t)\,G(t_f\!-\!t')\,G(t\!-\!t_0)\right] J(t) J(t')\;. \label{eq:action for generating function for harmonic oscillator}
\end{align}
Is that formula in Feynman and Hibbs \cite{feynmanhibbs}? Yes. Did any of my illustrious professors deign to make me learn it? No.
\subsubsection{Alternative method: Path integral over fluctuations up to free-particle factor}
A standard alternative to imposing the composition law on the overall factor for the oscillator path integral is to calculate that factor up to a frequency-independent factor using mode regularization. I will review that, too. 
\\\\
First divide by 
\begin{equation}
	Z_0 \equiv \left. Z(0,0|0) \right|_{m\,=\,0}\;,
\end{equation}
which is the free-particle amplitude. Define an operator $\hat \op$ whose matrix elements in time are the pertinent derivative operators:
\begin{equation}\label{eq:d^2/dt^2}
	\bra t|\hat\op|t'\ket \equiv \left(\frac{d^2}{dt^2} + m^2\right) \del(t\!-\!t')\;,\;\; \bra t|\op_0|t'\ket  \equiv \left. \bra t|\hat\op|t'\ket \right|_{m\,=\,0}\;.
\end{equation}
The ratio of generating functions is then \footnote{\label{ft: footnote about notation}A typical physicist would write that as $\frac{Z(0,0|0)}{Z_0} = \sqrt{\frac{\det\left(\frac{d^2}{dt^2}\right)}{\det\left(\frac{d^2}{dt^2} + m^2\right)}}$. But just as I write $\Ds q(\cdot)$ instead of $\Ds q(t)$, I refuse to write expressions that should not be read literally. This is not about mathematical pseudorigor, which I detest, but rather about writing what you mean.}
\begin{equation}\label{eq:ratio of path integrals}
	\frac{Z(0,0|0)}{Z_0} = \sqrt{\frac{\det \hat \op_0}{ \det \hat \op}}\;.
\end{equation}
The determinant of an operator is the product of its eigenvalues. The eigenfunctions $f(t)$ of $\frac{d^2}{dt^2}$ that equal zero at $t = t_0$ have the form
\begin{equation}
	f(t) = A \sin\left[\w(t\!-\!t_0)\right]\;,
\end{equation}
with corresponding eigenvalues $\w^2$. Demanding $f(t_f) = 0$ then quantizes the frequencies:
\begin{equation}
	f(t_f) = A\sin(\w T) \equiv 0 \implies \w = \frac{\pi n}{T}\;,\;\; 0, \pm 1, \pm 2, ...\;.
\end{equation}
Taking only the positive-frequency solutions gives the ratio of determinants as follows:
\begin{equation}
	\frac{\det \hat\op_0}{\det \hat\op} = \prod_{n\,=\,1}^\infty \frac{1}{1-\frac{T^2 m^2}{\pi^2 n^2}} = \left. \frac{\pi z}{\sin(\pi z)}\right|_{z\,=\,\frac{m T}{\pi}} = \frac{m T}{\sin(mT)} = \frac{T}{G(T)}\;.
\end{equation}
At this point I have the unsourced path integral up to a factor:\footnote{Aha, says the heckler who read footnote~\ref{ft: footnote about notation}: In Eq.~(\ref{eq:path integral up to a factor}) you have a $T$ on the right-hand side but not on the left-hand side, hypocrite! No, no. In Eq.~(\ref{eq:path integral up to a factor}), $T$ is a variable on both sides of the equation; it is just suppressed on the left-hand side because I did not feel like writing it. In contrast, the expression ``$\det(\frac{d^2}{dt^2})$'' makes no sense in the first place, because $t$ does not appear in it. It is an uncivilized attempt to express ``$\det(\frac{d^2}{d\cdot^2})$'' or what I wrote in Eqs.~(\ref{eq:d^2/dt^2}) and~(\ref{eq:ratio of path integrals}).} 
\begin{equation}\label{eq:path integral up to a factor}
	Z(0,0|0) = Z_0\sqrt{\frac{T}{G(T)}}\;.
\end{equation}
See Appendix~\ref{sec:free particle} for two ways of calculating $Z_0$. 
\vfill
\pagebreak
\section{Schwinger-Keldysh path integral}
The basic formula is Eq.~(\ref{eq:keldysh}), which I repeat below for convenience:
\begin{equation} \label{eq:keldysh2}
	Z(J,J') = \INT dq_f \INT dq_0 \INT dq'_0\;\rho(q_0,q'_0)\;Z(q_0,q_f|J)\,Z(q'_0,q_f|J')^*\;.
\end{equation}
Calculating that for even the oscillator in its ground state took me some time. I interrupt this broadcast for a public service announcement about the order of integration.  
%\\\\
\subsection{Rumination}\label{sec:preliminary commentary}
When I first tried to verify the results from Feynman and Vernon, I looked at Eq.~(\ref{eq:keldysh2}) and thought, well, if the state is pure---i.e., if $\hat\rho = |\psi\ket \bra \psi|$---then the double integral over the initial field configurations will factorize:
\begin{align}
	\INT dq_0  \INT dq'_0\;&\rho(q_0,q'_0)\;Z(q_0,q_f|J)\,Z(q'_0,q_f|J')^* \nonumber\\
	&= \left(\INT dq_0 \;\bra q_0|\psi\ket \,Z(q_0,q_f|J)\right) \left(\INT dq'_0 \;\bra q'_0|\psi\ket \,Z(q'_0,q_f|J')\right)^*. \nonumber
\end{align}
And if the state is ground, like beef, then $|\psi\ket$ will be Gaussian. So just do the one integral, copy it with a conjugation and prime, then collect everything for the remaining integral over the final field configuration. Go to Loloan, have a gimlet.  
\\\\
What a roller-coaster ride that turned out to be. I did it, and I will show you how to do it, but it is not easy. 
\\\\
The simpler method is to forget about the density matrix, go back to Eq.~(\ref{eq:keldysh2}), and first do the integral over $q_f$. Why? Revisit Eq.~(\ref{eq:action for generating function for harmonic oscillator}): Since the final field configuration is required to be the same, the quadratic terms in $S_{\text{cl}}(q_0,q_f|J) - S_{\text{cl}}(q'_0,q_f|J')$ will drop out!\footnote{This was also noticed by Anglin \cite{Anglin1993}, undoubtedly among others.} The integral over $q_f$ will produce a delta function that constrains $q_0-q'_0$, leaving behind only an integral over $q_0+q'_0$. Only then will it matter which $\rho$ you choose, and the resulting integral will not be so bad. 
\\\\
In the end, I wanted to evaluate $Z(J,J')$ using both methods anyway (at least for the oscillator in its ground state), both to check my work (and triple-check the reference) and to show that the order of integration does not matter. It is also important, for the sake of making progress, to accept that you will never know the right path until you just try one \cite{catmull}. 
\\\\
For most of the examples, I will present the more complicated method. Partly because I do not feel like redoing everything, and partly to save you the trouble.
\vfill
\pagebreak
\subsection{Ground state}\label{sec:ground state}
Back to your regularly scheduled programming. The ground-state wavefunction of the harmonic oscillator is 
\begin{equation}\label{eq:ground state}
	\bra q|0\ket = \left(\tfrac{m}{\pi}\right)^{\fourth} e^{-\half m q^2}\;.
\end{equation}
Preparing the oscillator in its ground state means selecting a density matrix
\begin{equation}\label{eq:ground-state density matrix}
	\rho(q,q') = \bra q|0\ket \bra 0|q'\ket = \sqrt{\tfrac{m}{\pi}}\;e^{-\half m(q^2+q'^2)}\;.
\end{equation}
\subsubsection{Integrals over initial field configurations}
The integral over the initial condition factorizes:
\begin{equation}\label{eq:initial condition factorizes}
	\INT dq_0\INT dq'_0 \;\rho(q_0,q'_0)\;Z(q_0,q_f|J)\,Z(q'_0,q_f|J')^* = \Psi(q_f|J)\,\Psi(q_f|J')^*\;,
\end{equation}
where
\begin{align}
	\Psi(q_f|J) &= \INT dq_0 \;\bra q_0|0\ket\, Z(q_0,q_f|J) = \left(\frac{m}{\pi}\right)^{\fourth} \frac{1}{\sqrt{2\pi i G(T)}}\INT dq_0\;e^{-\half m q_0^2 + i S_{\text{cl}}(q_0,q_f|J)}\;.
\end{align}
Extract the part that does not depend on $q_0$:
\begin{equation} \label{eq:Psi for ground state}
	\Psi(q_f|J) = \left(\frac{m}{\pi}\right)^{\fourth} \frac{1}{\sqrt{2\pi i G(T)}}\, e^{\,i \Scl(0,q_f|J)}\; \Omega(q_f|J)\;,
\end{equation}
with
\begin{align}
	\Omega(q_f|J) &= \INT dq_0\;e^{-\half m q_0^2 + i\frac{1}{2G(T)}\left\{ \dot G(T) q_0^2 +2q_0\left[\int_{t_0}^{t_f} \!\!dt\, G(t_f-t)J(t)-q_f\right]\right\} } \nonumber\\
	&= \INT dq_0\;e^{\,\frac{i}{2G(T)} \left\{ \left[\dot G(T) + im G(T)\right] q_0^2 + 2q_0\left[\int_{t_0}^{t_f} \!\!dt\, G(t_f-t)J(t)-q_f\right]\right\}} \nonumber\\
	&= \INT dq_0\;e^{\,\frac{i e^{\,i m T}}{2G(T)} \left\{ q_0^2 + 2q_0 e^{-imT}\left[\int_{t_0}^{t_f} \!\!dt\, G(t_f-t)J(t)-q_f\right]\right\}} \qquad\left(\dot G(T) + im G(T) = e^{\,imT}\right) \nonumber\\
%	&= e^{-\frac{ie^{\,imT}}{2G(T)} \left\{ e^{-imT} \left[\int_{t_0}^{t_f} \!\!dt\, G(t_f-t)J(t)-q_f\right]\right\}^2 } \INT dq_0\;e^{-\frac{1}{2i G(T) e^{-imT}} \left\{ q+e^{-imT}\left[\int_{t_0}^{t_f} \!\!dt\, G(t_f-t)J(t)-q_f\right]\right\}^2 } \nonumber\\
	&= e^{-i\frac{e^{-imT}}{2G(T)} \left[\int_{t_0}^{t_f} \!\!dt\, G(t_f-t)J(t)-q_f\right]^2} \sqrt{2\pi i G(T) e^{-imT}}\;. \label{eq:Phi for ground state}
\end{align}
The $\sqrt{2\pi i G(T)}$ from Eq.~(\ref{eq:Phi for ground state}) cancels the $1/\sqrt{2\pi i G(T)}$ from Eq.~(\ref{eq:Psi for ground state}), and the $\sqrt{e^{-imT}}$ from Eq.~(\ref{eq:Phi for ground state}) cancels out of the $\Omega(q_f|J)\Omega(q_f|J')^*$ implied by Eq.~(\ref{eq:initial condition factorizes}). The generating function is
\begin{align}
	&Z(J,J') = \INT \!dq_f \INT \!dq_0 \INT \!dq'_0\; \rho(q_0,q'_0)\, Z(q_0,q_f|J)\, Z(q'_0,q_f|J')^* \nonumber\\
	&= \INT \!dq_f\; \Psi(q_f|J)\,\Psi(q_f|J')^* \nonumber\\
	&= \sqrt{\frac{m}{\pi}} \INT dq_f\;e^{\,i\left[ \Scl(0,q_f|J)-\Scl(0,q_f|J') \right]} \frac{\Omega(q_f|J)}{\sqrt{2\pi i G(T)}} \frac{\Omega(q_f|J')^*}{\sqrt{-2\pi i G(T)}} \nonumber\\
	&= \sqrt{\frac{m}{\pi}} \INT dq_f\;e^{\,i\left[ \Scl(0,q_f|J)-\Scl(0,q_f|J') \right]} e^{-i\frac{e^{-imT}}{2G(T)} \left[\int_{t_0}^{t_f} \!\!dt\, G(t_f-t)J(t)-q_f\right]^2} e^{+i\frac{e^{+imT}}{2G(T)} \left[\int_{t_0}^{t_f} \!\!dt\, G(t_f-t)J'(t)-q_f\right]^2} \nonumber\\
	&\equiv \sqrt{\frac{m}{\pi}} \INT dq_f\;e^{\,\frac{i}{2G(T)} I(q_f|J,J')}\;.
\end{align}
\subsubsection{Integral over final field configuration}
Time to organize that $I(q_f|J,J')$:
\begin{align}
	&I(q_f|J,J') = \dot G(T)q_f^2+2q_f\int_{t_0}^{t_f}\!\! dt\; G(t\!-\!t_0)J(t) - 2\doubleint \Ta(t\!-\!t') G(t_f\!-\!t)G(t'\!-\!t_0)J(t)J(t') \nonumber\\
	&-e^{-imT}\left[q_f - \int_{t_0}^{t_f}\!\! dt\;G(t_f\!-\!t)J(t)\right]^2 \nonumber\\
	&-\dot G(T)q_f^2 - 2q_f\int_{t_0}^{t_f}\!\! dt\; G(t\!-\!t_0) J'(t) + 2\doubleint\Ta(t\!-\!t')G(t_f\!-\!t)G(t'\!-\!t_0)J'(t)J'(t') \nonumber\\
	&+ e^{+imT}\left[q_f - \int_{t_0}^{t_f}\!\! dt\; G(t_f\!-\!t)J'(t)\right]^2 \nonumber\\
	&\nonumber\\
	&= 2im G(T)\, q_f^2 + 2q_f\int_{t_0}^{t_f}\!\! dt\left\{\left[G(t\!-\!t_0)\!+\!e^{-imT}G(t_f\!-\!t)\right]J(t) - \left[G(t\!-\!t_0)\!+\!e^{\,imT}G(t_f\!-\!t)\right]J'(t)\right\} \nonumber\\
	&-\int_{t_0}^{t_f}\!\!\! dt\!\int_{t_0}^{t_f}\!\!\! dt' \left\{ \Big[\Ta(t\!-\!t')\,G(t_f\!-\!t)G(t'\!-\!t_0)\!+\!\Ta(t'\!-\!t)\,G(t_f\!-\!t')G(t\!-\!t_0) \Big.\right. \nonumber\\
	&\qquad\qquad\qquad\qquad\qquad\qquad\qquad\qquad\qquad\qquad\qquad\Big.+\;G(t_f\!-\!t)G(t_f\!-\!t')e^{-imT}\Big] J(t)J(t') \nonumber\\
	&\qquad\qquad\;\;\;\;\;\,-\Big[\Ta(t\!-\!t')G(t_f\!-\!t)G(t'\!-\!t_0)+\Ta(t'\!-\!t)G(t_f\!-\!t')G(t\!-\!t_0) \Big. \nonumber\\
	&\qquad\qquad\qquad\qquad\qquad\qquad\qquad\qquad\qquad\qquad\qquad\left.\Big.+G(t_f\!-\!t)G(t_f\!-\!t')e^{+imT}\Big] J'(t)J'(t') \right\}\;.
\end{align}
Defining
\begin{equation}\label{eq:alpha(t) for ground state}
	\al(t) \equiv G(t\!-\!t_0) + e^{\,imT}G(t_f\!-\!t)
\end{equation}
and
\begin{equation}
	j(J,J') \equiv \int_{t_0}^{t_f}\!\! dt\left[\al(t)^* J(t) - \al(t)J'(t)\right]\;,
\end{equation}
I will need to calculate the following integral:
\begin{equation}
	\INT dq_f\;e^{\,\frac{i}{2G(T)}\left[2imG(T) q_f^2 + 2j(J,J')q_f\right]} = \sqrt{\frac{\pi}{m}}\;e^{-\frac{1}{4mG(T)^2}j(J,J')^2}\;.
\end{equation}
The overall factor cancels, and the influence phase $\Phi(J,J') \equiv -i\ln Z(J,J')$ is
\begin{align}\label{eq:influence phase for oscillator unsimplified}
	&\Phi(J,J') = \frac{1}{2G(T)}\doubleint\left\{ \frac{i}{2m G(T)}\left[\al(t)^* J(t)-\al(t)J'(t)\right]\left[\al(t')^*J(t')-\al(t')J'(t')\right]\right. \nonumber\\
	&-\left[\Ta(t\!-\!t')G(t_f\!-\!t)G(t'\!-\!t_0)+\Ta(t'\!-\!t)G(t_f\!-\!t')G(t\!-\!t_0)+G(t_f\!-\!t)G(t_f\!-\!t')e^{-imT}\right] J(t)J(t') \nonumber\\
	&\left.+\left[\Ta(t\!-\!t')G(t_f\!-\!t)G(t'\!-\!t_0)+\Ta(t'\!-\!t)G(t_f\!-\!t')G(t\!-\!t_0)+G(t_f\!-\!t)G(t_f\!-\!t')e^{+imT}\right] J'(t)J'(t')\phantom{\frac{1}{2}}\!\!\!\right\}\;.
\end{align}
Now for simplification. I will simplify the $JJ$ term by hand then let you program the rest into Mathematica. 
%\\\\
\subsubsection{Simplification of the $JJ$ term}
For $t>t'$, the coefficient of $J(t)J(t')$ in the integrand of Eq.~(\ref{eq:influence phase for oscillator unsimplified}) is
\begin{align}\label{eq:coefficient of JJ for t>t' for oscillator in ground state}
	K(t,t') &\equiv -\left[G(t_f\!-\!t)G(t'\!-\!t_0)+G(t_f\!-\!t)G(t_f\!-\!t')e^{-imT}\right]+i\frac{\al(t)^*\al(t')^*}{2m G(T)} \nonumber\\
	&= \frac{i}{2m G(T)}\left[G(t\!-\!t_0)+e^{\,imT}G(t_f\!-\!t)\right]\left[G(t'\!-\!t_0)+e^{-imT}G(t_f\!-\!t')\right]\nonumber\\
	&= \frac{i}{2m G(T)}\left( R+iS\right)\;,
\end{align}
with
\begin{align}
	&R \equiv G(t\!-\!t_0)G(t'\!-\!t_0)+G(t_f\!-\!t)G(t_f\!-\!t')+\cos(mT)\left[ G(t_f\!-\!t)G(t'\!-\!t_0)+G(t_f\!-\!t')G(t\!-\!t_0) \right]\;, \nonumber\\
	&S \equiv \sin(mT)\left[G(t_f\!-\!t) G(t'\!-\!t_0)-G(t_f\!-\!t')G(t\!-\!t_0)\right]\;.
\end{align}
By writing $G(t) = \frac{1}{m}\sin(mt) = \frac{1}{2im}\left(e^{\,imt}-e^{-imt}\right)$, I find that $S$ reduces to
\begin{equation}
	S = -G(T)^2\, mG(t\!-\!t')\;.
\end{equation}
The expression $R$ requires more work. First, using $\sin(a$$-$$b) = \sin(a)\sin(b)$~$-$~$ \cos(a)\sin(b)$ with $a$~$=$~$m(t_f$$-$$t)$ and $b$~$=$~$m(t_f$$-$$t_0)$, I recognize that $G(t$$-$$t_0)$~$+$~$\cos(mT)G(t_f$$-$$t)$ $=$ $G(T)\cos[m(t_f$$-$$t)]$, and $G(t_f$$-$$t)$~$+$~$\cos(mT)G(t$$-$$t_0)$ $=$ $G(T)\cos[m(t$$-$$t_0)]$. So $R$ can be rewritten as
\begin{equation}
	R = \frac{G(T)}{m}\left\{ \cos[m(t_f\!-\!t)]\sin[m(t'\!-\!t_0)] + \cos[m(t\!-\!t_0)]\sin[m(t_f\!-\!t')] \right\}\;.
\end{equation}
Now I can use a trigonometric cyclic identity that I do not remember from high school:
\begin{equation}
	\cos(a\!-\!b)\sin(c\!-\!d)+\cos(b\!-\!d)\sin(a\!-\!c)+\cos(b\!-\!c)\sin(d\!-\!a) = 0\;.
\end{equation}
With $a = mt_f$, $b = mt$, $c = mt'$, and $d = mt_0$, I therefore get
\begin{equation}
	R = G(T)^2\, \dot G(t\!-\!t')\;.
\end{equation}
So the coefficient of $J(t)J(t')$ for $t > t'$ is
\begin{equation}
	K(t,t') = i\frac{G(T)}{2m}\left[\dot G(t\!-\!t')-i m G(t\!-\!t')\right] = G(T)\frac{i}{2m}e^{-im(t-t')}\;.
\end{equation}
For $t < t'$, the coefficient of $J(t)J(t')$ is
\begin{align}
	K'(t,t') &\equiv -\left[G(t_f\!-\!t')G(t\!-\!t_0)+G(t_f\!-\!t)G(t_f\!-\!t')e^{-imT}\right]+i\frac{\al(t)^*\al(t')^*}{2m G(T)} \nonumber\\
	%&= \frac{i}{2m G(T)}\left[G(t'\!-\!t_0) + e^{\,imT} G(t_f\!-\!t') \right]\left[ G(t\!-\!t_0)+e^{-imT} G(t_f\!-\!t)\right]\;.
	&=K(t',t) = G(T)\frac{i}{2m}e^{+im(t-t')}\;.
\end{align}
So the coefficient of $J(t)J(t')$ is
\begin{align}
	\Ta(t\!-\!t')K(t,t')+\Ta(t'\!-\!t) K'(t,t') &= G(T)\frac{i}{2m}\left[\Ta(t\!-\!t') e^{-im(t-t')}+\Ta(t'\!-\!t)e^{\,im(t-t')}\right] \nonumber\\
	&= G(T)\frac{i}{2m}e^{-im|t-t'|}\;.
\end{align}
%\\\\
Having derived that explicitly, I will state the results. 
\subsubsection{Influence phase}
First, some notation and terminology. The positive-frequency Wightman function, or ``greater Green's function,'' is\footnote{My conventions for Wightman functions differ by factors of $i$ from those of Birrell and Davies \cite{BD}.}
\begin{equation}\label{eq:G_>(t) for ground state}
	G_>(t) \equiv \frac{i}{2m}\,e^{-im t}\;,
\end{equation}
and the negative-frequency Wightman function, or ``lesser Green's function,'' is
\begin{equation}\label{eq:G_<(t) for ground state}
	G_<(t) \equiv \frac{i}{2m}\,e^{+im t} = -G_>(t)^*\;.
\end{equation}
The Feynman function is
\begin{equation}\label{eq:G_F(t) for ground state}
	G_F(t) \equiv \Ta(t)\,G_>(t) + \Ta(-t)\,G_<(t) = \frac{i}{2m}\,e^{-im|t|}\;,
\end{equation}
and the Dyson function is
\begin{equation}\label{eq:G_D(t) for ground state}
	G_D(t) \equiv -\left[\Ta(t)\,G_<(t) + \Ta(-t)\,G_>(t)\right] = -\frac{i}{2m}\,e^{+im|t|} = G_F(t)^*\;.
\end{equation}
In terms of those, the influence phase is
\begin{align}\label{eq:ground state influence phase}
	\Phi(J,J') &= \half \int_{t_0}^{t_f}\!\! dt\int_{t_0}^{t_f}\!\! dt'\; \Big[G_F(t\!-\!t')J(t)J(t')-G_D(t\!-\!t')J'(t)J'(t') \Big. \nonumber\\
	&\qquad \qquad \qquad \Big. -G_<(t\!-\!t') J(t)J'(t') - G_>(t\!-\!t') J'(t)J(t') \Big]\;.
\end{align}
That is what you should have expected. For example, in terms of the Feynman function $G_F(t)$, the Feynman expectation value is
\begin{align}
	\tr\left[\mathcal T\!\left(\hat q(t_1)\hat q(t_2)\right)e^{\,it_0\hat H}\hat\rho\,e^{-it_0\hat H}\right] &= -\left. \frac{\del^2 Z(J,J')}{\del J(t_1)\del J(t_2)}\right|_{J\,=\,J'\,=\,0} = -i\left.\frac{\del^2\Phi(J,J')}{\del J(t_1)\del J(t_2)}\right|_{J\,=\,J'\,=\,0} \nonumber\\
	&= -i G_F(t_1\!-\!t_2)\;.
\end{align}
More generally, because all of my intuition about effective field theory comes from path integrals, I \textit{define} $G_F$, $G_D$, $G_<$, and $G_>$ as the coefficients of $JJ$, $J'J'$, $JJ'$, and $J'J$ in the influence phase. 
\subsection{Other order of integration}\label{sec:other order}
As I mentioned in Sec.~\ref{sec:preliminary commentary}, despite the way I wrote Eq.~(\ref{eq:keldysh2}) I could have instead performed the ``final'' integration first:
\begin{equation}
	Z(J,J') = \INT dq_0 \INT dq'_0\;\rho(q_0,q'_0)\;Y(q_0,q'_0|J,J')\;,
\end{equation}
with
\begin{equation}\label{eq:Y}
	Y(q_0,q'_0|J,J') \equiv \INT dq_f\;Z(q_0,q_f|J)\,Z(q'_0,q_f|J')^*\;.
\end{equation}	
Since the $q_f^2$ term in Eq.~(\ref{eq:action for generating function for harmonic oscillator}) does not depend on $q_0$ or $J$, the quadratic terms will drop out of the product of amplitude-generating functions in Eq.~(\ref{eq:Y}):
\begin{equation}
	Z(q_0,q_f|J)\,Z(q'_0,q_f|J')^* = \frac{1}{2\pi G(T)} e^{\,i\left[ \Scl(q_0,q_f|J) - \Scl(q'_0,q_f|J')\right]}\;.
\end{equation}
Reorganizing the action as
\begin{align}
	\Scl(q_0,q_f|J) &= \frac{1}{2G(T)} \left\{ \dot G(T)q_f^2 + 2\left[\singleint G(t\!-\!t_0)\,J(t)-q_0\right]q_f + F(q_0|J) \right\}\;,
\end{align}
with
\begin{align}\label{eq:q_f-independent part}
	F(q_0|J) &= \dot G(T)q_0^2 + 2q_0\singleint G(t_f\minus t)\,J(t) - 2\doubleint \Ta(t\minus t')\,G(t_f\minus t)G(t'\minus t_0)J(t)J(t')\;,
\end{align}
I find for the difference in actions:
\begin{align}
	\Scl(q_0,q_f|J)\minus \Scl(q_0',q_f|J) &= \frac{1}{G(T)}\left[\singleint G(t\minus t_0)\left(J(t)\minus J'(t)\right)-q_0+q_0'\right]q_f \nonumber\\
	&+ \frac{1}{2G(T)}\left[F(q_0|J)\minus F(q_0'|J')\right].
\end{align}
Recalling that $\INT dq\,e^{\,ikq/a} = 2\pi a\,\del(k)$ (with $a$ assumed positive), I learn that the integral in Eq.~(\ref{eq:Y}) will constrain the difference in initial field configurations:
\begin{align}
	Y(q_0,q_0'|J,J') &= \frac{1}{2\pi G(T)} e^{\,\frac{i}{2G(T)}\left[F(q_0|J)-F(q_0'|J')\right]}\INT dq_f\;e^{\,\frac{i}{G(T)}\left[\singleint G(t\minus t_0)\left(J(t)\minus J'(t)\right)-q_0+q_0'\right] q_f} \nonumber\\
	&= e^{\,\frac{i}{2G(T)}\left[F(q_0|J)-F(q_0'|J')\right]}\;\del\left[\singleint G(t\minus t_0)\left(J(t)\minus J'(t)\right)-q_0+q_0'\right]\;.
\end{align}
The generating function becomes
\begin{align}
	Z(J,J') &= \INT dq_0'\;\rho\!\left(q_0'+\!\singleint G(t\minus t_0)\left(J(t)\minus J'(t)\right),\,q_0'\right) \times \nonumber\\
	&\qquad\qquad \qquad\qquad\qquad e^{\,\frac{i}{2G(T)}\left[F(q_0'+\singleint G(t - t_0)\left(J(t)-J'(t)\right)|J) - F(q_0'|J') \right]}\;.
\end{align}
Shift the integration variable to symmetrize the integrand:
\begin{equation}\label{eq:symmetrize q0 integration}
	q_0' \equiv q - \half k\;,\implies q_0'+k = q+\half k\;;\;\text{with}\; k \equiv \singleint G(t\minus t_0)\left(J(t)\minus J'(t)\right)\;.
\end{equation}
The generating function is now\footnote{Experts in the Schwinger-Keldysh formalism are probably fuming that I am not yet working in the sum and difference basis. Cool your jets and wait till Sec.~\ref{sec:iepsilon}.}
\begin{align}
	Z(J,J') &= \INT dq\;\rho\!\left( q+\frac{k}{2},\,q-\frac{k}{2}\right)\;e^{\,\frac{i}{2G(T)}\left[F(q+\frac{k}{2}|J) - F(q-\frac{k}{2}|J')\right]}\;.
\end{align}
%\\\\ 
From Eq.~(\ref{eq:q_f-independent part}) and $(q+\frac{k}{2})^2-(q-\frac{k}{2})^2 = 2qk$, I find
%\\\\
\begin{align}
	&F(q+\frac{k}{2}|J) - F(q-\frac{k}{2}|J') = 2q\singleint\left[\dot G(T)G(t\minus t_0)+G(t_f\minus t)\right]\left[J(t)\minus J'(t)\right] + B(J,J')\;,
\end{align}
with the $q$-independent part being
\begin{align}
	B(J,J') &= \doubleint G(t_f\minus t)G(t'\minus t_0)\left[J(t)\plus J'(t)\right]\left[J(t')\minus J'(t')\right] \nonumber\\
	&-2\doubleint \Ta(t\minus t') G(t_f\minus t)G(t'\minus t_0)\left[J(t)J(t')\minus J'(t)J'(t')\right]\;.
\end{align}
The generating function is now
\begin{align}\label{eq:intermediate form of Z from alternative order of integration}
	Z(J,J') &= e^{\,\frac{i}{2G(T)}B(J,J')} \times\nonumber\\
	&\!\INT \!\!dq\;\;\rho\!\left(q+\half\!\singleint G(t\minus t_0)\left[J(t)\!-\!J'(t)\right],\,q-\half\! \singleint G(t\minus t_0)\left[J(t)\!-\!J'(t)\right]\right) \nonumber\\
	&\qquad \times e^{\,\frac{i}{G(T)}q\singleint\left[\dot G(T)G(t- t_0) + G(t_f- t)\right]\left[J(t)-J'(t)\right]}\;.
\end{align}
\subsubsection{Ground state}
Now it is time to pick $\rho$. I will complete this calculation only for $\hat\rho = |0\ket \bra 0|$, which is why this whole business is not in its own section devoted to the alternative order of integration. With $\bra q|0\ket = (m/\pi)^{1/4}e^{-\half m q^2}$ as before, and with
\begin{equation}
	(q+\frac{k}{2})^2 + (q-\frac{k}{2})^2 = 2q^2 + \frac{k^2}{2}\;,
\end{equation}
I find
\begin{equation}
	\rho(q+\frac{k}{2},q-\frac{k}{2}) = \sqrt{\frac{m}{\pi}}\;e^{-\fourth m k^2} e^{-m q^2}\;.
\end{equation}
So the generating function is
\begin{equation}
	Z(J,J') = e^{\,\frac{i}{2G(T)}B(J,J')} \sqrt{\frac{m}{\pi}}e^{-\fourth m k^2} I(J,J')\;,
\end{equation}
with
\begin{equation}
	I(J,J') = \INT dq\;e^{-mq^2 + \frac{i}{G(T)}q\ell}\;,\;\; \ell = \singleint\left[\dot G(T)G(t\minus t_0)+G(t_f\minus t)\right]\left[J(t)\minus J'(t)\right]\;.
\end{equation}
Completing the square and performing the Gaussian integral gives
\begin{equation}
	\Phi(J,J') = -i\ln Z(J,J') = \frac{i}{4G(T)}\left[2B(J,J') + im G(T)k^2 + \frac{i}{m G(T)}\ell^2\right]\;,
\end{equation}
with $k = \singleint G(t\minus t_0)\left[J(t)\minus J'(t)\right]$ from Eq.~(\ref{eq:symmetrize q0 integration}). 
\\\\
On to simplification. First, $B(J,J')$. I trust that after symmetrizing in $t$ and $t'$, and by treating the $t > t'$ and $t<t'$ cases separately, you will obtain the form
\begin{align}
	B(J,J') &= \half\doubleint \left[G(t_f\minus t')G(t\minus t_0) - G(t_f\minus t)G(t'\minus t_0)\right] \nonumber\\
	&\times \left\{\Ta(t\minus t')\left[J(t)\minus J'(t)\right]\left[J(t')\plus J'(t')\right] - \Ta(t'\minus t)\left[J(t)\plus J'(t)\right]\left[J(t')\minus J'(t')\right] \right\}.
\end{align} 
I trust you further to combine that with
\begin{align}
	\ell^2 + m^2 G(T)^2 k^2 &= \doubleint \left\{\phantom{\frac{1}{2}}\!\!\!\!\left[\dot G(T)G(t\minus t_0) + G(t_f\minus t)\right]\left[\dot G(T)G(t'\minus t_0) + G(t_f\minus t')\right]\right. \nonumber\\
	&\left.+m^2G(T)^2 G(t\minus t_0)G(t'\minus t_0)\!\!\!\phantom{\frac{1}{2}}\right\}\left[J(t)\minus J'(t)\right] \left[J(t')\minus J'(t')\right]\;,
\end{align}
by which I mean take the resulting coefficients of $J(t)J(t')$, $J(t)J'(t')$, $J'(t)J(t')$, and $J'(t)J'(t')$, and insert them into Mathematica. The essential simplification is the following:
\begin{align}
	&\left[\dot G(T) G(t\minus t_0) + G(t_f\minus t)\right]\left[\dot G(T)G(t'\minus t_0)+G(t_f\minus t')\right] + m^2G(T)^2 G(t\minus t_0)G(t'\minus t_0) \nonumber\\
	&\qquad +im G(T)\left[ G(t_f\minus t')G(t\minus t_0) - G(t_f\minus t)G(t'\minus t_0)\right] = G(T)^2\;e^{\,im(t-t')}\;.
\end{align}
With that, you will arrive at Eq.~(\ref{eq:ground state influence phase}) with the definitions given in that subsection. 
\pagebreak
\subsection{Thermal state}\label{sec:thermal state}
Ludwig E. Boltzmann, I choose you:
\begin{equation}
	\hat\rho = N\,e^{-\beta \hat H}\;,\;\; N = 1/\tr(e^{-\beta \hat H})\;.
\end{equation}
As Feynman and Vernon pointed out, the position-space representation of $e^{-\beta \hat H}$ is an imaginary-time transition amplitude:
\begin{equation}
	\bra q_2| e^{-\beta \hat H}|q_1\ket = \left.Z(q_1,q_2|0)\right|_{t_2-t_1 \to -i\beta} = \frac{1}{\left[2\pi i G(-i\beta)\right]^{1/2}} e^{\,\frac{i}{2G(-i\beta)}\left[\dot G(-i\beta)(q_1^2+q_2^2)-2q_1q_2\right]}\;.
\end{equation}
Fix $N$ by demanding $\tr(\hat\rho) = 1$:
\begin{align}
	\INT dq\; \bra q|\hat\rho|q\ket &= \frac{N}{\left[2\pi i G(-i\beta)\right]^{1/2}}\INT dq\;e^{\,\frac{i}{G(-i\beta)} \left[\dot G(-i\beta)-1\right]q^2 } = \frac{N}{\left\{2\left[\dot G(-i\beta)-1\right]\right\}^{1/2}} \equiv 1 \nonumber\\
	&\implies N = \left\{2\left[\dot G(-i\beta)-1\right]\right\}^{1/2}\;.
\end{align}
So the thermal density matrix in the field basis is\footnote{Simplify the prefactor if you want to, but that is the most convenient way to write it.}
\begin{equation}\label{eq:thermal density matrix}
	\rho(q_1,q_2) = \left(\tfrac{\dot G(-i\beta)-1}{i\pi G(-i\beta)}\right)^{1/2} e^{\,\frac{i}{2G(-i\beta)} \left[\dot G(-i\beta)(q_1^2+q_2^2)-2q_1q_2\right] }\;.
\end{equation}
%%%%
\subsubsection{Integrals over initial field configurations}
Once again, the basic formula is Eq.~(\ref{eq:keldysh}):
\begin{equation}
	Z(J,J') = \INT dq_f \INT dq_0 \INT dq'_0\;\rho(q_0,q'_0)\;Z(q_0,q_f|J)\,Z(q'_0,q_f|J')^*\;.
\end{equation}
This time, however, the double integral over the initial field configurations does not factorize, so it is worth passing to a 2-by-2 matrix representation:
\begin{equation}
	\vec Q \equiv \ml q_0\\ q'_0 \mr \implies \rho(q_0,q'_0) = \left(\tfrac{\dot G(-i\beta)-1}{i\pi G(-i\beta)}\right)^{1/2} e^{\,\frac{i}{2G(-i\beta)} \vec Q^T\ml \dot G(-i\beta) &-1 \\ -1 &\dot G(-i\beta) \mr \vec Q}\;.
\end{equation}
The generating function is
\begin{align}
	Z(J,J') &= \left(\tfrac{\dot G(-i\beta)-1}{i\pi G(-i\beta)}\right)^{1/2}\tfrac{1}{2\pi G(T)}\, e^{\,\frac{i}{2G(T)}(-2)\doubleint \Ta(t-t')G(t_f-t)G(t'-t_0)\left[ J(t)J(t')-J'(t)J'(t') \right]}\;\times \nonumber\\
	&\INT dq_f\;e^{\,\frac{i}{2G(T)}(+2)\int_{t_0}^{t_f}\!\! dt\;q_f\;G(t-t_0)\left[J(t)-J'(t)\right]}I(q_f|J,J')\;,
\end{align}
where
\begin{equation}
	I(q_f|J,J') = \int_{R_2} d^2 Q\;e^{\,\frac{i}{G(T)}\left(\half \vec Q^T M \vec Q + \vec j^T\vec Q\right)}\;,
\end{equation}
with
\begin{equation}
	M = \ml \dot G(T) + \frac{\dot G(-i\beta)}{G(-i\beta)} G(T) & -\frac{G(T)}{G(-i\beta)} \\ -\frac{G(T)}{G(-i\beta)} & -\dot G(T) + \frac{\dot G(-i\beta)}{G(-i\beta)} G(T) \mr\;,\;\;\vec j = \ml \int_{t_0}^{t_f}\!\! dt\;G(t_f\!-\!t) J(t)-q_f \\ -\left[\int_{t_0}^{t_f}\!\! dt\;G(t_f\!-\!t)J'(t)-q_f\right] \mr\;.
\end{equation}
The Gaussian integral produces
\begin{equation}
	I(q_f|J,J') = 2\pi G(T)\;e^{-\frac{i}{2G(T)} \vec j^T M^{-1} \vec j}\;,
\end{equation}
with
\begin{equation}
	M^{-1} = \ml \dot G(T)-\frac{\dot G(-i\beta)}{G(-i\beta)} G(T) & - \frac{G(T)}{G(-i\beta)} \\ -\frac{G(T)}{G(-i\beta)} & -\left[\dot G(T) + \frac{\dot G(-i\beta)}{G(-i\beta)} G(T)\right] \mr\;.
\end{equation}
Therefore: 
\begin{align}
	&\vec j^T M^{-1} \vec j = \left[\dot G(T)-\frac{\dot G(-i\beta)}{G(-i\beta)}G(T)\right]\doubleint G(t_f\!-\!t)G(t_f\!-\!t') J(t) J(t') \nonumber\\
	&-\left[\dot G(T)+\frac{\dot G(-i\beta)}{G(-i\beta)}G(T)\right]\doubleint G(t_f\!-\!t)G(t_f\!-\!t') J'(t)J'(t')\nonumber\\
	&+2\frac{G(T)}{G(-i\beta)}\doubleint G(t\!-\!t_f)G(t_f\!-\!t')J(t)J'(t')-2G(T)\left(\tfrac{\dot G(-i\beta)-1}{G(-i\beta)}\right)q_f^2 \nonumber\\
	&-2q_f\int_{t_0}^{t_f}\!\! dt\;G(t_f\!-\!t)\left\{ \left[\dot G(T)-\left(\tfrac{\dot G(-i\beta)-1}{G(-i\beta)}\right)G(T)\right]J(t) -\left[\dot G(T) + \left(\tfrac{\dot G(-i\beta)-1}{G(-i\beta)}\right) G(T) \right]J'(t) \right\}\;.
\end{align}
With that, the generating function becomes
\begin{align}
	&Z(J,J') = \left(\tfrac{\dot G(-i\beta)-1}{i\pi G(-i\beta)}\right)^{1/2} e^{\,\frac{i}{2G(T)}(-2)\doubleint \Ta(t-t') G(t_f-t)G(t'-t_0)\left[ J(t)J(t')-J'(t)J'(t')\right]} \times \nonumber\\
	&e^{-\frac{i}{2G(T)} \doubleint G(t_f-t)G(t_f-t')\left\{ \left[\dot G(T)-\frac{\dot G(-i\beta)}{G(-i\beta)} G(T) \right]J(t)J(t') - \left[\dot G(T)+\frac{\dot G(-i\beta)}{G(-i\beta)}G(T)\right] J'(t)J'(t') \right\} } X(J,J')\;,
\end{align}
where
\begin{align}
	X(J,J') &= \INT dq_f\;e^{\,i\left(\frac{\dot G(-i\beta)-1}{G(-i\beta)}\right) q_f^2 + \frac{i}{G(T)}q_f\int_{t_0}^{t_f}\!\! dt\left[\al(t)^* J(t) - \al(t) J'(t)\right]}\;,
\end{align}
with
\begin{equation}
	\al(t) = G(t\!-\!t_0) + \left[ \dot G(T) + \left(\frac{\dot G(-i\beta)-1}{G(-i\beta)}\right) G(T) \right]G(t_f\!-\!t) \;.
\end{equation}
Note that when $\beta \to \infty$, $\al(t) \to G(t$$-$$t_0) + e^{\,imT} G(t_f$$-$$t)$, recovering the $\al(t)$ from Eq.~(\ref{eq:alpha(t) for ground state}). 
\subsubsection{Integral over final field configuration}
With
\begin{equation}
	j(J,J') \equiv \int_{t_0}^{t_f}\!\! dt\left[\al(t)^* J(t)-\al(t) J'(t)\right]\;,
\end{equation}
I get
\begin{equation}
	X(J,J') = \left( \frac{i\pi G(-i\beta)}{\dot G(-i\beta)-1}\right)^{1/2} e^{-\frac{iG(-i\beta)}{[\dot G(-i\beta)-1] 4G(T)^2} j(J,J')^2}\;.
\end{equation}
The overall factor cancels, and I get the influence phase $\Phi(J,J') \equiv -i\ln Z(J,J')$ in the form
\begin{align}\label{eq:unsimplified thermal influence phase}
	&\Phi(J,J') =\nonumber\\
	& -\tfrac{G(-i\beta)}{[\dot G(-i\beta)-1]4G(T)^2} \left\{ \int_{t_0}^{t_f}\!\! dt\left[\al(t)^*J(t)-\al(t)J'(t)\right]\right\}^2\!\!-\tfrac{1}{G(-i\beta)}\!\doubleint G(t_f\!-\!t)G(t_f\!-\!t')J(t)J'(t')  \nonumber\\
	&-\tfrac{1}{2G(T)}\!\!\doubleint\!\!\left[ \Ta(t\!-\!t')G(t_f\!-\!t)G(t'\!-\!t_0)+\Ta(t'\!-\!t)G(t_f\!-\!t')G(t\!-\!t_0) \right]\left[J(t)J(t')-J'(t)J'(t')\right] \nonumber\\
	&-\tfrac{1}{2G(T)} \!\!\int_{t_0}^{t_f}\!\! \!\!dt\!\!\int_{t_0}^{t_f}\!\! \!\!dt'\; \!G(t_f\!-\!t)G(t_f\!-\!t')\!\left\{\left[\dot G(T)\!-\!\tfrac{\dot G(-i\beta)}{G(-i\beta)}G(T)\right]\!\!J(t)J(t')\!-\!\!\left[\dot G(T)\!+\!\tfrac{\dot G(-i\beta)}{G(-i\beta)}G(T)\right]\!\!J'(t)J'(t')\!\right\}\!.
\end{align}
\subsubsection{Influence phase}
Now I ask Mathematica to simplify Eq.~(\ref{eq:unsimplified thermal influence phase}), term by term, frustration by frustration. The result is:
\begin{align}
	\Phi(J,J') &= \frac{i}{4m \sinh\left(\frac{\beta m}{2}\right)} \doubleint\left\{ \cos\left[m\left(|t\!-\!t'|+i\tfrac{\beta}{2}\right)\right] J(t) J(t') \right. \nonumber\\
	&\left.+ \cos\left[ m\left(|t\!-\!t'|-i\tfrac{\beta}{2}\right)\right] J'(t)J'(t')-2\cosh\left(\tfrac{\beta m}{2}\right) \cos\left[m\left(t\!-\!t'\right)\right]J(t)J'(t')\right\}\;.
\end{align} Denoting by $\Phi_0$ the ground-state expression from Eq.~(\ref{eq:ground state influence phase}), and defining the function
\begin{equation}
	\Del(t) \equiv \frac{1}{e^{\beta m}\!-\!1}\; \frac{i}{m}\cos(mt)\;,
\end{equation}
I recover the influence phase $\Phi(J,J') = -i\ln Z(J,J')$ as Feynman and Vernon expressed it:
\begin{align}
	\Phi(J,J') &= \Phi_0(J,J') + \half\int_{t_0}^{t_f}\!\! dt \int_{t_0}^{t_f}\!\! dt'\; \Del(t\!-\!t')\left[J(t)\!-\!J'(t)\right]\left[J(t')\!-\!J'(t')\right]\;.
\end{align}
The thermal correction respects Eqs.~(\ref{eq:influence phase property 1}) and~(\ref{eq:influence phase property 2}).
\pagebreak
\subsection{First excited state}\label{sec:first excited state}
%%%%
I sat at a desk in Pasadena and thought, you know, why not try the first excited state:
\begin{equation}
	\bra q|1\ket = \sqrt{2m}\left(\tfrac{m}{\pi}\right)^{1/4} q\;e^{-\half m q^2} = \sqrt{2m}\,q\,\bra q|0\ket\;.
\end{equation}
An oscillator in its first excited state would have a density matrix
\begin{equation}
	\rho(q,q') = \bra q|1\ket \bra 1|q'\ket = 2m\,qq'\,\bra q|0\ket \bra 0|q'\ket = 2m\sqrt{\tfrac{m}{\pi}}\,e^{-\half m (q^2+q'^2) + \ln(qq')}\;.
\end{equation}
That is not quadratic! It is, however, a hopping good time. Begin with the integral
\begin{align}
	&\psi_1(q_f|J) \equiv \INT dq_0\; \bra q_0|1\ket \;e^{\,iS_{\text{cl}}(q_0,q_f|J)} \nonumber\\
	&= \sqrt{2m}\left(\frac{m}{\pi}\right)^{1/4} e^{\,\frac{i}{2G(T)}\left[ \dot G(T)q_f^2+2q_f\int_{t_0}^{t_f}\!\! dt\;G(t-t_0)J(t)-2\doubleint \Ta(t-t')G(t_f-t)G(t'-t_0)J(t)J(t') \right]}\chi_1(q_f|J)\;,
\end{align}
where
\begin{align}
	&\chi_1(q_f|J) \equiv \INT dq_0\; q_0 \; e^{-\half m q_0^2 + \frac{i}{2G(T)}\left\{\dot G(T) q_0^2 + 2q_0\left[\int_{t_0}^{t_f}\!\! dt\;G(t_f-t)J(t)-q_f\right]  \right\}} \nonumber\\
	&= i G(T)\frac{\pa}{\pa q_f} \underbrace{\INT dq_0\;e^{\frac{i}{2G(T)}\left\{\left[\dot G(T)+im G(T) \right] q_0^2 +2q_0\left[ \int_{t_0}^{t_f}\!\! dt\;G(t_f-t)J(t) - q_f \right]\right\}}}_{\equiv\;\chi_0(q_f|J)}\;.
\end{align}
With $\dot G(T)+im G(T) = e^{\,imT}$, completing the square gives
\begin{align}
	\chi_0(q_f|J) &= \sqrt{2\pi i G(T) e^{-imT}}\;e^{-\frac{i e^{-imT}}{2G(T)}\left[\int_{t_0}^{t_f}\!\! dt\;G(t_f-t)J(t) - q_f\right]^2 }\;.
\end{align}
So
\begin{align}
	\chi_1(q_f|J) &= e^{-imT}\sqrt{2\pi i G(T) e^{-imT}} \left[q_f-\int_{t_0}^{t_f}\!\! dt\;G(t\!-\!t_f) J(t)\right] e^{-\frac{i e^{-imT}}{2G(T)} \left[q_f-\int_{t_0}^{t_f}\!\! dt\;G(t_f-t)J(t)\right]^2}\;.
\end{align}
Inserting that into $\psi_1(q_f|J)$ and complex-conjugating the result, I get
\begin{align}
	&\INT dq_f\;\psi_1(q_f|J)\,\psi_1(q_f|J')^* \nonumber\\
	&= 2m\left(\frac{m}{\pi}\right)^{1/2}2\pi G(T) e^{\,\frac{i}{2G(T)}\doubleint (-2)\Ta(t-t')G(t_f-t)G(t'-t_0)\left[J(t)J(t')-J'(t)J'(t')\right]} I(J,J')\;,
\end{align}
where
\begin{align}
	I(J,J') &\equiv \INT dq_f \left[q_f\!-\!\int_{t_0}^{t_f}\!\! dt\;G(t_f\!-\!t)J(t)\right]\left[q_f\!-\!\int_{t_0}^{t_f}\!\! dt'\;G(t_f\!-\!t')J'(t') \right] e^{\,\frac{i}{2G(T)} X(q_f|J,J')}\;,
\end{align}
with
\begin{align}
	X(q_f|J,J') &\equiv 2q_f\int_{t_0}^{t_f}\!\! dt\;G(t\!-\!t_0)\left[J(t)\!-\!J'(t)\right]\nonumber\\
	& -e^{-imT}\left[q_f\!-\!\int_{t_0}^{t_f}\!\! dt\;G(t_f\!-\!t)J(t)\right]^2+e^{\,imT}\left[q_f\!-\!\int_{t_0}^{t_f}\!\! dt\;G(t_f\!-\!t) J'(t)\right]^2\nonumber\\
	&\nonumber\\
	&= 2im G(T)\,q_f^2 + 2q_f\int_{t_0}^{t_f}\!\! dt\left[\al(t)^* J(T)-\al(t)J'(t)\right]\nonumber\\
	&-\doubleint G(t_f\!-\!t)G(t_f\!-\!t')\left[e^{-imT}J(t)J(t')-e^{\,imT}J'(t)J'(t')\right]\;,
\end{align}
where
\begin{equation}
	\al(t) \equiv G(t\!-\!t_0)+e^{\,imT} G(t_f\!-\!t)\;.
\end{equation}
Therefore,
\begin{align}
	I(J,J') &= e^{-\frac{i}{2G(T)} \doubleint G(t_f-t)G(t_f-t')\left[e^{-imT}J(t)J(t')-e^{\,imT}J'(t)J'(t')\right] } K(J,J')\;,
\end{align}
where
\begin{align}
	&K(J,J') \equiv\!\! \INT\!\! dq_f\left[q_f\!-\!\!\int_{t_0}^{t_f}\!\! \!\!dt\;G(t_f\!-\!t)J(t)\right]\left[q_f\!-\!\!\int_{t_0}^{t_f}\!\! \!\!dt'\;G(t_f\!-\!t')J'(t')\right] \times \nonumber\\
	&\qquad\qquad\qquad \qquad\qquad\qquad\qquad \qquad\qquad\qquad \qquad e^{-mq_f^2+\frac{iq_f}{G(T)} \int_{t_0}^{t_f}\!\! dt\;\left[\al(t)^*J(t)-\al(t)J'(t)\right]} \nonumber\\
	&=\left\{ \left[-iG(T)\pa_\mu\right]^2\!-\!\int_{t_0}^{t_f}\!\! \!\! dt\;G(t_f\!-\!t)\left[-iG(T)\pa_\mu\right] \right.\nonumber\\
	&\qquad\qquad\qquad\left.\left.\!+\doubleint  G(t_f\!-\!t)G(t_f\!-\!t')J(t)J'(t')\right\}M(J,J';\mu)\right|_{\mu\,=\,0}\;,
\end{align}
with
\begin{align}
	M(J,J';\mu) &\equiv \INT dq_f\;e^{-mq_f^2 +\frac{i}{G(T)}\left\{\mu+\int_{t_0}^{t_f}\!\! dt\left[ \al(t)^*J(t)-\al(t)J'(t)\right]\right\}q_f} = \left(\frac{\pi}{m}\right)^{1/2} e^{-\frac{j^2}{4mG(T)^2}}\;,
\end{align}
where $j \equiv \mu+\int_{t_0}^{t_f}\!\! dt\left[ \al(t)^*J(t)-\al(t)J'(t)\right]$. Using $\pa_\mu j = 1$ and $\pa_\mu^2 j = 0$, I find 
\begin{align}
	&\pa_\mu \left(e^{-\frac{j^2}{4mG(T)^2}}\right) = -\frac{j}{2mG(T)^2}e^{-\frac{j^2}{4mG(T)^2}}\;,\nonumber\\
	&\pa_\mu^2 \left(e^{-\frac{j^2}{4mG(T)^2}}\right) = -\frac{1}{2m G(T)^2}\left[1-\frac{j^2}{2mG(T)^2}\right] e^{-\frac{j^2}{4mG(T)^2}}\;.
\end{align}
So
\begin{align}
	K(J,J') &= \left.\left(\frac{\pi}{m}\right)^{1/2} N e^{-\frac{j^2}{4mG(T)^2}} \right|_{\mu\,=\,0}\;,
\end{align}
where
\begin{align}
	N &\equiv \frac{1}{2m}\left[1-\frac{j^2}{2m G(T)^2}\right] \!- \frac{ij}{2mG(T)} \int_{t_0}^{t_f}\!\!\!\! dt\;G(t_f\!-\!t)\left[J(t)\!+\!J'(t)\right] \nonumber\\
	&\qquad\qquad\qquad\qquad\qquad\qquad\left.+\!\int_{t_0}^{t_f}\!\!\!\! dt \!\int_{t_0}^{t_f}\!\!\!\! dt'\, G(t_f\!-\!t)G(t_f\!-\!t') J(t)J'(t')\right|_{\mu\,=\,0} \nonumber\\
	&= \frac{1}{2m}\left\{1+\doubleint\left[A(t,t')^*J(t)J(t')+A(t,t')J'(t)J'(t')+B(t,t')J(t)J'(t')\right]\right\}\;,
\end{align}
with the following functions:
\begin{align}
	&A(t,t') \equiv -\frac{\al(t) \al(t')}{2m G(T)^2} + i\frac{\al(t) G(t_f-t') + \al(t') G(t_f-t)}{2G(T)} \nonumber\\
	&\;\;\;\qquad = -\frac{1}{2m}\cos[m(t-t')]\;, \nonumber\\
	&B(t,t') \equiv \frac{\al(t)^*\al(t')}{m G(T)^2}-\frac{i}{G(T)}\left[\al(t)^*G(t_f\!-\!t')\!-\!\al(t')G(t_f\!-\!t)\right] + 2m G(t_f\!-\!t)G(t_f\!-\!t')\nonumber\\
	&\;\;\;\qquad  = \frac{1}{m}\cos[m(t\!-\!t')]\;.
\end{align}
The factor $N$ simplifies to
\begin{align}
	N &= \frac{1}{2m}\left\{1-\frac{1}{2m}\doubleint \cos[m(t-t')]\left[J(t)\!-\!J'(t)\right]\left[J(t')\!-\!J'(t')\right]\right\}\;.
\end{align}
Finally I can put everything together to obtain the generating function:
\begin{align}
	Z_1(J,J') &\equiv \INT dq_f \INT dq_0 \INT dq_0' \bra q_0|1\ket \bra 1|q_0'\ket Z(q_0,q_f|J)Z(q_0',q_f|J')^* \nonumber\\
	&= \frac{1}{2\pi G(T)}\INT dq_f\; \psi_1(q_f|J)\,\psi_1(q_f|J')^* \nonumber\\
	&\nonumber\\
	&= 2m\left(\frac{m}{\pi}\right)^{1/2} e^{\,\frac{i}{2G(T)} \doubleint (-2)\Ta(t-t') G(t_f-t)G(t'-t_0)\left[J(t)J(t')-J'(t)J'(t')\right]}I(J,J') \nonumber\\
	&\nonumber\\
	&= 2m\left(\frac{m}{\pi}\right)^{1/2} e^{\,\frac{i}{2G(T)} \doubleint (-2)\Ta(t-t') G(t_f-t)G(t'-t_0)\left[J(t)J(t')-J'(t)J'(t')\right]} \times \nonumber\\
	&e^{-\frac{i}{2G(T)} \doubleint G(t_f-t)G(t_f-t')\left[ e^{-imT}J(t)J(t')-e^{\,imT}J'(t)J'(t')\right]} K(J,J') \nonumber\\
	&\nonumber\\
	& = 2m N Z_0(J,J')\;,
\end{align}
where $Z_0(J,J') = e^{\,i\Phi_0(J,J')}$ is the generating function for the oscillator in its ground state. So the influence function $\Phi_1(J,J') \equiv -i \ln Z_1(J,J')$ is\footnote{One potentially interesting exercise would be to calculate $\Phi_n$ for a density matrix $\hat\rho = |n\ket\bra n|$, then use
	\begin{equation}
		e^{-\beta \hat H} = \sum_{n\,=\,0}^{\infty} e^{-\beta m (n+\half)} |n\ket \bra n|
	\end{equation}
	to arrive at the thermal influence phase. I thought about trying that, but then I looked up the series form of the Hermite polynomials and said forget it.}
\begin{equation}
	\Phi_1(J,J') = \Phi_0(J,J') - i\ln\left\{1-\frac{1}{2m}\doubleint \cos[m(t-t')]\left[J(t)\!-\!J'(t)\right]\left[J'(t)\!-\!J'(t')\right]\right\}\;.
\end{equation}
The Wightman correlator is
\begin{align}
	&\tr\left(\hat q(t_2)\hat q(t_1) \,e^{\,i t_0\hat H}\hat\rho_1\, e^{-it_0\hat H}\right) = \left.\frac{\del^2 Z(J,J')}{\del J(t_1)\del J(t_2)}\right|_{J\,=\,J'\,=\,0} \nonumber\\
	&\qquad\qquad = \tr\left(\hat q(t_2)\hat q(t_1)\,e^{\,it_0\hat H}\hat\rho_0\,e^{-it_0\hat H}\right) + \frac{1}{m}\cos\left[m(t_1\!-\!t_2)\right]\;.
\end{align}
The operator calculation for general $n$ is straightforward and leads to\footnote{This is a translationally invariant problem, so $t_0$ drops out.}
\begin{equation}
	\bra n|\hat q(t_2)\hat q(t_1)|n\ket = \frac{1}{2m}e^{-im(t_2-t_1)} + \frac{n}{m}\cos\left[m(t_2\!-\!t_1)\right]\;,
\end{equation}
which agrees with the above path-integral result for $n = 1$. 

\section{Larkin-Ovchinnikov path integral}
%%%%
As with the ``other order of integration'' method from Sec.~\ref{sec:other order} for evaluating the Schwinger-Keldysh path integral, there is much to simplify in the Larkin-Ovchinnikov path integral before choosing a density matrix. For convenience, I will repeat the basic formula, Eq.~(\ref{eq:LO}): 
\begin{align}
	&Z(J,J',J'',J''') \equiv \nonumber\\
	&\qquad \INT dq_0 \, dq'_0 \, dq''_0 \, dq_f \, dq'_f\;\rho(q_0,q''_0)\;Z(q_0,q_f|J)\, Z(q'_0,q_f|J')^*\,Z(q'_0,q'_f|J'')\, Z(q''_0,q'_f|J''')^*\;.
\end{align}
First I will use Eqs.~(\ref{eq:generating function for harmonic oscillator}) and~(\ref{eq:action for generating function for harmonic oscillator}) to evaluate
\begin{equation}
	A(q_0,q_i|J,J') \equiv \INT dq_f\;Z(q_0,q_f|J)Z(q_i,q_f|J')^*\;.
\end{equation}
The difference in classical actions is
\begin{align}
	S_{\text{cl}}(q_0,q_f|J) &- S_{\text{cl}}(q_i,q_f|J') = \frac{1}{2G(T)}\left\{ \dot G(T)(q_0^2-q_i^2) +2\int_{t_0}^{t_f}\!\! dt\;G(t_f\!-\!t)\left[q_0 J(t)-q_i J'(t)\right]\right. \nonumber\\
	&-2\int_{t_0}^{t_f}\!\! dt\int_{t_0}^{t_f}\!\! dt'\;\Ta(t\!-\!t')\,G(t_f\!-\!t)G(t'\!-\!t_0)[J(t)J(t')-J'(t)J'(t')] \nonumber\\
	&\left. +2\left[q_i-q_0+\int_{t_0}^{t_f}\!\!dt\;G(t\!-\!t_0)\left(J(t)-J'(t)\right)\right]q_f\right\}\;.
\end{align}
The only dependence on $q_f$ is linear, so the integral produces a delta function. Since $\INT dq\;e^{\,\frac{i}{G(T)}\mu q} = 2\pi G(T) \del(\mu)$ (assuming positive $G(T)$, which all of the previous formulas implicitly require), the integral evaluates to
\begin{align}
	A(q_0,q_i|J,J') &= \del\!\left(q_i-q_0+\int_{t_0}^{t_f}\!\!dt\;G(t\!-\!t_0)\left[J(t)-J'(t)\right] \right)\;e^{\,\frac{i}{2G(T)}B(q_0,q_i|J,J')}\;,
\end{align} 
with
\begin{align}
	B(q_0,q_i|J,J') &= \dot G(T)\left(q_0^2-q_i^2\right) +2\int_{t_0}^{t_f}\!\! dt\;G(t_f\!-\!t)\left[q_0 J(t)-q_i J'(t)\right] \nonumber\\
	&-2\int_{t_0}^{t_f}\!\! dt\int_{t_0}^{t_f}\!\! dt'\;\Ta(t\!-\!t')\,G(t_f\!-\!t)G(t'\!-\!t_0)[J(t)J(t')-J'(t)J'(t')] \;.
\end{align}
Similarly, 
\begin{align}
	A(q_i,q_0'|J'',J''') &= \INT dq_f'\;Z(q_i,q_f'|J'')Z(q_0',q_f'|J''')^* \nonumber\\
	&= \del\left(q_0'-q_i+\int_{t_0}^{t_f}\!\! dt\;G(t\!-\!t_0)\left[J''(t)-J'''(t)\right]\right)\;e^{\,\frac{i}{2G(T)} B(q_i,q_0'|J'',J''')}\;,
\end{align}
with
\begin{align}
	B(q_i,q_0'|J'',J''') &= \dot G(T)\left(q_i^2-q_0'^2\right) + 2\int_{t_0}^{t_f}\!\! dt\;G(t_f\!-\!t)\left[q_i J''(t) - q_0' J'''(t)\right] \nonumber\\
	&- 2\int_{t_0}^{t_f}\!\! dt \int_{t_0}^{t_f}\!\! dt'\;\Ta(t\!-\!t')\,G(t_f\!-\!t)G(t'\!-\!t_0)\left[J''(t)J''(t')-J'''(t)J'''(t')\right]\;.
\end{align}
Because of the delta functions, the integrals over $q_0$ and $q_0'$ can be done immediately:
\begin{align}
	&Z(J,J',J'',J''') = \INT dq_i \INT dq_0\INT dq_0'\;\rho(q_0,q_0')\;A(q_0,q_i|J,J')\,A(q_i,q_0'|J'',J''') \nonumber\\
	&\qquad = \left.\INT dq_i\;\rho(q_0,q_0')\;e^{\,\frac{i}{2G(T)}\left[ B(q_0,q_i|J,J') + B(q_i,q_0'|J'',J''') \right] } \right|_{\begin{matrix} q_0 = q_i+\int_{t_0}^{t_f}\!\! dt\;G(t\!-\!t_0)\left[J(t)\!-\!J'(t)\right],\\ q_0' = q_i-\int_{t_0}^{t_f}\!\! dt\;G(t\!-\!t_0)\left[J''(t)\!-\!J'''(t)\right] \end{matrix}}\;.
\end{align}
Time to simplify the argument of the exponential. First, simplify $q_0^2-q_0'^2$:
\begin{align}
	&q_0^2 - q_0'^2 = 2q_i\int_{t_0}^{t_f}\!\! dt\;G(t\!-\!t_0)\left[J(t)\!-\!J'(t)\!+\!J''(t)\!-\!J'''(t)\right] \nonumber\\
	&+ \doubleint G(t\!-\!t_0)\,G(t'\!-\!t_0)\left\{\left[J(t)\!-\!J'(t)\right]\left[J(t')\!-\!J'(t')\right] - \left[J''(t)\!-\!J'''(t)\right]\left[J''(t')\!-\!J'''(t')\right]\right\}\;.
\end{align}
The $q_i^2$ terms have canceled---they will come solely from the density matrix. With that, I find
\begin{align}\label{eq:B+B}
	&B(q_0,q_i|J,J') + B(q_i,q_0'|J'',J''') = 2q_i \int_{t_0}^{t_f}\!\! dt\left[\dot G(T)G(t\!-\!t_0)+G(t_f\!-\!t)\right]\left[J(t)\!-\!J'(t)\!+\!J''(t)\!-\!J'''(t)\right] \nonumber\\
	&+\dot G(T)\int_{t_0}^{t_f}\!\! dt \int_{t_0}^{t_f}\!\! dt'\;G(t\!-\!t_0)G(t'\!-\!t_0)\left\{\left[J(t)\!-\!J'(t)\right]\left[J(t')\!-\!J'(t')\right]-\left[J''(t)\!-\!J'''(t)\right]\left[J''(t')\!-\!J'''(t')\right]\right\}\nonumber\\
	&+C(J,J',J'',J''')\;,
\end{align}
where to anticipate simplifying further I have separated a quantity
\begin{align}\label{eq:gammaJJ'J''J'''}
	&C(J,J',J'',J''') \equiv \nonumber\\
	&-2\doubleint\Ta(t\!-\!t')\,G(t_f\!-\!t)G(t'\!-\!t_0)\left[J(t)J(t')\!-\!J'(t)J'(t')\!+\!J''(t)J''(t')\!-\!J'''(t)J'''(t')\right] \nonumber\\
	&+2\doubleint G(t_f\!-\!t)G(t'\!-\!t_0)\left[J(t)J(t')\!-\!J'''(t)J'''(t')\!-\!J(t)J'(t')\!+\!J'''(t)J''(t')\right]\;.
\end{align}
For $t>t'$, the $JJ$ and $J'''J'''$ terms cancel, and for $t < t'$ the first line is absent. So another way to write Eq.~(\ref{eq:gammaJJ'J''J'''}) is
\begin{align}
	C(J,J',J'',J''') &= 2\doubleint G(t_f\!-\!t)G(t'\!-\!t_0)\left\{ \Ta(t\!-\!t')\left[ J'(t)J'(t')-J''(t)J''(t') \right] \right. \nonumber\\
	&\left. +\Ta(t'\!-\!t)\left[J(t)J(t')-J'''(t)J'''(t')\right] - J(t)J'(t')+J'''(t)J''(t')\right\}\;.
\end{align}
The $JJ$, $J'J'$, $J''J''$, and $J'''J'''$ terms are symmetric in $t$ and $t'$, so I will use that to symmetrize the coefficients of those terms in the integrand. I will keep the $JJ'$ and $J'''J''$ terms as written. I get:
\begin{align}
	&C(J,J',J'',J''') = \nonumber\\
	&\doubleint \left\{ \left[ \Ta(t\!-\!t')G(t_f\!-\!t')G(t\!-\!t_0)+\Ta(t'\!-\!t)G(t_f\!-\!t)G(t'\!-\!t_0)\right] \left[J(t)J(t')-J'''(t)J'''(t')\right]\right. \nonumber\\
	&+\left[\Ta(t\!-\!t')G(t_f\!-\!t)G(t'\!-\!t_0)+\Ta(t'\!-\!t)G(t_f\!-\!t')G(t\!-\!t_0)\right]\left[J'(t)J'(t')-J''(t)J''(t')\right]\nonumber\\
	&\left.+2\,G(t_f\!-\! t)G(t'\!-\!t_0)\left[J'''(t)J''(t')-J(t)J'(t')\right]\right\}\;.
\end{align}
Combining that with the other $q_i$-independent terms from Eq.~(\ref{eq:B+B}) gives
\begin{align}
	&B(q_0,q_i|J,J') + B(q_i,q_0'|J'',J''') = \nonumber\\
	&\qquad 2q_i \int_{t_0}^{t_f}\!\! dt\left[\dot G(T)G(t\!-\!t_0)+G(t_f\!-\!t)\right]\left[J(t)\!-\!J'(t)\!+\!J''(t)\!-\!J'''(t)\right]+D(J,J',J'',J''')\;,
\end{align}
with
\begin{align}\label{eq:D for ground-state 4pt}
	&D(J,J',J'',J''') \equiv \doubleint\Big\{ 2\left[G(t_f\!-\!t)+\dot G(T)G(t\!-\!t_0) \right]G(t'\!-\!t_0)\; \times \Big. \nonumber\\
	&\qquad\qquad\qquad\qquad\qquad\qquad\qquad\qquad\qquad\qquad\qquad\qquad\qquad\left[J'''(t)J''(t')-J(t)J'(t')\right] \nonumber\\
	&+\!\left[\Ta(t\!-\!t')G(t_f\!-\!t')G(t\!-\!t_0)\!+\!\Ta(t'\!-\!t)G(t_f\!-\!t)G(t'\!-\!t_0)\!+\!\dot G(T)G(t\!-\!t_0)G(t'\!-\!t_0)\right] \times \nonumber\\
	&\qquad\qquad\qquad\qquad\qquad\qquad\qquad\qquad\qquad\qquad\qquad\qquad\qquad\left[J(t)J(t')\!-\!J'''(t)J'''(t')\right] \nonumber\\
	&+\!\left[ \Ta(t\!-\!t')G(t_f\!-\!t)G(t'\!-\!t_0)\!+\!\Ta(t'\!-\!t)G(t_f\!-\!t')G(t\!-\!t_0)\!+\!\dot G(T)G(t\!-\!t_0)G(t'\!-\!t_0) \right]\times\nonumber\\
	&\Big.\qquad\qquad\qquad\qquad\qquad\qquad\qquad\qquad\qquad\qquad\qquad\qquad\qquad\left[J'(t)J'(t')\!-\!J''(t)J''(t')\right]\Big\}\;.
\end{align}
To proceed I will need to specify $\rho$. 
\subsection{Ground state}
%%%%
I will need to simplify the oscillator ground-state density matrix from Eq.~(\ref{eq:ground state}) using the expressions for $q_0$ and $q_0'$ produced by the delta functions. The wavefunctions are:
\begin{align}
	&\left. \bra q_0|0\ket \right|_{q_0\,=\,q_i+\int_{t_0}^{t_f}\!\! dt\;G(t-t_0)\left[J(t)-J'(t)\right]} = \left(\frac{m}{\pi}\right)^{1/4} e^{-\half m \left[q_i^2+2q_i\int_{t_0}^{t_f} dt\;G(t-t_0)[J(t)-J'(t)]\right]}\times \nonumber\\
	&\qquad \qquad \qquad \qquad \qquad e^{-\half m\doubleint G(t-t_0)G(t'-t_0)[J(t)-J'(t)][J(t')-J'(t')]}\;,\nonumber\\
	&\nonumber\\
	&\left. \bra q_0'|0\ket\right|_{q_0'\,=\,q_i - \int_{t_0}^{t_f} dt\;G(t-t_0)[J''(t)-J'''(t)]} = \left(\frac{m}{\pi}\right)^{1/4} e^{-\half m \left[q_i^2-2q_i\int_{t_0}^{t_f}dt\;G(t-t_0)[J''(t)-J'''(t)] \right]} \times \nonumber\\
	&\qquad \qquad \qquad \qquad \qquad e^{-\half m\doubleint G(t-t_0)G(t'-t_0)[J''(t)-J'''(t)][J''(t')-J'''(t')]}\;.
\end{align}
So the density matrix is
\begin{align}
	\rho(q_0,q_0') &= \left(\frac{m}{\pi}\right)^{1/2} e^{-m\left\{ q_i^2+q_i\int_{t_0}^{t_f} dt\;G(t-t_0)[J(t)-J'(t)-J''(t)+J'''(t)] \right\}} \times \nonumber\\
	&e^{-\half m\doubleint G(t-t_0)G(t'-t_0)\left\{ [J(t)-J'(t)][J(t')-J'(t')] + [J''(t)-J'''(t)][J''(t')-J'''(t')] \right\}}\;.
\end{align}
There is the $q_i^2$ term, and with it the requisite factor of $-m$ to give a Gaussian integral that will cancel the overall factor of $(m/\pi)^{1/2}$. The coefficient of $q_i$ from $\rho$ and $\frac{i}{2G(T)}(B+B)$ leads to 
\begin{align}
	j(J,J',J'',J''')&\equiv \int_{t_0}^{t_f}\!\! dt\left\{ \frac{-i}{mG(T)}\left[\dot G(T)G(t\!-\!t_0)+G(t_f\!-\!t)\right]\left[J(t)\!-\!J'(t)\!+\!J''(t)\!-\!J'''(t)\right] \right. \nonumber\\
	&\left. +G(t\!-\!t_0)\left[J(t)\!-\!J'(t)\!-\!J''(t)\!+\!J'''(t)\right]\phantom{\frac{-i}{2}}\!\!\!\!\!\!\!\!\! \right\} \nonumber\\
	&\nonumber\\
	&= \int_{t_0}^{t_f}\!\! dt\Bigg\{ \left[G(t\!-\!t_0)-\frac{i}{m G(T)}\left(\dot G(T)G(t\!-\!t_0)+G(t_f\!-\!t)\right) \right]\left[J(t)\!-\!J'(t)\right] \Bigg. \nonumber\\
	&-\Bigg. \underbrace{\left[G(t\!-\!t_0)+\frac{i}{m G(T)}\left(\dot G(T) G(t\!-\!t_0)+G(t_f\!-\!t)\right) \right]}_{\begin{matrix}\equiv\;\al(t) \end{matrix}} \left[J''(t)\!-\!J'''(t)\right]\Bigg\} \nonumber\\
	&= \int_{t_0}^{t_f}\!\! dt\left\{\al(t)^*\left[J(t)\!-\!J'(t)\right]-\al(t)\left[J''(t)\!-\!J'''(t)\right]\right\}\;.
\end{align}
Completing the square and doing the integral over $q_i$ gives the generating function
\begin{equation}
	Z(J,J',J'',J''') = e^{\,i\Phi(J,J',J'',J''')}
\end{equation}
with generalized influence phase
\begin{align}\label{eq:4pt ground-state influence phase before simplification}
	&\Phi(J,J',J'',J''') = \frac{1}{2G(T)} D(J,J',J'',J''')-i\frac{m}{4}j(J,J',J'',J''')^2 \nonumber\\
	&+i\frac{m}{2}\doubleint G(t\!-\!t_0)G(t'\!-\!t_0)\Big\{ [J(t)\!-\!J'(t)][J(t')\!-\!J'(t')] \Big. \nonumber\\
	&\Big.\qquad\qquad\qquad\qquad\qquad\qquad\qquad\qquad\qquad + [J''(t)\!-\!J'''(t)][J''(t')\!-\!J'''(t')] \Big\}\;.
\end{align}
\subsubsection{Simplification}
Now I will need to systematically collect and simplify each type of term ($JJ$, $JJ'$, and so on). The square of $j$ gives
\begin{align}\label{eq:j^2 for ground-state 4pt}
	j^2 = \doubleint&\left\{\al(t)^*\al(t')^*\left[J(t)\!-\!J'(t)\right]\left[J(t')\!-\!J'(t')\right] \right. \nonumber\\
	&+\al(t)\al(t')\left[J''(t)\!-\!J'''(t)\right]\left[J''(t')\!-\!J'''(t')\right]\nonumber\\
	&-\al(t)^*\al(t')\left[J(t)\!-\!J'(t)\right]\left[J''(t')\!-\!J'''(t')\right]\nonumber\\
	&\left.-\al(t)\al(t')^*\left[J''(t)\!-\!J'''(t)\right]\left[J(t')\!-\!J'(t')\right]\right\}\;.
\end{align}
I want to explain a few things about simplifying that. 
\\\\
First, focus on the last two lines of Eq.~(\ref{eq:j^2 for ground-state 4pt}) and compare them to the second line of Eqs.~(\ref{eq:4pt ground-state influence phase before simplification}) and to~(\ref{eq:D for ground-state 4pt}). Notice something? The last two lines of Eq.~(\ref{eq:j^2 for ground-state 4pt}) have no counterparts anywhere else in the generalized influence phase. So $\al(t)^*\al(t')$ had better simplify by itself, and indeed it does:
\begin{equation}\label{eq:alpha(t)*alpha(t')}
	\al(t)^*\al(t') = \frac{1}{m^2}\,e^{\,im(t-t')}\;.
\end{equation}
By ``simplifies'' here I mean that all dependence on $t_0$ and $t_f$ cancels out, and the result depends on $t$ and $t'$ only as $t-t'$ (i.e., it is translationally invariant). In contrast, $\al(t)\al(t')$ does not simplify in that sense:
\begin{equation}
	\al(t)\al(t') = -\,\frac{1}{m^2}\,e^{-im(t+t'-2t_0)}\;.
\end{equation}
This lack of simplification is expected, because there are other terms in the influence phase that will combine with $J''(t)J''(t')$, etc.
\\\\
The second thing I want to explain is how to simplify one of the source-diagonal terms, say the $JJ$ one. I have already symmetrized Eq.~(\ref{eq:D for ground-state 4pt}) in $t$ and $t'$ to anticipate a relatively straightforward collecting of terms. For $t>t'$, the symmetrized coefficient of $J(t)J(t')$ is
\begin{align}\label{eq:O_>}
	\op_>(t,t') &\equiv \frac{1}{2G(T)}\left[\dot G(T) G(t\!-\!t_0)G(t'\!-\!t_0)+G(t_f\!-\!t')G(t\!-\!t_0)\right]\nonumber\\
	&\qquad \qquad \qquad \qquad +i\frac{m}{2}G(t\!-\!t_0)G(t'\!-\!t_0)-i\frac{m}{4}\al(t)^*\al(t')^*\;.
\end{align}
I begin by returning to $\al(t)\al(t')$:
\begin{align}
	&\al(t)\al(t') = \nonumber\\
	&\;\;G(t\!-\!t_0) G(t'\!-\!t_0) - \frac{1}{m^2G(T)^2}\left[\dot G(T)G(t\!-\!t_0)+G(t_f\!-\!t)\right]\left[\dot G(T)G(t'\!-\!t_0)+G(t_f\!-\!t')\right] \nonumber\\
	&\;\;+\frac{i}{m G(T)}\left[2\dot G(T) G(t\!-\!t_0)G(t'\!-\!t_0) + G(t\!-\!t_0)G(t_f\!-\!t')+G(t_f\!-\!t)G(t'\!-\!t_0)\right]\;.
\end{align}
Conjugating that and combining it with the other terms in Eq.~(\ref{eq:O_>}) produces the following real and imaginary parts:
\begin{align}
	\re[\op_>(t,t')] &= \frac{1}{4G(T)}\left[G(t_f\!-\!t')G(t\!-\!t_0)-G(t_f\!-\!t)G(t'\!-\!t_0)\right] = \frac{1}{4}G(t\!-\!t')\;,
\end{align}
\begin{align}
	\im[\op_>(t,t')] &= \frac{m}{4}\Bigg\{ \left[1+\frac{\dot G(T)^2}{m^2G(T)^2}\right] G(t\!-\!t_0)G(t'\!-\!t_0) \Bigg. \nonumber\\
	&+\Bigg.\frac{1}{m^2 G(T)^2} \left[ G(t_f\!-\!t)G(t_f\!-\!t')+\dot G(T)\left[ G(t\!-\!t_0)G(t_f\!-\!t')+G(t_f\!-\!t)G(t'\!-\!t_0) \right] \right]\Bigg\}\;.
\end{align}
Since $\dot G(T)^2 +m^2 G(T)^2 = \cos^2 (mT)+\sin^2(mT) = 1$, I can factor out an overall $m^2 G(T)^2$ in the denominator. Then I can use the trigonometric identity
\begin{equation}
	\cos(a\!-\!b)\sin(a\!-\!c)-\sin(a\!-\!b)\cos(a\!-\!c)=\sin(b\!-\!c)
\end{equation}
as follows: 
\begin{align}
	&G(t'\!-\!t_0) + \dot G(T) G(t_f\!-\!t') = G(T)\dot G(t_f\!-\!t')\;,\nonumber\\
	&G(t_f\!-\!t') + \dot G(T) G(t'\!-\!t_0) = G(T) \dot G(t'\!-\!t_0)\;.
\end{align}
With all of that, the imaginary part becomes: 
%%%
\begin{align}
	\im[\op_>(t,t')] &= \frac{1}{4m G(T)^2}\left\{G(t\!-\!t_0)\left[G(t'\!-\!t_0)+\dot G(T) G(t_f\!-\!t')\right] \right. \nonumber\\
	&\left.\qquad \qquad \;\;\;\;\; + G(t_f\!-\!t)\left[G(t_f\!-\!t')+\dot G(T) G(t'\!-\!t_0)\right]\right\} \nonumber\\
	&= \frac{1}{4m G(T)}\left\{ G(t\!-\!t_0)\dot G(t_f\!-\!t') + G(t_f\!-\!t)\dot G(t'\!-\!t_0)\right\}\nonumber\\
	&= \frac{1}{4m}\dot G(t\!-\!t')\;.
\end{align}
Putting the real and imaginary parts together, I find
\begin{align}
	\op_>(t,t') &= \frac{1}{4} G(t\!-\!t') + \frac{i}{4m}\dot G(t\!-\!t') = \frac{i}{4m}e^{-im(t-t')}\;.
\end{align}
That, remember, is only the coefficient of $J(t)J(t')$ for $t > t'$. But now that I have derived that much, I am confident that you can derive the rest (or ask Mathematica). 
\\\\
The final result for the generalized influence phase will indeed have the form in Eq.~(\ref{eq:general form of LK phase}) with the Green's functions in Eqs.~(\ref{eq:G_>(t) for ground state})-(\ref{eq:G_D(t) for ground state}). 
%%%%
\subsection{Thermal state}
%%%%
The thermal density matrix from Eq.~(\ref{eq:thermal density matrix}) is
\begin{equation}
	\rho(q_0,q_0') = \left(\tfrac{\dot G(-i\beta)-1}{i\pi G(-i\beta)}\right)^{1/2}e^{\,\frac{i}{2G(-i\beta)}\left[\dot G(-i\beta)(q_0^2+q_0'^2)-2q_0q_0'\right]}\;.
\end{equation}
Inserting
\begin{align}
	&q_0 = q_i + \int_{t_0}^{t_f}\!\! dt\;G(t\!-\!t_0)\left[J(t)\!-\!J'(t)\right]\;,\nonumber\\
	&q_0' = q_i - \int_{t_0}^{t_f}\!\! dt\;G(t\!-\!t_0)\left[J''(t)\!-\!J'''(t)\right]
\end{align}
gives
\begin{align}
	&\dot G(-i\beta)(q_0^2+q_0'^2)-2q_0q_0' = \nonumber\\
	&2\left[\dot G(-i\beta)\!-\!1\right]\left\{q_i^2 +  q_i\int_{t_0}^{t_f}\!\! dt\;G(t\!-\!t_0)\left[J(t)\!-\!J'(t)\!-\!J''(t)\!+\!J'''(t)\right]\right\} + E(J,J',J'',J''')\;,
\end{align}
where
\begin{align}
	E(J,J',J'',J''') &\equiv \doubleint G(t\!-\!t_0)G(t'\!-\!t_0)\left\{\dot G(-i\beta) \left[\left(J(t)\!-\!J'(t)\right)\left(J(t')\!-\!J'(t')\right) \right.\right. \nonumber\\
	&\left.\left. +\left(J''(t)\!-\!J'''(t)\right)\left(J''(t')\!-\!J'''(t')\right) \right] +2\left[J(t)\!-\!J'(t)\right]\left[J''(t')\!-\!J'''(t')\right]\phantom{\dot G(-i\beta)}\!\!\!\!\!\!\!\!\!\!\!\!\!\!\!\!\!\!\!\!\right\}\;.
\end{align}
The generating function is
\begin{equation}
	Z(J,J',J'',J''') = \left(\tfrac{\dot G(-i\beta)-1}{i\pi G(-i\beta)}\right)^{1/2} e^{\,\frac{i}{2G(T)}D} e^{\,\frac{i}{2G(-i\beta)} E}I\;,
\end{equation}
where the integral to be done is
\begin{align}
	I &= \INT dq_i\;e^{\,\frac{i}{G(-i\beta)}[\dot G(-i\beta)-1]\left\{q_i^2+q_i\int_{t_0}^{t_f}\!\!dt\;G(t-t_0)[J(t)-J'(t)-J''(t)+J'''(t)]\right\}} \times \nonumber\\
	&e^{\,\frac{i}{G(T)}q_i\int_{t_0}^{t_f}\!\! dt\;[\dot G(T)G(t-t_0)+G(t_f-t)][J(t)-J'(t)+J''(t)-J'''(t)]}\;.
\end{align}
I will complete the square as before, defining the thermal version of $\al(t)$ as
\begin{equation}
	\al(t) \equiv G(t\!-\!t_0)-\left(\frac{G(-i\beta)}{\dot G(-i\beta)-1}\right)\left(\frac{\dot G(T)G(t-t_0)+G(t_f-t)}{G(T)}\right)\;.
\end{equation}
Note that $\frac{G(-i\beta)}{\dot G(-i\beta)-1} = -\frac{i}{m}\coth(\frac{m\beta}{2})$ is imaginary, so indeed I retain the structure
\begin{equation}
	j \equiv \int_{t_0}^{t_f}\!\! dt\left\{\al(t)^*\left[J(t)\!-\!J'(t)\right]-\al(t)\left[J''(t)\!-\!J'''(t)\right]\right\}
\end{equation}
from the zero-temperature case. Completing the square gives
\begin{equation}
	I = \left(\tfrac{i\pi G(-i\beta)}{\dot G(-i\beta)-1}\right)^{1/2} e^{-i\frac{\dot G(-i\beta)-1}{G(-i\beta)}\frac{j^2}{4} }\;.
\end{equation}
So the generating function has the form $Z = e^{\,i\Phi}$ with generalized influence phase
\begin{equation}
	\Phi(J,J',J'',J''') = \frac{D}{2G(T)}+\frac{E}{2G(-i\beta)} -\left(\frac{\dot G(-i\beta)-1}{4G(-i\beta)}\right) j^2\;.
\end{equation}
\subsubsection{Result}
Simplification happens largely as before. This time I have used Mathematica to simplify everything and will not pretend otherwise. But there is still something to say. As an example, I will focus on the $JJ$ term, which ends up being
\begin{align}
	\Phi_{JJ\text{ term}} &= \frac{i}{4m\sinh\left(\frac{m\beta}{2}\right)}\doubleint\left\{\Ta(t\!-\!t')\cos\left[m\left(t\!-\!t'+i\frac{\beta}{2}\right)\right] \right. \nonumber\\
	&\left.+ \Ta(t'\!-\!t)\cos\left[m\left(t\!-\!t'-i\frac{\beta}{2}\right)\right]\right\}J(t)J(t')\;.
\end{align}
Consider the zero-temperature limit:
\begin{equation}
	\left.\frac{\cos[m(t-t'+i\frac{\beta}{2})]}{\sinh(\frac{m\beta}{2})}\right|_{\beta \to \infty} = e^{-im(t-t')}\;.
\end{equation}
Subtracting that from the nonzero-temperature expression gives a Bose-Einstein factor:
\begin{equation}
	\frac{\cos[m(t-t'+i\frac{\beta}{2})]}{\sinh(\frac{m\beta}{2})}-e^{-im(t-t')} = \frac{2}{e^{\beta m}-1}\cos[m(t\!-\!t')]\;.
\end{equation}
That is real and symmetric in $t$ and $t'$. Subtracting the zero-temperature contribution from the $JJ$ term gives
\begin{equation}
	\Phi_{JJ\text{ term}} - \left.\Phi_{JJ\text{ term}}\right|_{\beta\, =\, \infty} = \frac{1}{e^{\beta m}-1}\doubleint \frac{i}{2m}\cos\left[m(t\!-\!t')\right] J(t)J(t')\;. 
\end{equation}
Repeating that process for everything will give the final result for the generalized influence phase:
\begin{align}
	\Phi(J,J',J'',J''') &= \Phi_0(J,J',J'',J''') + \frac{1}{e^{\beta m}-1}\doubleint \frac{i}{2m}\cos\left[m(t\!-\!t')\right] \times \nonumber\\
	&\left\{\left[J(t)\!-\!J'(t)\!+\!J''(t)\!-\!J'''(t)\right]\left[J(t')\!-\!J'(t')+J''(t')\!-\!J'''(t')\right] \right\}\;.
\end{align}
\section{Quenched oscillator}\label{sec:quench}
%%%%
Expectation values, expectation values, expectation values. That, not dynamics, is what this is about. Expectation values.
\\\\
Understand? 
\\\\
Congratulations, you are ready for a dynamical problem. At some initial time, I change the oscillator frequency instantaneously from one value to another. The Hamiltonian will be
\begin{equation}
	\hat H(t) = \half \hat p^2 + \half m(t)^2 \hat q^2\;,\;\; m(t) = \left\{ \begin{matrix} m_0\;,\;\; t < t_0 \\ m\;,\;\; t > t_0 \end{matrix} \right.\;.\label{eq:quenched oscillator hamiltonian}
\end{equation}
Okay, I lied: Even that is not really dynamical. Written that way, the Hamiltonian changes in time. But the perspective offered by the generating function is that Eq.~(\ref{eq:quenched oscillator hamiltonian}) merely describes evolution with the $t > t_0$ constant-frequency Hamiltonian---it is just that the system is in an excited state. 
\\\\
A genuinely dynamical version of this idea would be to take a smooth profile in $m(t)$ that approaches constant values at $t_0$ and $t_f$; for an example over an infinite interval, see Jordan \cite{jordan1986}. I will content myself with Eq.~(\ref{eq:quenched oscillator hamiltonian}).
%%%%
\subsection{Operator calculations}
Before diving into the path-integral calculation, it is worth reviewing the operator setup.
\subsubsection{Bogoliubov transformation}\label{sec:bogoliubov}
%%%%
For $t > t_0$, the Hamiltonian is 
\begin{equation}
	\hat H(t>t_0) = \half \hat p^2 + \half m^2 \hat q^2 \equiv \hat H_{\OUT}\;.
\end{equation}
In terms of 
\begin{equation}
	\hat a_{\OUT} \equiv \sqrt{\frac{m}{2}}\left(\hat q + \frac{i}{m}\hat p\right)\;,
\end{equation}
the Hamiltonian is
\begin{equation}
	\hat H_\OUT = m\left(\hat a_{\OUT}^\da \hat a_{\OUT}+\half \hat 1\right)\;.
\end{equation}
For $t < t_0$, the Hamiltonian is
\begin{equation}
	\hat H(t<t_0) = \half \hat p^2 + \half m_0^2 \hat q^2 \equiv \hat H_\IN\;.
\end{equation}
In terms of
\begin{equation}
	\hat a_\IN \equiv \sqrt{\frac{m_0}{2}} \left(\hat q + \frac{i}{m_0}\hat p\right)\;,
\end{equation}
the Hamiltonian is
\begin{equation}
	\hat H_\IN = m_0\left(\hat a_\IN^\da a_\IN + \half \hat 1\right)\;.
\end{equation}
When $m_0 \neq m$, the operators $\hat a_\OUT$ and $\hat a_\IN$ are different. Concretely, for complex numbers $A$ and $B$, the operators can be expressed in terms of each other as
\begin{equation}\label{eq:bogoliubov}
	\hat a_\OUT \equiv A\;\hat a_\IN + B\;\hat a_\IN^\da\;.
\end{equation}
%%%
To solve for $A$ and $B$, introduce the ground states $|0_\OUT\ket$ and $|0_\IN\ket$. They satisfy
\begin{align}
	&\hat a_\OUT |0_\OUT\ket = 0\;,\;\; \hat a_\IN|0_\IN\ket = 0\;, \nonumber\\
	&\bra 0_\OUT| a_\OUT^\da = 0\;,\;\; \bra 0_\IN| a_\IN^\da = 0\;.
\end{align}
Therefore:
\begin{align}\label{eq:a_out |0_in>}
	&\hat a_\OUT |0_\IN\ket = B\; \hat a_\IN^\da |0_\IN\ket\;,\;\; \bra 0_\IN|\hat a_\OUT = A \bra 0_\IN| \hat a_\IN\;.
\end{align}
In terms of the 1-particle states
\begin{equation}
	|1_\OUT\ket \equiv \hat a_\OUT^\da|0_\OUT\ket\;,\;\; |1_\IN\ket \equiv \hat a_\IN^\da |0_\IN\ket\;,
\end{equation}
those imply
\begin{equation}
	B = \bra 1_\IN|\hat a_\OUT |0_\IN\ket\;,\;\; A = \bra 0_\IN|\hat a_\OUT |1_\IN\ket\;.
\end{equation}
But those are just formal expressions, so let me derive what they mean. To begin, I will review how to solve the condition $\hat a_\OUT|0_\OUT\ket = 0$ for the wavefunction $\bra q|0_\OUT\ket$. Acting on the equation $\hat a_\OUT|0_\OUT\ket = 0$ from the left with $\bra q|$ gives
\begin{align}
	\bra q|\hat a_\OUT|0_\OUT\ket &= \sqrt{\frac{m}{2}}\bra q|\,\hat q + \frac{i}{m}\hat p\,|0_\OUT\ket = \sqrt{\frac{m}{2}} \left(q+\frac{1}{m}\pa_q\right)\bra q|0_\OUT\ket = 0\;.
\end{align}
The solution is
\begin{equation}\label{eq:0out wavefunction unnormalized}
	\bra q|0_\OUT\ket = C\;e^{-\half m q^2}\;,
\end{equation}
with $C$ fixed by demanding $\bra 0_\OUT| 0_\OUT\ket = 1$:
\begin{align}
	\bra 0_\OUT|0_\OUT\ket = \INT dq\;\bra 0_\OUT|q\ket \bra q|0_\OUT\ket = |C|^2 \INT dq\;e^{-mq^2} = |C|^2\sqrt{\frac{\pi}{m}} \equiv 1 \implies |C| = \left(\frac{m}{\pi}\right)^{1/4}\;.
\end{align}
The overall phase in Eq.~(\ref{eq:0out wavefunction unnormalized}) is arbitrary; I will choose $C = |C|$. 
\\\\
The wavefunction for the 1-particle state is
\begin{align}
	\bra q|1_\OUT\ket &= \bra q|\hat a_\OUT^\da |0_\OUT\ket = \sqrt{\frac{m}{2}}\left(q-\frac{1}{m}\pa_q\right) \bra q|0_\OUT\ket =\sqrt{2m}\, q\, C e^{-\half mq^2}\;.
\end{align}
What I need are the ``in'' wavefunctions, which are
\begin{equation}\label{eq:"in" wavefunctions}
	\bra q|0_\IN\ket = \left(\frac{m_0}{\pi}\right)^{1/4}e^{-\half m_0 q^2}\;,\;\; \bra q|1_\IN\ket = \left(\frac{m_0}{\pi}\right)^{1/4} \sqrt{2m_0}\, q\, e^{-\half m_0 q^2}\;.
\end{equation}
Therefore, the coefficient $A$ is
\begin{align}
	A &= \bra 0_\IN|\hat a_\OUT|1_\IN\ket = \INT dq\; \bra 0_\IN|q\ket \sqrt{\frac{m}{2}}\left(q+\frac{1}{m}\pa_q\right) \bra q|1_\IN\ket \nonumber\\
	&= \sqrt{\frac{m}{2}} \sqrt{\frac{m_0}{\pi}} \sqrt{2m_0}\INT dq\;e^{-\half m_0 q^2}\left(q+\frac{1}{m}\pa_q\right)\left(q\, e^{-\half m_0 q^2}\right) \nonumber\\
	&= \frac{m_0}{\sqrt{\pi m}}\INT dq\left[\left(m\!-\!m_0\right) q^2 + 1\right]\,e^{-m_0 q^2} = \frac{m_0}{\sqrt{\pi m}}\left[-(m\!-\!m_0)\pa_{m_0}+1\right]\INT dq\;e^{-m_0 q^2} \nonumber\\
	&= \frac{m_0}{\sqrt{m}}\left[-(m\!-\!m_0)\pa_{m_0}+1\right] m_0^{-1/2} = \frac{m_0}{\sqrt{m}} \left[\frac{m-m_0}{2m_0^{3/2}}+\frac{1}{m_0^{1/2}} \right] = \frac{1}{2\sqrt{m m_0}}\left[m-m_0+2m_0\right] \nonumber\\
	&= \frac{m+m_0}{2\sqrt{m m_0}}\;.
\end{align}
By an analogous calculation, the coefficient $B$ is
\begin{equation}
	B = \frac{m-m_0}{2\sqrt{m m_0}}\;.
\end{equation}
%\\\\
That concludes baby's first Bogoliubov transformation. The coefficients satisfy $|A|^2 - |B|^2 = 1$; for $m_0 \to m$, they reduce to $A \to 1$ and $B \to 0$. 
\\\\
For later use, I want to record a physical interpretation of $B$. Return to Eq.~(\ref{eq:a_out |0_in>}) and consider the magnitude-squared of $\hat a_{\text{out}}|0_{\text{in}}\ket$. Since the number operator for ``out'' modes is $\hat N_{\text{out}} \equiv \hat a_{\text{out}}^\da \hat a_{\text{out}}$, the average number of quanta with frequency $m$ in the ground state of the oscillator with frequency $m_0$ is
\begin{equation}\label{eq:average number of quanta}
	\bra 0_\text{in}|\hat N_\text{out}|0_\text{in}\ket = |B|^2\;.
\end{equation}
%\\\\
\subsubsection{Fields in terms of creation/annihilation operators}
The Heisenberg-picture field operator for $t > t_0$ is
\begin{equation}\label{eq:q(t>t0)}
	\hat q(t>t_0) = \frac{1}{\sqrt{2m}}\left(e^{-imt}\, \hat a_\OUT + e^{\,imt}\, \hat a_\OUT^\da\right)\;.
\end{equation}
Inserting $a_\OUT = A \hat a_\IN + B\hat a_\IN^\da$ gives
\begin{align}
	&\hat q(0)|0_\IN\ket = \frac{1}{\sqrt{2m}}\left(A^*+B\right)|1_\IN\ket\;,\nonumber\\
	&\bra 0_\IN|\hat q(t) = \frac{1}{\sqrt{2m}} \left(e^{-imt}A+e^{imt}B^*\right)\bra 1_\IN|\;.
\end{align}
With the results for $A$ and $B$ I find
\begin{equation}
	e^{imt} A^* + e^{-imt} B = \frac{1}{\sqrt{m m_0}}\left[ m\,\cos(mt)+im_0\,\sin(mt)\right]\;.
\end{equation}
In terms of my old friend $G(t) \equiv \frac{1}{m}\sin(m t)$, I define 
\begin{equation}\label{eq:Gdot + im0 G}
	\zeta(t) \equiv \dot G(t) + im_0 G(t)\;,
\end{equation}
in terms of which 
\begin{equation}\label{eq:q(t>t0) alt}
	\hat q(t>t_0) = \frac{1}{\sqrt{2m_0}}\left[\z(t)^*\, \hat a_{\text{in}} + \z(t)\, \hat a_{\text{in}}^\da\right]\;.
\end{equation}
Eqs.~(\ref{eq:q(t>t0) alt}) and~(\ref{eq:q(t>t0)}) form a poignant expression of the Bogoliobov transformation. 
%\\\\
\subsubsection{Wightman correlators for $t_0 = 0$}
As somewhat of a warmup but mainly to make a point, I will briefly set $t_0 = 0$ and calculate a few 2-point correlation functions. 
\\\\
With Eqs.~(\ref{eq:q(t>t0)}) and~(\ref{eq:q(t>t0) alt}), I can calculate the following expectation value:
\begin{equation}\label{eq:wightman correlator special case}
	\bra 0_\IN|\hat q(t)\hat q(0)|0_\IN\ket = \frac{1}{4mm_0}\left[ e^{-imt}\left(m\!+\!m_0\right)+e^{\,imt}\left(m\!-\!m_0\right)\right]\;.
\end{equation}
I can also calculate the following transition amplitude:
\begin{equation}
	\bra 0_\OUT|\hat q(t)\hat q(0)|0_\IN\ket = \frac{\sqrt 2 \left(m m_0\right)^{1/4}}{\left(m+m_0\right)^{3/2}}\; e^{-imt}\;.
\end{equation}
I can also calculate, for reference, the ``out-out'' version of Eq.~(\ref{eq:wightman correlator special case}):
\begin{equation}\label{eq:wightman correlator special case out-out}
	\bra 0_\OUT |\hat q(t)\hat q(0)|0_\OUT\ket = \frac{1}{2m}\;e^{-imt}\;,
\end{equation}
consistent with the greater Green's function from Eq.~(\ref{eq:G_>(t) for ground state}).
\\\\
When $m_0 \to m$, all of those will reduce to the same thing. When $m_0 \neq m$, I know of no innate reason to prefer the expectation value over the transition amplitude. If the physics calls for an expectation value (whether ``in-in'' or ``out-out''), calculate an expectation value; otherwise, calculate a transition amplitude. The choice is yours. 
\subsubsection{Wightman expectation value for $t_0 \neq 0$}
What the Schwinger-Keldysh path integral calculates is
\begin{equation}\label{eq:general wightman correlator for quenched oscillator from operators}
	\tr\left(\hat q(t_2)\hat q(t_1)\, e^{\,it_0\hat H_{\text{out}}}\hat\rho \,e^{-it_0\hat H_{\text{out}}}\right)\;,
\end{equation}
with $t_{1,2} > t_0$. The physical setup is to put the system into the pre-quench ground state, 
\begin{equation}\label{eq:|0in><in0|}
	\hat\rho = |0_{\text{in}}\ket \bra 0_{\text{in}}|\;,
\end{equation}
and evolve the fields using the post-quench Hamiltonian, $\hat H_\OUT$. The mathematical setup should be general enough to admit a limit $t_0 \to -\infty$, which is what is always done when implementing an $i\e$ prescription. I will take $t_0$ negative but keep it finite. 
\\\\
I will calculate Eq.~(\ref{eq:general wightman correlator for quenched oscillator from operators}) in two ways: Easy and Hard. 
\\\\
The easy way is to first calculate\footnote{Not a state, just shorthand.}
\begin{equation}
	|\psi(t_1)\ket \equiv \hat q(t_1)\,e^{\,it_0\hat H_{\text{out}}}|0_{\text{in}}\ket = e^{\,it_0\hat H_{\text{out}}}e^{-it_0\hat H_{\text{out}}}\hat q(t_1)\,e^{\,it_0\hat H_{\text{out}}}|0_{\text{in}}\ket = e^{\,it_0\hat H_{\text{out}}}\hat q(t_1-t_0)|0_{\text{in}}\ket\;,
\end{equation}
where the last equality follows from $t_1 > t_0$. Using Eq.~(\ref{eq:q(t>t0) alt}), I find
\begin{equation}
	|\psi(t_1)\ket = e^{\,it_0\hat H_{\text{out}}}\frac{1}{\sqrt{2m_0}}\,\z(t_1\!-\!t_0)\,|1_{\text{in}}\ket\;.
\end{equation}
Therefore, Eq.~(\ref{eq:general wightman correlator for quenched oscillator from operators}) becomes:
\begin{align}\label{eq:easy way}
	\tr\left(\hat q(t_2)\hat q(t_1)\,e^{\,it_0\hat H_{\text{out}}}|0_{\text{in}}\ket \bra 0_{\text{in}}|e^{-it_0\hat H_{\text{out}}}\right) &= \bra 0_{\text{in}}|e^{-it_0\hat H_{\text{out}}}\hat q(t_2)\hat q(t_1)\,e^{\,it_0\hat H_{\text{out}}}|0_{\text{in}}\ket\nonumber\\
	& = \bra \psi(t_2)|\psi(t_1)\ket \nonumber\\
	&= \frac{1}{2m_0}\,\z(t_1\!-\!t_0)\,\z(t_2\!-\!t_0)^*\;.
\end{align}
For $m_0 = m$, $\z(t) = e^{\,imt}$, and I recover the translationally invariant result $G_<(t,t') = \frac{i}{2m} e^{\,im(t-t')} = G_<(t\!-\!t')$. Stage cleared.
\\\\
Now for the hard way.
\\\\
In terms of the excited states,
\begin{equation}
	|n_\OUT\ket \equiv \frac{1}{\sqrt{n!}}(\hat a_\OUT^\da)^n|0_\OUT\ket\;,
\end{equation}
the correlation function can be written as follows:
\begin{align}
	&\tr\left(\hat q(t_2)\hat q(t_1)\,e^{\,it_0H_{\text{out}}}\hat\rho\,e^{-it_0H_{\text{out}}}\right) = \sum_{n\,=\,0}^\infty \bra n_\OUT| \hat q(t_2)\hat q(t_1)\,e^{\,it_0H_{\text{out}}}\hat\rho\,e^{-it_0H_{\text{out}}} |n_\OUT\ket \nonumber\\
	&= \sum_{n\,=\,0}^\infty e^{\,it_2 E_n^\OUT}e^{-it_0 E_n^\OUT} \bra n_\OUT|\hat q e^{-it_2\hat H_\OUT} \hat q(t_1) e^{\,it_0\hat H_\OUT}\hat\rho |n_\OUT\ket\nonumber\\
	&= \sum_{n,n',n''\,=\,0}^\infty e^{\,it_2 E_n^\OUT}e^{-it_0 E_n^\OUT}e^{-it_2E_{n''}^\OUT}e^{\,it_0 E_{n'}^\OUT}\bra n_\OUT|\hat q|n''_\OUT\ket \!\!\!\!\!\!\underbrace{\bra n''_\OUT|\hat q(t_1)|n'_\OUT\ket}_{e^{\,it_1(E_{n''}^\OUT-E_{n'}^\OUT)}\bra n''_\OUT|\hat q|n'_\OUT\ket} \!\!\!\!\!\!\bra n'_\OUT|\hat\rho|n_\OUT\ket \nonumber\\
	&= \sum_{n,n',n''\,=\,0}^\infty e^{\,it_2(E_n^\OUT-E_{n''}^\OUT)+it_1(E_{n''}^\OUT-E_{n'}^\OUT)+it_0(E_{n'}^\OUT-E_n^\OUT)} \bra n_\OUT|\hat q|n''_\OUT\ket \bra n''_\OUT |\hat q|n'_\OUT\ket \bra n'_\OUT|\hat\rho|n_\OUT\ket\;.
\end{align}
Since $t_0 < 0$, the $\hat q$ appearing above is defined in terms of $\hat a_\OUT$:
\begin{equation}
	\hat q = \hat q(0) = \frac{1}{\sqrt{2m}}\left(\hat a_\OUT + \hat a_\OUT^\da\right)\;.
\end{equation}
So the matrix elements of the field operator are the usual sort of thing:
\begin{align}
	&\bra\hat n_\OUT |\hat q| n''_\OUT\ket = \frac{1}{\sqrt{2m}}\left(\sqrt{n''}\;\del_{n,n''-1} + \sqrt{n''\!+\!1}\;\del_{n,n''+1}\right)\;,\nonumber\\
	&\bra\hat n''_\OUT |\hat q| n'_\OUT\ket = \frac{1}{\sqrt{2m}}\left(\sqrt{n'}\;\del_{n'',n'-1} + \sqrt{n'\!+\!1}\;\del_{n'',n'+1}\right)\;.
\end{align}
The density matrix in Eq.~(\ref{eq:|0in><in0|}) has matrix elements
\begin{equation}
	\bra n'_\OUT|\hat\rho|n_\OUT\ket = \bra n'_\OUT|0_\IN\ket \bra 0_\IN|n_\OUT\ket\;.
\end{equation}
Calculating the overlap $\bra 0_\IN | n_\OUT\ket$ is the hard part. In terms of the Hermite polynomials
\begin{equation}
	\mathcal H_n(x) \equiv (-1)^n e^{\,x^2}\pa_x^n (e^{-x^2})\;,
\end{equation}
the wavefunction for the $n^{\text{th}}$ excited state is
\begin{equation}
	\bra q|n_\OUT\ket = \frac{1}{\sqrt{2^n n!}}\left(\frac{m}{\pi}\right)^{1/4}e^{-\half m q^2} \mathcal H_n(\sqrt m\,q)\;.
\end{equation}
So the overlap I need is
\begin{equation}
	\bra 0_\IN | n_\OUT\ket = \frac{(mm_0)^{1/4}}{\sqrt \pi}\frac{1}{\sqrt{2^n n!}}\INT dq\;e^{-\half(m+m_0)q^2}\mathcal H_n(\sqrt m\, q)\;.
\end{equation}
Because the Hermite polynomials have definite parity, that overlap is nonzero only for even $n$. For $n = 0, 2, 4, 6, 8$, Mathematica provides the ratios
\begin{align}
	r_n \equiv \frac{\bra 0_\IN|n_\OUT\ket}{\bra 0_\IN|0_\OUT\ket} &= \left( 1,\;\frac{1}{\sqrt 2}\left(\frac{m\!-\!m_0}{m\!+\!m_0}\right),\;\sqrt{\frac{3}{4}}\left(\frac{m\!-\!m_0}{m\!+\!m_0}\right)^2,\;\sqrt{\frac{5}{16}}\left(\frac{m\!-\!m_0}{m\!+\!m_0}\right)^3,\;\sqrt{\frac{35}{128}}\left(\frac{m\!-\!m_0}{m\!+\!m_0}\right)^4 \right)\;.
\end{align}
Apparently those are related to the central binomial coefficients,\footnote{I thank Jacob Lin for figuring that out.} leading to the closed-form expression
\begin{equation}
	r_n = \left(\frac{n!}{2^n\left[\left(\frac{n}{2}\right)!\right]^2}\right)^{1/2}\left(\frac{m\!-\!m_0}{m\!+\!m_0}\right)^{n/2}\;,\;\; n\;=\;0,2,4,6,8,...\;.
\end{equation}
In terms of that, the density-matrix elements are
\begin{equation}
	\bra n'_\OUT|\hat\rho|n_\OUT\ket = \left\{ \begin{matrix} |\bra 0_\IN|0_\OUT\ket|^2 \; r_{n'} r_n\qquad \text{for }n\text{ and }n'\text{ even}\\ 0\qquad \qquad\;\;\text{otherwise} \end{matrix} \right.\;\;.
\end{equation}
The overlap between ground states is
\begin{equation}
	\bra 0_\IN | 0_\OUT \ket = \sqrt{\frac{2}{m+m_0}}\left(mm_0\right)^{1/4}\;.
\end{equation}
With that, the rest is just algebra and shifting summation indices. One thing I will note is
\begin{equation}
	r_{2k+2} = \sqrt{\frac{2k+1}{2(k+1)}}\left(\frac{m\!-\!m_0}{m\!+\!m_0}\right) r_{2k}\;.
\end{equation}
I will need the following two infinite sums:
\begin{align}
	&S_0 \equiv \sum_{k\,=\,0}^\infty 2k\; r_{2k}^2 = \frac{(m\!-\!m_0)^2(m\!+\!m_0)}{8(mm_0)^{3/2}}\;,\;\; S_1 \equiv \sum_{k\,=\,0}^\infty(2k+1)r_{2k}^2 = \frac{(m\!+\!m_0)^3}{8(mm_0)^{3/2}}\;.
\end{align}
Putting everything together eventually gives
\begin{align}\label{eq:wightman correlator for quenched oscillator in ground state}
	\tr\left(\hat q(t_2)\hat q(t_1)\,e^{\,it_0H_{\text{out}}}\hat\rho\,e^{-it_0H_{\text{out}}}\right) &= \frac{1}{8m^2m_0}\left\{(m\!+\!m_0)^2\, e^{-im(t_2-t_1)} + (m\!-\!m_0)^2\, e^{\,im(t_2-t_1)} \right. \nonumber\\
	&\left. +2(m^2\!-\!m_0^2)\cos[m(t_2\!+\!t_1-2t_0)]\right\}.
\end{align}
When $t_2 = t$ and $t_1 = t_0 = 0$, that reduces to Eq.~(\ref{eq:wightman correlator special case}). For general $t_0$, $t_1$, and $t_2$, I simplify Eq.~(\ref{eq:wightman correlator for quenched oscillator in ground state}) using $\z(t)$ from Eq.~(\ref{eq:Gdot + im0 G}) by working with $\tau_{1,2} \equiv m(t_{1,2}\!-\!t_0)$ and using $\cos(\tau_2\!-\!\tau_1) = \cos(\tau_2)\cos(\tau_1) + \sin(\tau_2)\sin(\tau_1)$ and $\sin(\tau_2\!-\!\tau_1) = \sin(\tau_2)\cos(\tau_1) - \cos(\tau_2)\sin(\tau_1)$. I will once again arrive at Eq.~(\ref{eq:easy way}), which this time I will express in terms of the lesser Green's function:
\begin{equation}\label{eq:G_< for quenched oscillator}
	G_<(t,t') = i\tr\left(\hat q(t_2) \hat q(t_1)\; e^{\,i\hat H t_0} \hat\rho\, e^{-i\hat H t_0}\right) = \frac{i}{2 m_0}\, \z(t\!-\!t_0)\, \z(t'\!-\!t_0)^*\;.
\end{equation}
Was hard mode pointless? Probably. 
\subsubsection{Out-of-time-ordered correlator}
Motivated by the steps leading to Eq.~(\ref{eq:easy way}), I will consider the following object:
\begin{equation}
	|\psi(t_1,t_2)\ket \equiv \hat q(t_2)\hat q(t_1)\,e^{\,it_0\hat H_{\text{out}}}|0_{\text{in}}\ket \;.
\end{equation}
Inserting factors of $\hat 1 = e^{\,it_0\hat H_{\text{out}}}e^{-it_0\hat H_{\text{out}}}$ before and after $\hat q(t_2)$, recalling that $t_{1,2} > t_0$, and using Eq.~(\ref{eq:q(t>t0) alt}), I find
\begin{align}
	|\psi(t_1,t_2)\ket = e^{\,it_0\hat H_{\text{out}}} \frac{\z(t_1\minus t_0)}{2m_0}\left[ \z(t_2\minus t_0)^*\,|0_{\text{in}}\ket + \sqrt{2}\,\z(t_2\minus t_0)\,|2_{\text{in}}\ket\right]\;.
\end{align} 
Therefore:\footnote{I am using $t_2' = t_3$ and $t_1' = t_4$ this time because the first line expresses the ``scattering matrix'' interpretation of the OTOC \cite{thooft1996, stringy_effects}.}
\begin{align}
	&\tr\left( \hat q(t_1')\hat q(t_2')\hat q(t_2) \hat q(t_1)\,e^{\,it_0\hat H_{\text{out}}} |0_{\text{in}}\ket \bra 0_{\text{in}}|e^{-it_0\hat H_{\text{out}}}\right) = \bra \psi(t_1',t_2')|\psi(t_1,t_2)\ket \nonumber\\
	&= \frac{\z(t_1\minus t_0)\z(t_1'\minus t_0)^*}{(2m_0)^2}\left[ \z(t_2'\minus t_0)\,\bra 0_{\text{in}}| + \sqrt{2}\,\z(t_2'\minus t_0)^*\,\bra 2_{\text{in}}|\right] \left[ \z(t_2\minus t_0)^*\,|0_{\text{in}}\ket + \sqrt{2}\,\z(t_2\minus t_0)\,|2_{\text{in}}\ket\right] \nonumber\\
	&= \frac{\z(t_1\minus t_0)\z(t_1'\minus t_0)^*}{4m_0^2}\left[\z(t_2'\minus t_0) \z(t_2\minus t_0)^* + 2\z(t_2'\minus t_0)^* \z(t_2\minus t_0)\right]\;.\label{eq:otoc from operators unsimplified}
\end{align}
From Eq.~(\ref{eq:G_< for quenched oscillator}), I know that $\z(t\minus t_0)\z(t'\minus t_0) = -2im_0\,G_<(t,t')$. That factor of 2 in the second term of Eq.~(\ref{eq:otoc from operators unsimplified}) suggests that I write $2 = 1+1$ and pair the $\z$ and $\z^*$s in two different ways to get two different product of $G_<$s. So I get:
\begin{align}
	&\tr\left( \hat q(t_1')\hat q(t_2')\hat q(t_2) \hat q(t_1)\,e^{\,it_0\hat H_{\text{out}}} |0_{\text{in}}\ket \bra 0_{\text{in}}| e^{-it_0\hat H_{\text{out}}}\right) \nonumber\\
	&= -\left[ G_<(t_1,t_2) G_<(t_2',t_1') + G_<(t_1,t_1') G_<(t_2,t_2') + G_<(t_1,t_2')G_<(t_2,t_1')\right]\;, \label{eq:otoc from operators}
\end{align}
consistent with Eq.~(\ref{eq:otoc in terms of G_<}).
\subsection{Schwinger-Keldysh path integral}
After that recapitulation of the operator formalism, I am ready to evaluate the path integral.
\subsubsection{Integrals over initial field configurations}
Given the ``in" wavefunction $\bra q|0_{\text{in}}\ket = \left(\tfrac{m_0}{\pi}\right)^{1/4} e^{-\half m_0 q^2}$ from Eq.~(\ref{eq:"in" wavefunctions}) and the density matrix $\hat\rho = |0_{\text{in}}\ket \bra 0_{\text{in}}|$, I express the generating function in its factorized form:
\begin{align}
	Z(J,J') &= \INT dq_f\;\Psi(q_f|J)\,\Psi(q_f|J')^*\;,
\end{align}
with
\begin{align}
	&\Psi(q_f|J) \equiv \INT dq_0\;\bra q_0|0_{\text{in}}\ket\, Z(q_0,q_f|J) \nonumber\\
	&= \left(\tfrac{m_0}{\pi}\right)^{1/4} \tfrac{1}{\sqrt{2\pi i G(T)}}\INT dq_0\;e^{\,i\left[\frac{1}{2}i m_0 q_0^2 + S_{\text{cl}}(q_0,q_f|J)\right]} \nonumber\\
	&= \left(\tfrac{m_0}{\pi}\right)^{1/4} \tfrac{1}{\sqrt{2\pi i G(T)}} e^{\,\frac{i}{2G(T)}\left[\dot G(T) q_f^2 + 2q_f\int_{t_0}^{t_f}\!\!dt \,G(t-t_0)J(t) -2\doubleint \Ta(t-t') G(t_f-t)G(t'-t_0)J(t)J(t') \right]} \chi(q_f|J)\;,
\end{align}
where [again, let $\z(t) \equiv \dot G(t) + im_0 G(t)$, as in Eq.~(\ref{eq:Gdot + im0 G})]:
\begin{align}
	\chi(q_f|J) &\equiv \INT dq_0\;e^{\,\frac{i}{2G(T)}\Big\{\left[\dot G(T)+i m_0 G(T)\right] q_0^2 + 2q_0\left[\int_{t_0}^{t_f}\!\! dt\,G(t_f-t)J(t)-q_f\right]\Big\}} \nonumber\\
	&= \INT dq_0\;e^{\,\frac{i\z(T)}{2G(T)} \underbrace{\Big\{q_0^2 + 2q_0\z(T)^{-1}\left[\int_{t_0}^{t_f}\!\! dt\,G(t_f-t)J(t)-q_f\right]\Big\}}_{\left(q_0+\z(T)^{-1}\left[\int_{t_0}^{t_f}\!\! dt\,G(t_f-t)J(t)-q_f\right] \right)^2 - \z(T)^{-2}\left[\int_{t_0}^{t_f}\!\! dt\,G(t_f-t)J(t)-q_f\right]^2}} \nonumber\\
	&= \sqrt{2\pi i G(T) \z(T)^{-1}}\;e^{-\frac{i\z(T)^{-1}}{2G(T)}\left[\int_{t_0}^{t_f}\!\! dt\,G(t_f-t) J(t) - q_f\right]^2}\;.
\end{align}
\subsubsection{Integral over final field configuration}
Integrating over the initial field configurations leaves me with
\begin{align}
	Z(J,J') &= \left(\tfrac{m_0}{\pi}\right)^{1/2}\tfrac{1}{|\z(T)|} \INT dq_f\;e^{\,\frac{i}{2G(T)} I(q_f|J,J')}\;,
\end{align}
where
\begin{align}
	I(q_f|J,J') &= \left[-\z(T)^{-1}+\z(T)^{*-1}\right]q_f^2 + 2q_f\int_{t_0}^{t_f}\!\! dt\left[\al(t)^* J(t) - \al(t) J'(t)\right] \nonumber\\
	&-\doubleint\left\{[2\Ta(t\!-\!t') G(t_f\!-\!t)G(t'\!-\!t_0)+\z(T)^{-1}G(t_f\!-\!t)G(t_f\!-\!t')] J(t)J(t')\right. \nonumber\\
	&\left. -[2\Ta(t\!-\!t')G(t_f\!-\!t)G(t'\!-\!t_0)+\z(T)^{*-1} G(t_f\!-\!t)G(t_f\!-\!t')]J'(t)J'(t') \right\}\;,
\end{align}
with
\begin{equation}
	\al(t) = G(t\!-\!t_0) + \z(T)^{*-1} G(t_f\!-\!t)\;.
\end{equation}
Since $\z(T) = \dot G(T) + i m_0 G(T)$, I find
\begin{equation}
	-\z(T)^{-1}+\z(T)^{*-1} = \frac{2i m_0 G(T)}{|\z(T)|^2}\;,
\end{equation}
which is good, because that is the factor required from the Gaussian integral to cancel out the overall factor in $Z(J,J')$. With $j(J,J') \equiv \int_{t_0}^{t_f}dt\left[\al(t)^* J(t) - \al(t) J'(t)\right]$ as usual, I find
\begin{equation}
	\INT dq_f e^{\,\frac{i}{2G(T)}\left\{\left[-\z(T)^{-1}+\z(T)^{*-1}\right]q_f^2 + 2q_f j\right\}} = \sqrt{\tfrac{\pi |\z(T)|^2}{m_0}}\;e^{-\frac{|\z(T)|^2}{4m_0 G(T)^2}j^2}\;.
\end{equation}
So the influence phase $\Phi(J,J') \equiv -i\ln Z(J,J')$ is
\begin{align}
	\Phi(J,J') &= \frac{1}{2G(T)}\left\{ \frac{i|\z(T)|^2}{2m_0 G(T)} j(J,J')^2\right. \nonumber\\
	&-\doubleint \left[2\Ta(t\!-\!t') G(t_f\!-\!t)G(t'\!-\!t_0) + \z(T)^{-1} G(t_f\!-\!t)G(t_f\!-\!t')\right] J(t)J(t') \nonumber\\
	&\left. + \doubleint \left[2\Ta(t\!-\!t') G(t_f\!-\!t)G(t'\!-\!t_0) + \z(T)^{*-1} G(t_f\!-\!t)G(t_f\!-\!t')\right] J'(t) J'(t') \right\}\;. \label{eq:obstinate}
\end{align}
That should (and does) reduce to Eq.~(\ref{eq:influence phase for oscillator unsimplified}) when $m_0 = m$. 
\subsubsection{Simplification}
Mathematica did not comply when I asked it to simplify Eq.~(\ref{eq:obstinate}), so I will reorganize that expression and try, try again. Consider the $JJ$ term:
\begin{equation}
	\Phi_{JJ\text{ term}} = \frac{1}{2G(T)}\doubleint \left[ R(t,t') + i S(t,t')\right] J(t)J(t')\;,
\end{equation}
with $R+iS$ written in its symmetrized form:
\begin{align}
	&R(t,t') + iS(t,t') \equiv \frac{i|\z(T)|^2}{2m_0 G(T)} \al(t)^* \al(t')^* \nonumber\\
	&-\left[\Ta(t\!-\!t') G(t_f\!-\!t)G(t'\!-\!t_0) + \Ta(t'\!-\!t)G(t_f\!-\!t') G(t\!-\!t_0) + \z(T)^{-1} G(t_f\!-\!t)G(t_f\!-\!t')\right]\;.
\end{align}
After some uninteresting algebra,\footnote{As opposed to the rest of it, right? See, I did exercise editorial discretion.} I obtain the following form of the imaginary part:
\begin{align}
	S(t,t') &= \frac{1}{2m_0 G(T)}\left\{|\z(T)|^2 G(t\!-\!t_0) G(t'\!-\!t_0) + G(t_f\!-\!t)G(t_f\!-\!t') \right. \nonumber\\
	& \left.+ \dot G(T)\left[G(t\!-\!t_0)G(t_f\!-\!t') + G(t_f\!-\!t)G(t'\!-\!t_0) \right]\right\}\;.
\end{align}
Even in that form, Mathematica remained obstinate unless I first told it to set $m = 1$. At any rate, the simplified form is
\begin{equation}
	S(t,t') = \frac{G(T)}{4m_0} \left\{\left(1+\frac{m_0^2}{m^2}\right) \cos[m(t\!-\!t')] +\left(1-\frac{m_0^2}{m^2}\right)\cos[m(t\!+\!t'\!-\!2t_0)]\right\}\;.
\end{equation}
The real part turns out to be much simpler:
\begin{equation}
	R(t,t') = \frac{G(T)}{2}\left[\Ta(t\!-\!t') G(t\!-\!t') + \Ta(t'\!-\!t)G(t'\!-\!t)\right] = \half G(T) G(|t\!-\!t'|)\;.
\end{equation}
So, finally, I get
\begin{align}
	&R(t,t') + iS(t,t') = \nonumber\\
	&\qquad \frac{G(T)}{2}\left\{G(|t\!-\!t'|) + \frac{i}{2m_0}\left[\left(1+\frac{m_0^2}{m^2}\right)\dot G(t\!-\!t') + \left(1-\frac{m_0^2}{m^2}\right)\dot G(t\!+\!t'\!-\!2t_0)\right]\right\}\;.
\end{align}
When $m_0 = m$, that correctly reduces to $R+iS = G(T) \frac{i}{2m} e^{-im|t-t'|}$.
\\\\
Expressing the $J'J'$ term as
\begin{equation}
	\Phi_{J'J' \text{ term}} = \frac{1}{2G(T)}\doubleint\left[R'(t,t') + i S'(t,t')\right] J'(t)J'(t')
\end{equation}
and simplifying along analogous lines, I find that
\begin{equation}
	R'+iS' = -R+iS = -(R-iS) = -(R+iS)^*\;,
\end{equation}
consistent with Eqs.~(\ref{eq:first constraint})-(\ref{eq:third constraint}).
\\\\
The off-diagonal part is the easiest to simplify:
\begin{align}
	\Phi_{JJ' \text{ term}} &= -\frac{1}{2G(T)}\frac{i|\z(T)|^2}{2m_0 G(T)}\doubleint \left[ \al(t)^* \al(t') J(t)J'(t') + \al(t)\al(t')^* J'(t)J(t')\right]\;.
\end{align}
The coefficient of $J(t)J'(t')$ simplifies to
\begin{equation}
	\al(t)^*\al(t') = \frac{G(T)^2}{|\z(T)|^2}\left\{\cos[m(t\!-\!t_0)]+i\frac{m_0}{m}\sin[m(t\!-\!t_0)]\right\}\left\{\cos[m(t'\!-\!t_0)]-i\frac{m_0}{m}\sin[m(t'\!-\!t_0)]\right\}\;.
\end{equation}
So the  influence phase has the expected form
\begin{align}\label{eq:expected form of influence phase}
	\Phi(J,J') &= \half \doubleint\left[G_F(t,t') J(t)J(t') - G_D(t,t')J'(t)J'(t') \right. \nonumber\\
	&\left.-G_>(t,t') J(t)J'(t')-G_<(t,t')J'(t)J(t')\right]\;,
\end{align}
with
\begin{equation}
	G_F(t,t') = \frac{1}{2}\left\{G(|t\!-\!t'|)+\frac{i}{2m_0} \left[\left(1+\frac{m_0^2}{m^2}\right)\dot G(t\!-\!t') + \left(1-\frac{m_0^2}{m^2}\right) \dot G(t\!+\!t'\!-\!2t_0) \right] \right\}\;,
\end{equation}
$G_D(t,t') = G_F(t,t')^*$, and
\begin{equation}\label{eq:lesser green's function for quenched oscillator}
	%G_<(t,t') = \frac{i}{2m_0}\left\{\cos[m(t\!-\!t_0)]+i\frac{m_0}{m}\sin[m(t\!-\!t_0)]\right\}\left\{\cos[m(t'\!-\!t_0)]-i\frac{m_0}{m}\sin[m(t'\!-\!t_0)]\right\}\;,
	G_<(t,t') = \frac{i}{2m_0}\left[\dot G(t\!-\!t_0)+i m_0 G(t\!-\!t_0)\right]\left[\dot G(t'\!-\!t_0)-i m_0 G(t'\!-\!t_0)\right]\;,
\end{equation}
$G_>(t,t') = -G_<(t,t')^*$. To verify that those $G_>$ and $G_F$ are consistent, calculate the following:
\begin{align}
	F(t,t') &\equiv \Ta(t\!-\!t') G_>(t,t') + \Ta(t'\!-\!t)G_<(t,t') \nonumber\\
	&\nonumber\\
	&= \frac{i}{2m_0}\left\{\Ta(t\!-\!t')\left[\dot G(t\!-\!t_0)-im_0 G(t\!-\!t_0)\right]\left[\dot G(t'\!-\!t_0)+im_0G(t'\!-\!t_0)\right]\right. \nonumber\\
	&\left. +\Ta(t'\!-\!t)\left[\dot G(t\!-\!t_0)+im_0G(t\!-\!t_0)\right]\left[\dot G(t'\!-\!t_0)-im_0G(t'\!-\!t_0)\right] \right\} \nonumber\\
	&\nonumber\\
	&=\frac{i}{2m_0}\left\{ \underbrace{\left[\Ta(t\!-\!t')+\Ta(t'\!-\!t)\right]}_{\,=\,1} \left[ \dot G(t\!-\!t_0)\dot G(t'\!-\!t_0) + m_0^2 G(t\!-\!t_0)G(t'\!-\!t_0)\right] \right. \nonumber\\
	&\left. -\left[\Ta(t\!-\!t')-\Ta(t'\!-\!t)\right] im_0 \underbrace{\left[G(t\!-\!t_0)\dot G(t'\!-\!t_0)-\dot G(t\!-\!t_0)G(t'\!-\!t_0)\right]}_{\,=\,G(t-t')}\right\} \nonumber\\
	&\nonumber\\
	&=\frac{i}{2m_0}\left[ \dot G(t\!-\!t_0)\dot G(t'\!-\!t_0) + m_0^2 G(t\!-\!t_0)G(t'\!-\!t_0)\right]+\frac{1}{2} G(|t\!-\!t'|)\;.
\end{align}
The real part is real good. To finish simplifying the imaginary part, use $\cos(a\pm b) = \cos(a)\cos(b)\mp\sin(a)\sin(b)$ with $a = t-t_0$ and $b = t'-t_0$ to write
\begin{align}
	&\dot G(t-t_0)\dot G(t'-t_0) = \frac{1}{2}\left[\dot G(t\!+\!t'\!-\!2t_0) + \dot G(t\!-\!t')\right]\;,\nonumber\\
	&G(t-t_0)G(t'-t_0) = -\frac{1}{2m^2}\left[\dot G(t\!+\!t'\!-\!2t_0)-G(t\!-\!t')\right]\;.
\end{align}
Therefore, 
\begin{align}
	&\dot G(t\!-\!t_0)\,\dot G(t'\!-\!t_0) + m_0^2 G(t\!-\!t_0)\,G(t'\!-\!t_0) = \nonumber\\
	&\qquad \qquad \frac{1}{2}\left[\left( 1+\frac{m_0^2}{m^2} \right)\dot G(t\!-\!t') +\left(1-\frac{m_0^2}{m^2}\right) \dot G(t\!+\!t'\!-\!2t_0)\right]\;, 
\end{align}
in which case $F(t,t') = G_F(t,t')$. Win. 
\subsubsection{Energy in the field}
The intuitive picture I had before calculating the Schwinger-Keldysh path integral for the quenched oscillator is that I should be able to see the field in the ``wrong'' vacuum absorb or emit particles as it relaxes to the right one. I still do not know how to see that directly from Eqs.~(\ref{eq:expected form of influence phase}) and~(\ref{eq:lesser green's function for quenched oscillator}), but I do understand how to recover an averaged version of it.\footnote{I thank Beatrice Bonga and Justin Wilson for conversations about this.}
\\\\
Consider the Heisenberg-picture energy stored in the field operator (Lagrangian formalism, remember?):
\begin{equation}
	\hat{\mathscr H}(t>t_0) = \half [\pa_t\hat q(t)]^2 + \half m^2 \hat q(t)^2\;.
\end{equation}
The average energy in the field is then:
\begin{align}
	E(t) &\equiv \tr\left[\hat{\mathscr H}(t)\;e^{\,it_0\hat H_{\text{out}}} \,\hat\rho_{\text{in}}\, e^{-it_0\hat H_{\text{out}}}\right] \nonumber\\
	&= \lim_{t' \to t} \half \left(\pa_t\pa_{t'} + m^2\right)\tr\left[\hat q(t')\hat q(t)\;e^{\,it_0\hat H_{\text{out}}} \,\hat\rho_{\text{in}}\, e^{-it_0\hat H_{\text{out}}}\right] \nonumber\\
	&= \lim_{t' \to t} \half \left(\pa_t\pa_{t'} + m^2\right) \left[-i\frac{\del^2}{\del J(t)\del J'(t')}\Phi(J,J')\right] \nonumber\\
	&= -i\lim_{t' \to t} \half \left(\pa_t\pa_{t'} + m^2\right)G_<(t,t')\;,
\end{align}
with $G_<(t,t')$ given in Eq.~(\ref{eq:lesser green's function for quenched oscillator}). The result of taking those derivatives and setting $t' = t$ is
\begin{equation}\label{eq:average energy in quenched oscillator}
	E = \frac{m^2+m_0^2}{4m_0}\;.
\end{equation}
The average energy in the field remains constant, which is one manifestation of the comment I made below Eq.~(\ref{eq:quenched oscillator hamiltonian}). To interpret Eq.~(\ref{eq:average energy in quenched oscillator}) further, recall the Bogoliubov coefficients $A$ and $B$ from Sec.~\ref{sec:bogoliubov}:
\begin{equation}
	A = \frac{m+m_0}{2\sqrt{mm_0}}\;,\;\; B = \frac{m-m_0}{2\sqrt{mm_0}}\;.
\end{equation}
The energy in Eq.~(\ref{eq:average energy in quenched oscillator}) therefore has the form
\begin{equation}
	E = \left(|A|^2 + |B|^2\right)\frac{m}{2} = \left(|B|^2 + \half\right)m\;.
\end{equation}
That makes a lot of sense: The $\half$ is the contribution of the post-quench ground state, and $|B|^2$ is the average number of post-quench quanta contained in the pre-quench ground state [recall Eq.~(\ref{eq:average energy in quenched oscillator})]. 
\\\\
It is also interesting to take limits of Eq.~(\ref{eq:average energy in quenched oscillator}):
\begin{align}
	E \to \left\{ \begin{matrix} \frac{m_0}{4}\;\;\text{ for}\;\; m \to 0 \\\\
	\frac{m}{2}\;\;\text{ for}\;\; m_0 \to m \\\\ \frac{m^2}{4m_0}\;\;\text{ for}\;\;m_0 \to 0\end{matrix} \right.\;\;.
\end{align}
The first case shows that sending a particle into the void releases half of the ground-state energy; the second case is a sanity check, recovering the full ground-state energy. The third case shows that trying to confine an almost-free particle requires almost infinite energy. Poetic. 
%
%\pagebreak
%
\subsection{Larkin-Ovchinnikov path integral}
%%%%
A ground-state density matrix
\begin{equation}
	\hat\rho = |0_\IN\ket \bra 0_\IN|
\end{equation} 
will give the same $\rho(q_0,q_0')$ as for the unquenched case except for the replacement $m \to m_0$:
\begin{align}
	\rho(q_0,q_0') &= \left(\frac{m_0}{\pi}\right)^{1/2} e^{-m_0\left\{q_i^2+q_i\int_{t_0}^{t_f}\!\! dt\;G(t-t_0) \left[J(t)-J'(t)-J''(t)+J'''(t)\right]\right\}} \times \nonumber\\
	&e^{-\half m_0\doubleint G(t-t_0) G(t'-t_0) \left\{ \left[ J(t)-J'(t)\right]\left[J(t')-J'(t')\right]+\left[J''(t)-J'''(t)\right]\left[J''(t')-J'''(t')\right]\right\}}\;.
\end{align}
Completing the square and so on will proceed along the same lines as for the unquenched case, leading to the same kind of $j = \int_{t_0}^{t_f}dt\left\{\al(t)^*\left[J(t)-J'(t)\right]-\al(t)\left[J''(t)-J'''(t)\right]\right\}$ but with a modified $\al(t)$:
\begin{equation}\label{eq:alpha(t) for quench}
	\al(t) \equiv G(t\!-\!t_0)+\frac{i}{m_0G(T)}\left[\dot G(T)G(t\!-\!t_0)+G(t_f\!-\!t)\right]\;.
\end{equation}
The preliminary result is $Z(J,J',J'',J''') = e^{\,i\Phi(J,J',J'',J''')}$ with
\begin{align}
	&\Phi(J,J',J'',J''') = \frac{1}{2G(T)}D(J,J',J'',J''') -i\frac{m_0}{4}j(J,J',J'',J''')^2 \nonumber\\
	&+i\frac{m_0}{2}\doubleint G(t\!-\!t_0)G(t'\!-\!t_0)\Big\{\left[J(t)\!-\!J'(t)\right]\left[J(t')\!-\!J'(t')\right]  \Big. \nonumber\\
	&\qquad\qquad\qquad \qquad\qquad\qquad \qquad\qquad\qquad\Big.+ \left[J''(t)\!-\!J'''(t)\right]\left[J''(t')\!-\!J'''(t')\right]\Big\}.
\end{align}
Now I ask Mathematica to simplify each term. First, observe that\footnote{Mathematica required some coaxing to put the result into that form, at least as of version 12.0 on macOS 10.15.3. I had to first collect terms in $m$ and $m_0$, then work with the ratio $m/m_0$, then simplify. Concretely, after defining functions $\al[t]$ and $\al\text{Star}[t]$ for Eq.~(\ref{eq:alpha(t) for quench}) and its conjugate, I wrote:
	\begin{equation}
		\text{Collect[$\al$Star[t] $\al$[tp], \{m,m0\}, FullSimplify] /. \{m $\to$ $\mu$ m0\} // FullSimplify /. \{$\mu$ $\to$ $\frac{\text{m}}{\text{m0}}$\} // FullSimplify}
	\end{equation} 
}
\begin{align}
	\al(t)^*\al(t') &= \frac{1}{m^2 m_0^2}\,\big\{m\cos[m(t\!-\!t_0)]+i m_0\sin[m(t\!-\!t_0)]\big\} \times \nonumber\\
	&\qquad\qquad\qquad\big\{m\cos[m(t'\!-\!t_0)]-i m_0\sin[m(t'\!-\!t_0)]\big\}\;,
\end{align}
which correctly reduces to Eq.~(\ref{eq:alpha(t)*alpha(t')}) when $m_0 = m$. The final time $t_f$ was arbitrary and appropriately dropped out. But for this case, the initial time $t_0$ is physical: It is the time at which the oscillator's frequency changed. The problem is not translationally invariant, and the resulting correlation functions should depend on $t_0$. 
\\\\
Anticipating the final form of the influence phase, I define the functions
\begin{align}
	G_>(t,t') &\equiv \frac{i}{2m^2 m_0}\,\big\{m\cos[m(t\!-\!t_0)]-im_0\sin[m(t\!-\!t_0)]\big\} \times \nonumber\\
	&\qquad\qquad\qquad\big\{m \cos[m(t'\!-\!t_0)] + i m_0 \sin[m(t'\!-\!t_0)]\big\}\;,
\end{align}
and
\begin{align}
	G_<(t,t') &\equiv \frac{i}{2m^2 m_0}\,\big\{m \cos[m(t\!-\!t_0)] + i m_0 \sin[m(t\!-\!t_0)]\big\} \times \nonumber\\
	&\qquad\qquad\qquad \big\{m \cos[m(t'\!-\!t_0)] - im_0 \sin[m(t'\!-\!t_0)]\big\} \nonumber\\
	&= G_>(t',t) = -G_>(t,t')^*\;.
\end{align}
Given those two, I further define
\begin{equation}
	G_F(t,t') \equiv \Ta(t\!-\!t') G_>(t,t') + \Ta(t'\!-\!t)G_<(t,t')\;,
\end{equation}
and
\begin{equation}
	G_D(t,t') \equiv -\left[\Ta(t\!-\!t')G_<(t,t') + \Ta(t'\!-\!t)G_>(t,t')\right] = G_F(t,t')^*\;.
\end{equation}
In terms of those, the generalized influence phase will look exactly like Eq.~(\ref{eq:general form of LK phase}), which I will repeat using the now-required translationally noninvariant notation: 
\begin{align}
	&\Phi(J,J',J'',J''') = \half \doubleint\!\!\left\{ \phantom{\tfrac{1}{2}}\right. \nonumber\\
	&\qquad \;\;\;\;\, G_F(t,t')\left[J(t)J(t')+J''(t)J''(t')\right]-G_D(t,t')\left[J'(t)J'(t')+J'''(t)J'''(t')\right]\nonumber\\
	&\qquad -G_<(t,t')\left[J(t)J'(t')+J''(t)J'''(t')\right] -G_>(t,t')\left[J'(t)J(t')+J'''(t)J''(t')\right]\nonumber\\
	&\qquad +G_<(t,t')\left[J(t)J''(t')+J'(t)J'''(t') - J(t)J'''(t')-J'(t)J''(t')\right]  \nonumber\\
	&\qquad +G_>(t,t')\left[J''(t)J(t')+J'''(t)J'(t') - J'''(t)J(t')-J''(t)J'(t')\right] \nonumber\\
	&\left. \phantom{\tfrac{1}{2}}\right\}\;.
\end{align}
%%%%
Long live the quench.
\pagebreak
\section{Open systems}
%%%%
\begin{center}
	Drainage! Drainage, Eli, you boy.\\
	\qquad\qquad\qquad\qquad\qquad\qquad ---Daniel Plainview, \textit{There Will Be Blood}
\end{center}
%%%%
\subsection{State decay with conservation of probability}\label{sec:decay}
The most pedestrian way to model the decay of an excited state is to add a small imaginary part to the Hamiltonian. For example, if a single-particle wavefunction $\psi(\vec x)$ were an eigenstate of a nonhermitian Hamiltonian 
\begin{equation}\label{eq:nonhermitian H}
	\hat{\mathcal H} = \hat H - \half i \hat\G
\end{equation}
with complex energy $\mathcal E = E-\half i \G$, then the wavefunction at later times would be $\psi(t,\vec x) = \psi(\vec x)\, e^{-i\mathcal E t}$. The probability density would be $|\psi(t,\vec x)| = |\psi(\vec x)|^2\, e^{-\G t}$, and the state would decay with a lifetime $\G^{-1}$. Drainage. 
\\\\
But unlike oil prospectors, quantum subsystems return what they take. Consider the density matrix in a basis of instantaneous field eigenstates:\footnote{Notwithstanding some earlier footnotes.}
\begin{equation}
	\hat\rho(t) \equiv e^{+i\hat H t} \hat\rho\, e^{-i\hat H t}\;.
\end{equation}
That time-dependent operator satisfies the differential equation
\begin{equation}\label{eq:heisenberg eq for rho}
	\pa_t\hat\rho(t) = i (\hat H \hat \rho(t)-\hat\rho(t)\hat H)\;.
\end{equation}
Including decay as above would entail replacing the right-hand side of Eq.~(\ref{eq:heisenberg eq for rho}) by
\begin{equation}
	\pa_t\hat\rho(t) = i\left(\hat{\mathcal H}\hat\rho(t)-\hat\rho(t) \hat{\mathcal H}^\da\right)\;.
\end{equation}
As I said, this approach does not conserve probability. So let me try to compensate the loss of probability by transforming $\hat\rho$ with some other operator $\hat L$ while preserving $\hat\rho(t)^\da = \hat\rho(t)$:
\begin{equation}\label{eq:modified evolution}
	\pa_t\hat\rho(t) = i\left(\hat{\mathcal H}\hat\rho(t)-\hat\rho(t) \hat{\mathcal H}^\da\right) - \hat L \hat\rho(t) \hat L^\dagger\;.
\end{equation}
Imposing conservation of probability in the form $\pa_t \tr[\hat\rho(t)] \equiv 0$ will relate $\hat L$ to $\im (\hat{\mathcal H})$:
\begin{align}
	0 &= i\tr\left([\hat H,\hat\rho(t)]-\half i\{\hat \G,\hat\rho(t)\}\right)-\tr\left(\hat L\hat\rho(t)\hat L^\da\right) \nonumber\\
	&= \tr\left(\left(\hat \G-\hat L^\da \hat L\right)\hat\rho(t)\right)\;.
\end{align}
The goal is to conserve probability for any density matrix, so the above implies
\begin{equation}
	\hat \G = \hat L^\da \hat L\;.
\end{equation}
I included only a single $\hat L$ to streamline the presentation. Eq.~(\ref{eq:modified evolution}) could be generalized to
\begin{equation}\label{eq:further modified evolution}
	\pa_t\hat\rho(t) = i\left(\hat{\mathcal H}\hat\rho(t)-\hat\rho(t) \hat{\mathcal H}^\da\right) - \sum_{I\,=\,1}^{N_L}\hat L_I \hat\rho(t) \hat L_I^\dagger\;,
\end{equation}
in which case imposing $\pa_t\tr[\hat\rho(t)] = 0$ would imply
\begin{equation}\label{eq:gamma = L*L}
	\hat\G = \sum_{I\,=\,1}^{N_L} \hat L_I^\da \hat L_I\;.
\end{equation}
The evolution law in Eq.~(\ref{eq:further modified evolution}) with the nonhermitian Hamiltonian in Eq.~(\ref{eq:nonhermitian H}) constrained by Eq.~(\ref{eq:gamma = L*L}) is called Lindblad evolution.\footnote{This is usually written in terms of the Schrodinger-picture density matrix $\hat\rho_{\text{Sch}}(t)$ as $\pa_t\hat\rho_{\text{Sch}}(t) = -i\left(\hat{\mathcal H}\hat\rho_{\text{Sch}}(t) - \hat\rho_{\text{Sch}}(t) \hat{\mathcal H}\right)+\sum_{I\,=\,1}^{N_L}\hat L_I\hat\rho_{\text{Sch}}(t)\hat L_I^\da$. The discrepancy is reconciled by the trivial observation that $\hat\rho_{\text{Sch}}(t) = \hat\rho(-t)$. Also, only recently did I stumble upon yet another useful paper by Strunz, whose derivation of Lindblad evolution is like mine \cite{strunz_stochastic}.}
%%%%
%%%%
\subsection{Double-sided evolution}
%%%%
The objective is to construct a generating function for expectation values of Heisenberg-picture operators that evolve with Lindblad dynamics instead of Hamiltonian dynamics \cite{buchhold2015, buchhold2016}. To begin I will revisit the path-integral decomposition that led to the generating function $Z(J,J')$ in Eq.~(\ref{eq:keldysh}). 
\\\\
Recall the group-theoretic relation
\begin{equation}
	e^{\,i t\hat H} \hat q\, e^{-i t\hat H} = e^{\,it[\hat H,\cdot]}\hat q\;,
\end{equation}
which relates the group elements $U(t) \equiv e^{\,it\hat H}$ to infinitesimal transformations in the adjoint representation, $\del_t\hat q = it[\hat H,\hat q]$. What I want is to redo the path-integral decomposition, this time working directly with the adjoint representation. 
%\\\\
\begin{align}
	\tr\left(\hat q(t_1) \hat \rho(t_0)\right) &= \tr\left(e^{-it_f\hat H} \hat q(t_1)\hat\rho(t_0) e^{\,it_f\hat H}\right) = \tr\left(e^{-it_f\hat H}e^{\,it_1\hat H}\hat q e^{-it_1\hat H} e^{\,it_0\hat H} \hat\rho\, e^{-it_0\hat H} e^{it_f\hat H}\right) \nonumber\\
	&= \tr\left(e^{-i(t_f-t_1)\hat H} \hat q e^{-i(t_1-t_0)\hat H} \hat\rho\, e^{\,i(t_1-t_0)\hat H} e^{\,i(t_f-t_1)\hat H}\right) \nonumber\\
	&= \tr\left\{ e^{-i(t_f-t_1)[\hat H,\cdot]}\left( \hat q\, e^{-i(t_1-t_0)[\hat H,\cdot]}\hat\rho\right)\right\}\;.
\end{align}
In this presentation the evolution is less ``right-to-left-then-back-again'' and more ``outward.'' For now this is just an alternative way to organize the evolution, but it will be necessary to evolve this way for an open system. 
%%%%
\subsubsection{Path-integral decomposition}
%%%%
Expand the density matrix in the coordinate basis:
\begin{equation}
	\hat\rho = \INT dq_0 \INT dq_0'\;\rho(q_0,q_0')\;|q_0\ket \bra q_0'|\;.
\end{equation}
Writing $t_1-t_0 \equiv n\e$ as in Eq.~(\ref{eq:t1-t0=ne}), I want to perform a path-integral decomposition of $e^{-i(t_1-t_0)[\hat H,\cdot]}\left(|q_0\ket \bra q_0'|\right)$. I enclosed the matrix-basis element in parentheses to emphasize that the commutator acts \textit{on both sides}, which is why I call this ``double-sided'' evolution.\footnote{Operators like $e^{-i(t_1-t_0)[\hat H,\cdot]}$ that act in this way are sometimes called ``superoperators'' \cite{buchhold2016}.} To perform the path-integral decomposition in this presentation, I need to insert $\hat 1 = \INT dq\,|q\ket \bra q|$ simultaneously on the left and right:
\begin{align}\label{eq:double-sided matrix elements}
	&e^{-i\e [\hat H,\cdot]}\left(|q_0\ket \bra q_0'|\right) = \hat 1\; e^{-i\e [\hat H,\cdot]}\left(|q_0\ket \bra q_0'|\right)\; \hat 1 \nonumber\\
	&= \INT dq_1 \INT dq_1'\; |q_1\ket \bra q_1|\left[e^{-i\e [\hat H,\cdot]}\left(|q_0\ket \bra q_0'|\right)\right]|q_1'\ket \bra q_1'| \nonumber\\
	&= \INT dq_1 \INT dq_1'\; |q_1\ket \bra q_1|\left[\left(1-i\e[\hat H,\cdot]\right)\left(|q_0\ket \bra q_0'|\right)\right]|q_1'\ket \bra q_1'| \nonumber\\
	&= \INT dq_1 \INT dq_1'\; |q_1\ket \left[\bra q_1|q_0\ket \bra q_0'|q_1'\ket - i\e\left(\bra q_1|\hat H|q_0\ket \bra q_0'|q_1'\ket - \bra q_1|q_0\ket \bra q_0'|\hat H|q_1'\ket\right)\right]\bra q_1'|\;.
\end{align}
With that, and with $\hat H = \half \hat p^2 + V(\hat q)$, so that I can do the momentum integrals, evolution up to $t = t_1$ gives
\begin{align}
	&e^{-i(t_1-t_0)[\hat H,\cdot]}\hat\rho = \INT dq_0 dq_0'\;\rho(q_0,q_0')\;\times \nonumber\\
	&\INT \frac{dq_1 dq_1'}{2\pi \e}...\INT \frac{dq_n dq_n'}{2\pi \e}\;|q_n\ket  \;e^{\,i\e\sum_{j\,=\,1}^n\left(\half \dot q_{j-1}^2-V(q_{j-1})-\half \dot q_{j-1}'^2+V(q_{j-1}')\right)}\bra q_n'|\;.
\end{align}
Now I can act with $\hat q$ and perform the analogous decomposition of the operator $e^{-i(t_f-t_1)[\hat H,\cdot]}$ acting on $\hat q e^{-i(t_1-t_0)[\hat H,\cdot]}\hat\rho$ to get the same result as before. 
\\\\
The point is really the last line of Eq.~(\ref{eq:double-sided matrix elements}): When $\hat H$ acts on the left I get the matrix element $\bra q_1|\hat H|q_0\ket$, and when $\hat H$ acts on the right I get the matrix element $\bra  q_0'|\hat H|q_1'\ket$. 
%%%%
\subsubsection{Phase-space path integral for Lindblad evolution}
%%%%
%\\\\
Continue to focus on Eq.~(\ref{eq:double-sided matrix elements}). For Hamiltonian evolution, I had to calculate matrix elements of the form
\begin{align}
	C_{j,j-1} &\equiv \bra q_j|\left( e^{-i\e[\hat H,\cdot]}\left(|q_{j-1}\ket \bra q'_{j-1}|\right) \right)|q_j'\ket = \nonumber\\
	&= \bra q_j|\left(\left( \hat 1-i\e[\hat H,\cdot]\right)\left(|q_{j-1}\ket \bra q'_{j-1}|\right) \right)|q_j'\ket \nonumber\\
	&= \bra q_j|q'_j\ket -i\e \bra q_j| \left( \hat H|q_{j-1}\ket \bra q'_{j-1}| - |q_{j-1}\ket \bra q'_{j-1}|\hat H \right)|q'_j\ket \nonumber\\
	&=  \bra q_j|q'_j\ket -i\e \left( \bra q_j|\hat H|q_{j-1}\ket \bra q'_{j-1}|q'_j\ket - \bra q_j|q_{j-1}\ket \bra q'_{j-1}|\hat H|q'_j\ket \right)\;.
\end{align}
For Lindblad evolution, I will need to calculate the analogous matrix elements but with the replacement
\begin{align}
	[\hat H,\cdot]\left(|q_{j-1}\ket\bra q'_{j-1}|\right) &\to \hat{\mathcal H}|q_{j-1}\ket\bra q'_{j-1}|\;-\;|q_{j-1}\ket\bra q'_{j-1}|\hat{\mathcal H}^\da +i \sum_{I\,=\,1}^{N_L} \hat L_I |q_{j-1}\ket\bra q'_{j-1}| \hat L_I^\da\;.
\end{align}
The generating function in phase-space variables\footnote{I still call this the ``Hamiltonian formalism'' (i.e., $q$s and $p$s) with the understanding that it describes Lindblad evolution.} is therefore
\begin{align}
	Z(J,J') &= \!\INT \!\!dq_f \!\INT \!\!dq_0 \!\INT \!\!dq_0'\;\rho(q_0,q_0')\;\times\nonumber\\
	&\qquad\qquad \int_{q(t_0)\,=\,q_0}^{q(t_f)\,=\,q_f}\!\!\!\!\!\! \Ds q(\cdot) \int_{q'(t_0)\,=\,q'_0}^{q'(t_f)\,=\,q_f}\!\!\!\!\!\! \Ds q'(\cdot) \int \Ds p(\cdot) \int \Ds p'(\cdot)\;e^{\,i \mathcal I(q,p;q'\!,\,p'|J,J')}\;,
\end{align}
with
\begin{align}\label{eq:hamiltonian action}
	\mathcal I(q,p;q',p'|J,J') &= \int_{t_0}^{t_f}\!\!\!dt\left\{ p\dot q - \mathcal H(q,p|J)-\left[p'\dot q'-\mathcal H^*(q',p'|J')\right] -i\sum_{I\,=\,1}^{N_L} L_I(q,p)L_I^*(q',p') \right\}\;,
\end{align}
and $\mathcal H(q,p|J) = \mathcal H(q,p)-Jq$, $\mathcal H(q,p) = H(q,p)-\half i\G(q,p)$, with $\G(q,p) = \sum_{I\,=\,1}^{N_L}|L_I(q,p)|^2$.
%%%%
\section{Probability-conserving $i\e$ prescription}\label{sec:iepsilon}
Strunz \cite{strunz1997} performed the momentum integrals with Lindblad operators $\hat L_I = \al_I \hat q + \beta_I \hat p$ for general coefficients $\al_I$ and $\beta_I$. I want to focus on a particular case: 
\begin{equation}\label{eq:loss}
	\hat L = \sqrt{2m\e}\; \hat a\;,
\end{equation}
with $\hat a = \sqrt{\frac{m}{2}}\left(\hat q+\frac{i}{m}\hat p\right)$. This describes the loss of a single quantum to the bath \cite{buchhold2016}. The nonhermitian Hamiltonian is for this case is
\begin{align}
	\hat{\mathcal H} &= \hat H-\frac{i}{2}\hat \G = m\left[(1\!-\!i\e)\,\hat a^\da \hat a + \half \hat 1\right]\;.
\end{align}
That means it will correctly generalize the $i\e$ prescription to the formalism in which probability is conserved. Inserting Eq.~(\ref{eq:loss}) into Eq.~(\ref{eq:hamiltonian action}), performing the momentum integrals, and dropping terms of $O(\e^2)$ and higher leads to the generating function
\begin{align}
	&Z(J,J') = \nonumber\\
	&\INT \!\!dq_f \INT \!\!dq_0 \INT \!\!dq_0'\;\rho(q_0,q_0') \int_{q(t_0)\,=\,q_0}^{q(t_f)\,=\,q_f}\!\!\!\!\!\!\!\!\! \Ds q(\cdot) \int_{q'(t_0)\,=\,q'_0}^{q'(t_f)\,=\,q_f}\!\!\!\!\!\!\!\!\! \Ds q'(\cdot)\;e^{\,i\int_{t_0}^{t_f} dt\left[\la(q(t),\,q'(t))+J(t)q(t) +J'(t)q'(t)\right]}\;,
\end{align}
with Lagrangian 
\begin{align}\label{eq:ie lagrangian}
	\la(q,q') &= \half(1\!+\!i\e)\dot q^2-\half(1\!-\!i\e)m^2 q^2-\left[\half(1\!-\!i\e)\dot q'^2-\half(1\!+\!i\e)m^2q'^2\right] \nonumber\\
	&-i\e\left[\dot q \dot q' + m^2 qq'+im(\dot q q' \!-\! \dot q' q)\right]\;.
\end{align}
The first line is the Lagrangian that would result from the traditional $i\e$ prescription, as described in Sec.~\ref{sec:wrong iepsilon}; the second line is how to fix it. Note that $\la(q,q) = 0$, as required.\footnote{Thank you, General Buchhold.}
%\\\\
\subsection{Phenomenology vs. fundamentals}
As I qualified back in Sec.~\ref{sec:wrong iepsilon}, the prescription defined by Eq.~(\ref{eq:ie lagrangian}) is still phenomenological. It is still just a prescription, not a derivation from a fundamental model.\footnote{A particle-physics analogy is treating the proton mass as a free parameter instead of deriving it from the QCD coupling and the quark masses \cite{zeeQFT}. I will return to QCD for different reasons in Sec.~\ref{sec:end}.} 
\\\\
A derivation would entail the following: Take a closed system, such as a self-interacting field in a box, split the field into ``fast'' and ``slow'' modes, trace out the fast modes, and leave behind an effective Lindblad action for the slow modes. The parameter $\e$ would be given in terms of the fast/slow cutoff and the parameters of the underlying model. 
\\\\
This qualification continues, I think, the remarks in Sec.~10-5 of Feynman \& Hibbs \cite{feynmanhibbs} and explains why the work of Caldeira and Leggett does not solve the problem \cite{caldeira_leggett}.\footnote{I thank A.~Zee for pointing me to that section and for a discussion about this.} An unsatisfying attempt was given by Hu et al.~\cite{Hu1995b}\footnote{They set up everything correctly but then lament that separating ``a single field into the high and low momentum sectors'' would be ``cumbersome to carry out'' and instead ``consider two independent self-interacting scalar fields $\phi(x)$ depicting the system, and $\psi(x)$ depicting the bath.'' Gee whiz. And in their main example, they ``extend the range of all integrations in the Feynman diagrams to cover the whole momentum space,'' effectively assuming that ``$\phi$ and $\psi$ are two independent fields.'' Wonderful.} Perhaps I should be able to adapt the work of Calzetta et al.~\cite{Calzetta2001}, Lombardo and Mazzitelli \cite{Lombardo1996}, or Zanella and Calzetta \cite{calzetta_RG} to derive an effective $\e$ from $\phi^3$ theory or $\phi^4$ theory.\footnote{Maybe $\phi^3$ theory in $5+1$ dimensions, as advocated for pedagogical reasons by Srednicki \cite{srednicki2007}.} In another life. 
\vfill
\pagebreak
%%%%
\subsection{Step 1: Keldysh basis for Hamiltonian evolution}\label{sec:intro}
The last horse crosses the finish line: I will finally switch to the Keldysh basis.
\\\\
Start with the combined Lagrangian for the forward and backward fields:
\begin{equation}\label{eq:combined lagrangian}
	\la(q,q') = \half \dot q^2-\half m^2 q^2 + Jq - \left(\half \dot q'^2 - \half m^2 q'^2 + J'q'\right)\;.
\end{equation}
The Keldysh basis is defined by the sum and difference of the fields:\footnote{Others prefer the notation $q_+$ and $q_-$ to denote what I call $q$ and $q'$, and they instead call the sum and difference fields $q_{\text{classical}}$ and $q_{\text{quantum}}$ (or some variation thereof).}
\begin{align}
	q_\pm \equiv q\pm q'\;,\;\; J_\pm \equiv J\pm J'\;.
\end{align}
Then
\begin{align}
	& q = \frac{q_+ + q_-}{2}\;\text{ and }\;\; q' = \frac{q_+ - q_-}{2}\;,
\end{align}
and similarly for $J$ and $J'$. The pertinent combinations of fields are
\begin{equation}
	q^2\minus q'^2 = q_+q_-\;,\;\; Jq - J'q' = \half(J_+q_- \!+\! J_- q_+)\;.
\end{equation}
In terms of $q_\pm$, the Lagrangian in Eq.~(\ref{eq:combined lagrangian}) becomes
\begin{align}\label{eq:combined lagrangian in keldysh basis}
	\la_K(q_+,q_-) &\equiv \la\left(\tfrac{q_+ + q_-}{2}, \tfrac{q_+ - q_-}{2} \right) = \half \dot q_+ \dot q_- - \half m^2 q_+ q_- + \half \left(J_+ q_- + J_- q_+\right)\;.
\end{align}
Now drop the $K$ subscript, because even I have limits. The equations of motion are:
\begin{align}
	&\frac{\del S}{\del q_+} = -\half (\ddot q_- + m^2 q_-) + \half J_- = 0 \implies \ddot q_- + m^2 q_- = J_-\;, \\
	&\frac{\del S}{\del q_-} = -\half(\ddot q_+ + m^2 q_+) + \half J_+ = 0 \implies \ddot q_+ + m^2 q_+ = J_+\;.
\end{align}
The action evaluated on solutions of those equations is
\begin{align}
	S_{\text{cl}} &\equiv \!\left.\int_{t_0}^{t_f}\!\! dt\;\la \right|_{\text{those eqs.}}\!\!\!\!\! = \left.\half q_+ \dot q_-\right|_{t\,=\,t_0}^{t_f} \!+\!\left. \int_{t_0}^{t_f}\!\! dt\left[ -\half q_+\left(\ddot q_- + m^2 q_-\right) + \half\left(J_+ q_- + J_- q_+\right) \right]\right|_{\text{those eqs.}} \nonumber\\
	&= \left.\half q_+ \dot q_-\right|_{t\,=\,t_0}^{t_f} + \int_{t_0}^{t_f} dt\;\half J_+ q_-\;.
\end{align}
To verify that that reproduces the $S_{\text{cl}}$ from the original basis, note that
\begin{align}
	\left. q'\dot q - q \dot q' \right|_{t_0}^{t_f} = \int_{t_0}^{t_f}dt\; \tfrac{d}{dt}\left(q' \dot q - q \dot q'\right) = \int_{t_0}^{t_f} dt \left( \dot q' \dot q + q' \ddot q - \dot q \dot q' - q \ddot q'\right) = \int_{t_0}^{t_f} dt\left(q' \ddot q - q \ddot q'\right)\;.
\end{align}
Applying the equations of motion in the original basis (in the forms $\ddot q - J = -m^2 q$ and $\ddot q' - J' = -m^2 q'$), I find
\begin{align}
	\left.q'\dot q - q \dot q'\right|_{t\,=\,t_0}^{t_f} \!+\! \int_{t_0}^{t_f}\!\! dt\left(J' q \!-\! J q'\right) &= \int_{t_0}^{t_f}\!\! dt \left[ \,(J'\!-\!\ddot q\,') q +(\ddot q \!-\! J) q'\,\right] \nonumber\\
	&= \int_{t_0}^{t_f}\!\! dt \left(m^2 q' q - m^2 q q'\right) = 0\;.
\end{align}
So indeed inserting the definitions of $q_\pm$ and $J_\pm$ into the above $S_{\text{cl}}$ would reproduce $S_{\text{cl}} = \left.\half q\dot q \right|_{t\,=\,t_0}^{t_f} + \half \int_{t_0}^{t_f} dt\; J q - \left( \left.\half q'\dot q' \right|_{t\,=\,t_0}^{t_f} + \half \int_{t_0}^{t_f} dt\; J' q'\right)$.
%%%%
%%%%
\subsection{Step 2: Keldysh basis for Lindblad $i\e$ prescription}
%%%%
Inserting the Lagrangian from Eq.~(\ref{eq:feynman Z with wrong iepsilon}) into Eq.~(\ref{eq:keldysh Z with wrong iepsilon}) produces what I will call the ``old'' Lagrangian for the $i\e$ prescription: 
\begin{align}
	\la_{\text{old}} &\equiv \half(1\!+\!i\e) \dot q^2 - \half(1\!-\!i\e)m^2 q^2 - \left[\half(1\!-\!i\e) \dot q'^2 - \half(1\!+\!i\e) m^2 q'^2 \right] \\
	&= \half\left(\dot q_+ \dot q_- \!-\! m^2 q_+ q_-\right) + \fourth i\e \left[\dot q_+^2 \!+\! \dot q_-^2 + m^2\left(q_+^2 \!+\! q_-^2\right) \right]\;.\label{eq:old lagrangian for iepsilon}
\end{align}
The additional cross terms I get from Lindblad evolution are
\begin{align}
	\la_{\text{cross}} &\equiv -i\e\left[\dot q \dot q' + m^2 q q' + im\left(\dot q q' - \dot q' q\right)\right] \\ 
	&= -\fourth i\e \left[ \dot q_+^2 \!-\! \dot q_-^2 + m^2(q_+^2 \!-\! q_-^2) + 2im\left(q_+ \dot q_- \!-\! q_- \dot q_+\right)\right]\;.\label{eq:additional terms for iepsilon}
\end{align}
Adding that to Eq.~(\ref{eq:old lagrangian for iepsilon}) produces a ``new'' Lagrangian, completed and corrected:
\begin{align}\label{eq:new lagrangian for iepsilon}
	\la_{\text{new}} &\equiv \la_{\text{old}} + \la_{\text{cross}} = \half\left(\dot q_+ \dot q_- \!-\! m^2 q_+ q_-\right) +\half i\e \left[ \dot q_-^2 \!+\! m^2 q_-^2 + im\left(q_- \dot q_+ \!-\! q_+ \dot q_-\right)\right]\;.
\end{align}
Magnifique. To that I add the source term $Jq-J'q' = \half(J_- q_+ + J_+ q_-)$, as before. The new sourced equations of motion are
\begin{align}
	& \ddot q_- - 2\e m \dot q_- + m^2 q_- = J_-\;, \label{eq:equation for difference field}\\
	& \ddot q_+ + 2\e m \dot q_+ + m^2 q_+ = J_+ + 2i\e\left(2m^2 q_- - J_-\right) \equiv \hat J_+\;. \label{eq:equation for sum field}
\end{align}
%%%%
This is where the utility of the Keldysh basis becomes evident. The difference field is an \textit{unstable} oscillator, and the sum field is a \textit{damped} oscillator sourced by the difference field. 
%\\\\
\subsubsection{Action on classical solutions}
%%%
Now I will evaluate the action on classical solutions. With $\dot q_+ \dot q_- = (q_+ \dot q_-)^\cdot - q_+ \ddot q_-$, $\dot q_-^2 = (q_-\dot q_-)^\cdot - q_-\ddot q_-$, and $q_-\dot q_+ = (q_- q_+)^\cdot - \dot q_- q_+$, I find:
\begin{align}
	&S_{\text{cl}}^{(0)} \equiv \left.\int_{t_0}^{t_f}\!\! dt\;\la_{\text{new}}\right|_{\text{eqs.~of motion}}\\
	& = \left.\half\left(q_+\dot q_- \!+\!i\e q_-\dot q_- \!-\!\e m q_+q_-\right)\right|_{t\,=\,t_0}^{t_f} \nonumber\\
	&\qquad + \left.\half\! \int_{t_0}^{t_f}\!\! dt\left[ -q_+\left(\ddot q_- \!+\! m^2 q_- \!-\! 2\e m \dot q_-\right) + i\e q_-(-\ddot q_- \!+\! m^2 q_-)\right]\right|_{\text{eqs.~of motion}} \\
	&= \left. \half\left[\left(q_+ \!+\! i\e q_-\right)\dot q_- - \e m\, q_+ q_-\right]\right|_{t\,=\,t_0}^{t_f} + \half\int_{t_0}^{t_f}\!\! dt\left[ -q_+ J_- + i\e q_- \left(2m^2 q_- - J_- + O(\e) \right)\right] \\
	&= \left. \half\left[\left(q_+ \!+\! i\e q_-\right)\dot q_- - \e m\, q_+ q_-\right]\right|_{t\,=\,t_0}^{t_f} + \half \int_{t_0}^{t_f}\!\! dt\left[-(q_+ + i\e q_-)J_- + 2i\e m^2 q_-^2 \right] + O(\e^2)\;.
\end{align}
Adding the source terms, I get:
\begin{align}
	S_{\text{cl}} &= S_{\text{cl}}^{(0)} + \int_{t_0}^{t_f}\!\! dt\;\half\left(J_- q_+ + J_+ q_-\right) \\
	&= \left. \half\left[\left(q_+ \!+\! i\e q_-\right)\dot q_- \!-\! \e m\, q_+ q_-\right]\right|_{t\,=\,t_0}^{t_f} + \half \int_{t_0}^{t_f}\!\! dt \left[(J_+ \!-\! i\e J_-) q_- + 2i\e m^2 q_-^2\right]\;. \label{eq:sum field is not required}
\end{align}
%%%%
\subsubsection{Sourced damped/unstable oscillator}\label{sec:sourced damped oscillator}
The equation of motion for the sourced damped/unstable oscillator is
\begin{equation}\label{eq:damped oscillator}
	\ddot q(t) + \g\dot q(t) + m^2 q(t) = J(t)\;.
\end{equation}
I will first regale you with dimensional analysis:\footnote{I will not use this directly, but I found it invaluable for catching mistakes. Unless you are much smarter than I am, you will make mistakes.}
\begin{align}
	&\bullet \;\;S=\int dt\;\half \dot q^2 \sim t\frac{q^2}{t^2} \sim 1 \implies q \sim \sqrt t \nonumber\\
	&\bullet \;\;m \sim \frac{d}{dt} \sim \frac{1}{t} \nonumber\\
	&\bullet\;\; \g \frac{d}{dt} \sim \g \frac{1}{t} \sim \frac{d^2}{dt^2} \sim \frac{1}{t^2} \implies \g \sim \frac{1}{t} \nonumber\\
	&\bullet\;\; J \sim \frac{d^2}{dt^2}q \sim \frac{t^{1/2}}{t^2} \sim t^{-3/2}\;.
\end{align}
Then I will remind you of Sec.~\ref{sec:Green's function}, in which I explained my conception of the Green's function:
\begin{equation}\label{eq:damped green's function}
	\ddot \Gs(t) + \g\dot \Gs(t) + m^2 \Gs(t) = 0\;,\;\; \Gs(0) = 0\;,\;\;\dot \Gs(0) = 1\;.
\end{equation}
My reason for invoking calligraphy will become evident in a New York minute. Fourier-transforming the defining homogeneous equation in Eq.~(\ref{eq:damped green's function}) gives
\begin{equation}
	\left(-\w^2-i\g\w + m^2\right)\tilde \Gs(\w) = 0\;,
\end{equation}
which implies that $\tilde \Gs(\w) = 0$ or 
\begin{equation}
	\w = -i\tfrac{\g}{2}\pm\sqrt{m^2-\left(\tfrac{\g}{2}\right)^2}\;.
\end{equation}
In terms of cosines and sines, the solution is therefore
\begin{equation}
	\Gs(t) = e^{-\half \g t}\left[ A\cos\left(\sqrt{m^2-\fourth\g^2} t\right) + B\sin\left(\sqrt{m^2-\fourth\g^2}t\right)\right]\;.
\end{equation}
Imposing $G(0) = 0$ and $\dot G(0) = 1$ produces the standard result:
\begin{equation}
	\Gs(t) = \frac{e^{-\half\g t}}{\sqrt{m^2-\fourth\g^2}}\; \sin\!\left(\sqrt{m^2-\fourth\g^2} t\right)\;.
\end{equation}
The approximation I will use is 
\begin{equation}\label{eq:green's function for iepsilon}
	\Gs(t) \approx e^{-\half \g t}\, G(t)\;,\;\;G(t) = \frac{1}{m}\sin(m t)\;.
\end{equation}
That relation is why I chose the notation $\Gs(t)$ for this Green's function. After a colosseum of trial and error, I have determined that it is best to express everything in terms of the very same $G(t)$ from Eq.~(\ref{eq:green's function}). 
\\\\
Eq.~(\ref{eq:green's function for iepsilon}) establishes the approximation scheme for the $i\e$ prescription and the sign conventions for stability of the oscillator. Use Eqs.~(\ref{eq:damped oscillator}) and~(\ref{eq:green's function for iepsilon}) to interpret Eqs.~(\ref{eq:equation for difference field}) and~(\ref{eq:equation for sum field}).
\subsubsection{The difference field}
Given the Green's function in Eq.~(\ref{eq:green's function for iepsilon}) and the steps in Secs.~\ref{sec:homogeneous solution}-\ref{sec:complete solution}, I trust you to solve Eq.~(\ref{eq:equation for difference field}) for the difference field:
\begin{align}
	q_-(t) &= \frac{1}{G(T)} \Bigg[ q_{-0} e^{\,\e m(t-t_0)} G(t_f\!-\!t) \Bigg. \nonumber\\
	&\qquad \qquad \Bigg.+ \int_{t_0}^{t_f}\!\! dt'\,e^{\,\e m(t-t')} \Big(G(T)G(t\!-\!t')\Ta(t\!-\!t') - G(t\!-\!t_0) G(t_f\!-\!t') \Big) J_- (t') \Bigg],
\end{align}
with $G(t) \equiv \frac{1}{m}\sin(mt)$. The time-derivative is
\begin{align}
	\dot q_-(t) &= \frac{1}{G(T)}\left\{ q_{-0}\, e^{\,\e m(t-t_0)} \left[ -\dot G(t_f\!-\!t) + \e m G(t_f\!-\!t) \right]\right. \nonumber\\
	&\left. +\int_{t_0}^{t_f}\!\! dt'\,e^{\,\e m(t-t')} \left[ G(T) \dot G(t\!-\!t') \Ta(t\!-\!t')- \dot G(t\!-\!t_0) G(t_f\!-\!t') \right. \right. \nonumber\\
	&\left. \left. \phantom{\frac{a}{b}}+ \e m \left(G(T)G(t\!-\!t')\Ta(t\!-\!t') - G(t\!-\!t_0) G(t_f\!-\!t') \right)\right] J_-(t')\right\}\;. \label{eq:q_-(t)}
\end{align}
The values of the derivative at $t_0$ and $t_f$ are:
\begin{align}
	&\dot q_-(t_0) = \frac{-1}{G(T)} \left\{ q_{-0} \left[\dot G(T)-\e m G(T)\right] + \int_{t_0}^{t_f}\!\! dt\,e^{-\e m(t-t_0)} G(t_f\!-\!t) J_-(t)\right\}\;, \nonumber \\
	&\dot q_-(t_f) = \frac{1}{G(T)} \left\{ -q_{-0} \,e^{\,\e m T} + \int_{t_0}^{t_f}\!\! dt\, e^{\,\e m(t_f-t)} \left[G(T)\dot G(t_f-t) - \dot G(T) G(t_f-t)\right]J_-(t) \right\} \nonumber\\
	&\qquad = \frac{e^{\,\e m T}}{G(T)} \left\{ -q_{-0} + \int_{t_0}^{t_f}\!\! dt\, e^{-\e m(t-t_0)} G(t-t_0) J_-(t) \right\}\;. \label{eq:initial and final derivatives of difference field}
\end{align}
A blessing from Eq.~(\ref{eq:sum field is not required}) is that the explicit solution for the sum field is not required. 
\subsubsection{Simplify the action: Boundary terms}
Given Eq.~(\ref{eq:initial and final derivatives of difference field}), I can start to simplify Eq.~(\ref{eq:sum field is not required}). The boundary terms are:
\begin{align}
	&\left.\Big[(q_+ \!+\! i\e q_-)\, \dot q_- - \e m q_+ q_- \Big]\right|_{t\,=\,t_0}^{t_f} = 2q_f \dot q_-(t_f) - \left[(q_{+0}+i\e q_{-0}) \dot q_-(t_0) - \e m q_{+0}q_{-0}\right] \nonumber\\
	%&\nonumber\\
	&= \frac{2q_f e^{\,\e m T}}{G(T)}\left\{ -q_{-0} + \int_{t_0}^{t_f}\!\! dt\, e^{-\e m(t-t_0)} G(t-t_0) J_-(t) \right\} + \e m q_{+0} q_{-0} \nonumber\\
	&+\frac{q_{+0}+i\e q_{-0}}{G(T)}\left\{ q_{-0} \left[\dot G(T)-\e m G(T)\right] + \int_{t_0}^{t_f}\!\! dt\,e^{-\e m(t-t_0)} G(t_f\!-\!t) J_-(t)\right\} \nonumber\\
	&\nonumber\\
	&= \frac{2q_f e^{\,\e m T}}{G(T)}\left\{ -q_{-0} + \int_{t_0}^{t_f}\!\! dt\, e^{-\e m(t-t_0)} G(t-t_0) J_-(t) \right\} \nonumber\\
	&+ \frac{q_{+0}+i\e q_{-0}}{G(T)} \left\{q_{-0} \dot G(T) + \int_{t_0}^{t_f}\!\! dt\,e^{-\e m(t-t_0)} G(t_f\!-\!t) J_-(t)\right\}.
\end{align}
%
%\pagebreak
%
\subsubsection{Simplify the action: Integral over $q_f$}
There are no terms quadratic in $q_f$. If I do the integral over $q_f$, the terms linear in $q_f$ will result in a delta function that sets 
\begin{equation}\label{eq:q_-0}
	q_{-0} = \int_{t_0}^{t_f}\!\! dt\;e^{-\e m (t-t_0)}\, G(t\!-\!t_0)\, J_-(t)\;.
\end{equation}
So instead of trying to insert $q_-(t)$ into $S_{\text{cl}}$, I could first do the integral over $q_f$ and insert Eq.~(\ref{eq:q_-0}) into $q_-(t)$. That integral is
\begin{align}
	&\INT dq_f\;e^{\,i q_f\;\frac{e^{\e m T}}{G(T)}\left[ -q_{-0} + \int_{t_0}^{t_f}\!\! dt\, e^{-\e m(t-t_0)} G(t-t_0) J_-(t)\right]} = \nonumber\\
	&\qquad \qquad 2\pi G(T) e^{-\e m T}\; \del\!\left(q_{-0}-\int_{t_0}^{t_f}\!\! dt\; e^{-\e m(t-t_0)} G(t\!-\!t_0) J_-(t)\right)\;.
\end{align}
Instead of calculating the overall factor from first principles, this time I will just fix it by requiring $Z = 1$ when $J' = J$ [recall Eq.~(\ref{eq:Z(J,J)=1})]. In terms of $J_+$ and $J_-$, that would mean $Z = 1$ when $J_- = 0$. From now on I will write $Z_K(J_+,J_-) \equiv Z(\frac{J_+ + J_-}{2},\frac{J_+ - J_-}{2})$ and drop the $K$, as I already did after Eq.~(\ref{eq:combined lagrangian in keldysh basis}):
\begin{align}
	&Z(J_+,J_-) = N \INT dq_{+0} \INT dq_{-0} \;\rho\left(\tfrac{q_{+0}+q_{-0}}{2},\tfrac{q_{+0}-q_{-0}}{2}\right) \INT dq_f \;e^{\,i S_{\text{cl}}(q_{+0},q_{-0},q_f|J_+,J_-)} \nonumber\\
	&\nonumber\\ 
	&= 2\pi G(T) e^{-\e m T} N \;\times \nonumber\\
	&\qquad \left.\INT dq_{+0}\; \rho\left(\tfrac{q_{+0}+q_{-0}}{2},\tfrac{q_{+0}-q_{-0}}{2}\right)\;e^{\,i S_{\text{cl}}(q_{+0}, q_{-0}, 0|J_+,J_-)} \right|_{q_{-0}\,=\,\int_{t_0}^{t_f} dt\,e^{-\e m(t-t_0)} G(t-t_0) J_-(t)}\;.
\end{align}
\subsubsection{Simplify the action: Boundary terms, part 2}
With Eq.~(\ref{eq:q_-0}), I find
\begin{align}
	&\dot G(T) q_{-0} + \int_{t_0}^{t_f}\!\! dt\;e^{-\e m (t-t_0)} G(t_f\!-\!t)J_-(t) = \nonumber\\
	&\qquad \int_{t_0}^{t_f}\!\! dt\;e^{-\e m (t-t_0)}\left[ \dot G(T) G(t\!-\!t_0) + G(t_f\!-\!t)\right] J_-(t)\;,
\end{align}
an expression that looks reassuringly like something I found for $\e = 0$. At this point it is useful to organize terms in powers of $\e$ (without expanding any of the exponentials, as usual---powers of $\e $, not powers of $\e m t$). The $O(\e^0)$ part of the boundary terms is
\begin{align}
	&\frac{q_{+0}}{2G(T)}\! \left\{\dot G(T) q_{-0} + \!\int_{t_0}^{t_f}\!\! dt\;e^{-\e m (t-t_0)} G(t_f\!-\!t)J_-(t)\right\}\nonumber\\
	&\qquad = \frac{q_{+0}}{2G(T)}\int_{t_0}^{t_f}\!\! dt\;e^{-\e m (t-t_0)}\left[ \dot G(T) G(t\!-\!t_0) + G(t_f\!-\!t)\right] J_-(t)\;,
\end{align}
and the $O(\e)$ part is
\begin{align}
	&\frac{i\e q_{-0}}{2G(T)}\left\{\dot G(T) q_{-0} + \!\int_{t_0}^{t_f}\!\! dt\;e^{-\e m (t-t_0)} G(t_f\!-\!t)J_-(t)\right\} \nonumber\\
	&= \frac{i\e}{2G(T)} \left\{\int_{t_0}^{t_f}\!\! dt'\;e^{-\e m(t'-t_0)} G(t'\!-\!t_0) J_-(t') \right\}\int_{t_0}^{t_f}\!\! dt\;e^{-\e m (t-t_0)}\left[ \dot G(T) G(t\!-\!t_0) + G(t_f\!-\!t)\right] J_-(t) \nonumber\\
	&= \frac{i\e}{2G(T)}\doubleint e^{-\e m(t+t'-2t_0)} \left[\dot G(T) G(t\!-\!t_0) + G(t_f\!-\!t)\right]G(t'\!-\!t_0)\; J_-(t) J_-(t')\;.
\end{align}
\subsubsection{Simplify the action: Simplify $q_-(t)$}
That expression for $q_{-0}$ will greatly simplify $q_-(t)$:
\begin{align}
	&q_-(t) = \frac{1}{G(T)}\left\{ \left[ \int_{t_0}^{t_f}\!\! dt'\;e^{-\e m (t'-t_0)} G(t'\!-\!t_0) J_-(t')\right] e^{\e m (t-t_0)} G(t_f\!-\!t) \right. \nonumber\\
	&\qquad \qquad\qquad \qquad \left.+ \int_{t_0}^{t_f}\!\! dt'\;e^{\,\e m (t-t')}\left[G(T) G(t\!-\!t') \Ta(t\!-\!t') - G(t\!-\!t_0) G(t_f\!-\!t') \right] J_-(t')\right\} \nonumber\\
	&\nonumber\\
	&= \frac{1}{G(T)}\int_{t_0}^{t_f}\!\! dt'\;e^{\, \e m(t-t')}\left[G(T) G(t\!-\!t') \Ta(t\!-\!t') + \underbrace{G(t_f\!-\!t)G(t'\!-\!t_0) - G(t_f\!-\!t') G(t\!-\!t_0)}_{\,=\,-G(T)\,G(t-t')}\right] J_-(t') \nonumber\\
	&\nonumber\\
	&= -\int_{t_0}^{t_f}\!\! dt'\;e^{\,\e m(t-t')} G(t\!-\!t') \Ta(t'\!-\!t) J_-(t')\;. \label{eq:better q_-(t)}
\end{align}
Compare that to Eq.~(\ref{eq:q_-(t)}).
\subsubsection{Simplify the action: Remaining terms in action}
With Eq.~(\ref{eq:better q_-(t)}), I can simplify the nonboundary terms in the action:
\begin{align}
	S_{\text{cl}}^{\,\text{nonboundary}} &= \half \int_{t_0}^{t_f}\!\! dt\left[(J_+(t) - i\e J_-(t)) q_-(t) + 2i\e m^2 q_-(t)^2 \right]\;.
\end{align}
The $q_-^2$ term will require special attention, since there will be an intermediate integral:
\begin{align}
	&\int_{t_0}^{t_f}\!\!dt'' q_-(t'')^2\nonumber\\
	&\;\; = \int_{t_0}^{t_f}\!\! dt''\left[\int_{t_0}^{t_f}\!\! dt\;e^{\,\e m(t''-t)} G(t''\!-\!t)\Ta(t\!-\!t'')J_-(t) \right]\left[\int_{t_0}^{t_f}\!\! dt'\;e^{\,\e m(t''-t')} G(t''\!-\!t') \Ta(t'\!-\!t'') J_-(t')\right] \nonumber\\
	&\;\;= \doubleint e^{-\e m(t+t')} J_-(t) J_-(t') \int_{t_0}^{t_f}\!\! dt''\;e^{\,2\e m t''} G(t''\!-\!t)G(t''\!-\!t')\;\Ta(t\!-\!t'')\Ta(t'\!-\!t'')\;.
\end{align}
That product of step functions says that $t''$ must be less than $t$ \textit{and} less than $t'$, so the integral over $t''$ has two cases: $t > t'$ and $t < t'$. Considering each separately, I rewrite the integral as
\begin{align}
	I(t,t') &\equiv \int_{t_0}^{t_f}\!\! dt''\;e^{\,2\e m t''} G(t''\!-\!t)G(t''\!-\!t')\;\Ta(t''\!-\!t)\Ta(t''\!-\!t') \nonumber\\
	&= \Ta(t\!-\!t') \int_{t_0}^{t'}\!\! dt''\;e^{\,2\e m t''} G(t''\!-\!t) G(t''\!-\! t') + \Ta(t'\!-\!t) \int_{t_0}^{t}\!\! dt''\;e^{\,2\e m t''} G(t''\!-\!t) G(t''\!-\! t')\;.
\end{align}
That integral will have an overall factor of $\e^{-1}$, making its contribution to the action include $O(\e^0)$ and $O(\e)$ terms. The coefficients of $\Ta(t$$-$$t')$ and $\Ta(t'$$-$$t)$ are
\begin{align}
	&\int_{t_0}^{t'}\!\! dt''\;e^{\,2\e m t''} G(t''\!-\!t) G(t''\!-\! t') \nonumber\\
	&\qquad = \frac{1}{4m^3 \e}\left\{ \left(e^{2\e m t'}-e^{2\e m t_0}\right) \dot G(t\!-\!t')+\e m\left[ e^{2\e m t'} G(t\!-\!t')-e^{2\e m t_0} G(t\!+\!t'\!-\!2t_0)\right]\right\} + O(\e)\;, \\
	&\nonumber\\
	&\int_{t_0}^{t}\!\! dt''\;e^{\,2\e m t''} G(t''\!-\!t) G(t''\!-\! t') \nonumber\\
	&\qquad = \frac{1}{4m^3 \e}\left\{\left( e^{2\e m t}-e^{2\e m t_0} \right) \dot G(t\!-\!t') - \e m \left[e^{2\e m t} G(t\!-\!t') + e^{2\e m t_0} G(t\!+\!t'\!-\!2t_0)\right]\right\} + O(\e)\;.
\end{align}
Therefore,
\begin{align}
	&e^{-\e m(t+t')} I(t,t') \nonumber\\
	&= \frac{1}{4m^3\e}\left\{ \Ta(t\!-\!t')\!\left[ \left(e^{-\e m(t-t')} \!-\! e^{-\e m(t+t'-2t_0)}\right) \dot G(t\!-\!t') \right. \right. \nonumber\\
	&\qquad\qquad\qquad\qquad \left.+\e m\left(e^{-\e m(t-t')} G(t\!-\!t') \!-\! e^{-\e m(t+t'-2t_0)} G(t\!+\!t'\!-\!2t_0) \right)\right] \nonumber\\
	&\;\;\;\;\;\qquad +\Ta(t'\!-\!t)\!\left[\left(e^{\,\e m(t-t')} - e^{-\e m(t+t'-2t_0)}\right) \dot G(t\!-\!t')\right. \nonumber\\
	&\qquad\qquad\qquad\qquad \left.\left. - \e m\left(e^{\,\e m(t-t')} G(t\!-\!t') + e^{-\e m(t+t'-2t_0)}G(t\!+\!t'\!-\!2t_0) \right)\right] \right\} + O(\e).
\end{align}
\subsubsection{Simplify the action: Organization}
The way to organize all of that is to remember that there is still an integral over $q_{+0}$ to be done against the density matrix. So I will organize $S_{\text{cl}}$ into three terms: A part that depends on $q_{+0}$, a $q_{+0}$-independent part of $O(\e^0)$, and a $q_{+0}$-independent part of $O(\e)$. 
\begin{equation}
	S_{\text{cl}} = \frac{q_{+0}}{2G(T)}\int_{t_0}^{t_f}\!\! dt\;e^{-\e m(t-t_0)}\left[\dot G(T) G(t\!-\!t_0) + G(t_f\!-\!t)\right]J_-(t) + A(J_+,J_-) + i\e B(J_-)\;,
\end{equation}
with
\begin{align}
	&A(J_+,J_-) = -\half \doubleint e^{\e m(t-t')} G(t\!-\!t')\Ta(t'\!-\!t) J_+(t)J_-(t') \nonumber\\
	&+\frac{i}{4m}\int_{t_0}^{t_f}\!\!\!\! dt \! \int_{t_0}^{t_f}\!\!\!\! dt'\; \Big[\Ta(t\!-\!t')\left(e^{-\e m(t-t')} \!-\! e^{-\e m(t+t'-2t_0)}\right) \Big. \nonumber\\
	&\qquad\qquad \qquad\qquad\Big.+ \Ta(t'\!-\!t) \left( e^{\e m(t-t')} \!-\! e^{-\e m(t+t'-2t_0)} \right) \Big] \dot G(t\!-\!t') J_-(t) J_-(t') \nonumber\\
	&\nonumber\\
	&= -\half \doubleint e^{\e m(t-t')} G(t\!-\!t')\Ta(t'\!-\!t) J_+(t)J_-(t') \nonumber\\
	&+\frac{i}{4m}\doubleint \left[\Ta(t\!-\!t')\, e^{-\e m(t-t')} + \Ta(t'\!-\!t)\,e^{\e m(t-t')} - e^{-\e m(t+t'-2t_0)}\right] \dot G(t\!-\!t') J_-(t)J_-(t')\;,
\end{align}
and (do not forget the part from the boundary terms):
\begin{align}
	&B(J_-) = \frac{1}{2G(T)}\doubleint e^{-\e m(t+t'-2t_0)} \left[\dot G(T) G(t\!-\!t_0) + G(t_f\!-\!t)\right]G(t'\!-\!t_0)\; J_-(t) J_-(t') \nonumber\\
	&+\half \doubleint e^{\e m(t-t')} G(t\!-\!t')\Ta(t'\!-\!t) J_-(t) J_-(t') \nonumber\\
	&+\fourth \doubleint \left[\Ta(t\!-\!t') \left(e^{-\e m(t-t')}G(t\!-\!t') - e^{-\e m(t+t'-2t_0)}G(t\!+\!t'\!-\!2t_0) \right) \right. \nonumber\\
	&\qquad \qquad \qquad \qquad \left. - \Ta(t'\!-\!t)\left(e^{\e m(t-t')}G(t\!-\!t') + e^{-\e m(t+t'-2t_0)} G(t\!+\!t'\!-\!2t_0)\right)\right] J_-(t)J_-(t') \nonumber\\
	&\nonumber\\
	&=\frac{1}{2G(T)}\doubleint e^{-\e m(t+t'-2t_0)} \left[\dot G(T) G(t\!-\!t_0) + G(t_f\!-\!t)\right]G(t'\!-\!t_0)\; J_-(t) J_-(t') \nonumber\\
	&+\fourth \doubleint \Big[\left(\Ta(t\!-\!t')\,e^{-\e m(t-t')} \!+\! \Ta(t'\!-\!t)\,e^{\,\e m(t-t')}\right) G(t\!-\!t') \Big. \nonumber\\
	&\qquad\qquad\qquad\qquad \qquad\qquad\qquad\Big.- e^{-\e m(t+t'-2t_0)} G(t\!+\!t'\!-\!2t_0)\;\Big] J_-(t)J_-(t')\\
	&\nonumber\\
	&= \frac{1}{2G(T)}\doubleint e^{-\e(t+t'-2t_0)}\Big[\;\dot G(T) G(t\!-\!t_0)G(t'\!-\!t_0)+G(t_f\!-\!t)G(t'\!-\!t_0) \Big. \nonumber\\
	&\qquad\qquad\qquad\qquad \qquad\qquad\qquad\qquad \qquad\qquad \qquad \qquad \Big.-\half G(t\!+\!t'\!-\!2t_0)\;\Big] J_-(t)J_-(t') \nonumber\\
	&+\fourth \doubleint \left(\Ta(t\!-\!t')\,e^{-\e m(t-t')} \!+\! \Ta(t'\!-\!t)\,e^{\,\e m(t-t')}\right) G(t\!-\!t') J_-(t)J_-(t')\;.
\end{align}
Since $J_-(t)J_-(t')$ is symmetric in $t$ and $t'$, I should write $G(t_f$$-$$t)G(t'$$-$$t_0)$ $\to$ $\half[G(t_f$$-$$t)G(t'$$-$$t_0)$ $+$ $G(t_f$$-$$t')G(t$$-$$t_0)]$ and behold that
\begin{equation}
	2\dot G(T) G(t-t_0)G(t'-t_0) + G(t_f-t)G(t'-t_0)+G(t_f-t')G(t-t_0)-G(t+t'-2t_0) = 0\;.
\end{equation}
So the first line is zero. Meanwhile, the expression in the second line,
\begin{align}
	M(t,t') &\equiv \left(\Ta(t\!-\!t')\,e^{-\e m(t-t')} \!+\! \Ta(t'\!-\!t)\,e^{\,\e m(t-t')}\right) G(t\!-\!t')\;,
\end{align}
is antisymmetric:
\begin{align}
	M(t',t) &= \left(\Ta(t'\!-\!t) e^{-\e m(t'-t)} + \Ta(t\!-\!t')e^{\e m(t'-t)} \right)G(t'\!-\!t) \nonumber\\
	&= -\left(\Ta(t\!-\!t')e^{-\e m(t-t')} + \Ta(t'\!-\!t) e^{\e m(t-t')}\right) G(t\!-\!t') \nonumber\\
	&= -M(t,t') .
\end{align}
So $\doubleint M(t,t') J_-(t) J_-(t') = 0$, and I learn that $B(J_-) = 0$: The $O(\e)$ part is zero.
\\\\
At this stage the generating function is
\begin{align}
	&Z(J_+,J_-) = 2\pi G(T) e^{-\e m T}N e^{\,iA(J_+,J_-)}\; \times \nonumber\\
	&\INT dq\;\rho\left(\tfrac{q + \int_{t_0}^{t_f}\! dt\, e^{-\e m(t-t_0)} G(t-t_0) J_-(t)}{2},\tfrac{q-\int_{t_0}^{t_f}\! dt\, e^{-\e m(t-t_0)} G(t-t_0) J_-(t)}{2}\right)\;\times \nonumber\\
	&\qquad \;e^{\,\frac{i}{2G(T)}q \int_{t_0}^{t_f} \! dt\, e^{-\e m(t-t_0)}\left[\dot G(T) G(t-t_0) + G(t_f-t)\right] J_-(t)}. \label{eq:unnormalized generating function for iepsilon}
\end{align}
I suspect there is physics to be gleaned from that general form. 
\subsection{Specify the density matrix and complete the calculation}
I will complete this calculation only for the $\e = 0$ oscillator ground state:
\begin{equation}
	\rho(q,q') = \sqrt{\frac{m}{\pi}}\;e^{-\half m(q^2+q'^2)}\;.
\end{equation}
Now to fix the normalization factor in Eq.~(\ref{eq:unnormalized generating function for iepsilon}). Let $k = \int_{t_0}^{t_f} dt e^{-\e m(t-t_0)} G(t-t_0) J_-(t)$ and $j = \int_{t_0}^{t_f}dt e^{-\e m(t-t_0)} [\dot G(T) G(t-t_0) + G(t_f-t)]J_-(t)$.
%$(q+k)^2+(q-k)^2 = q^2 + k^2 + 2kq + q^2+k^2 - 2kq = 2(q^2+k^2)$. \\
%$\left(\frac{q+k}{2}\right)^2 + \left(\frac{q-k}{2}\right)^2 = \half (q^2+k^2)$ \\\\
%\\
The integral over $q$ is:
\begin{align}
	\INT dq\;\rho\left(\tfrac{q+k}{2},\tfrac{q-k}{2}\right)\;e^{\,\frac{ij}{2G(T)}q} &=  \sqrt{\frac{m}{\pi}} e^{-m k^2} \INT dq\;e^{-\frac{m}{4}\left(q^2 - \frac{2ij}{m G(T)}q\right)} \nonumber\\
	&= 2 e^{-\frac{m}{4} k^2} e^{-\frac{m}{4}\frac{j^2}{m^2 G(T)^2}}\;.
\end{align}
So the normalization factor should be
\begin{equation}
	N = \frac{1}{4\pi G(T) e^{-\e m T}}\;,
\end{equation}
leaving an influence phase
\begin{equation}
	\Phi(J_+,J_-) = A(J_+, J_-) + \frac{im k^2}{4} + \frac{ij^2}{4m G(T)^2}\;.
\end{equation}
\subsubsection{Simplify the influence phase}
\begin{align}
	&\Phi(J_+,J_-) = -\half \doubleint e^{\e m(t-t')} G(t\!-\!t')\Ta(t'\!-\!t) J_+(t) J_-(t') \nonumber\\
	&+\frac{i}{4m}\doubleint\left[\Ta(t\!-\!t')\,e^{-\e m(t-t')} + \Ta(t'\!-\!t)\,e^{\,\e m(t-t')} - e^{-\e m(t+t'-2t_0)}\right] \dot G(t\!-\!t') J_-(t) J_-(t') \nonumber\\
	&+\frac{i m}{4} \doubleint e^{-\e m (t+t'-2t_0)} G(t\!-\!t_0) G(t'\!-\!t_0) J_-(t) J_-(t') \nonumber\\
	&+\frac{i}{4m G(T)^2} \doubleint e^{-\e m (t+t'-2t_0)} \left[\dot G(T) G(t\!-\!t_0)+G(t_f\!-\!t)\right] \times \nonumber\\
	&\qquad\qquad\qquad\qquad\qquad\qquad\qquad\qquad\qquad \left[\dot G(T) G(t'\!-\!t_0)+G(t_f\!-\!t')\right] J_-(t) J_-(t')\;.
\end{align}
Everything in this problem is translationally invariant, so everything that depends on $t_0$ and $t_f$ had better cancel. Guess what:
\begin{align}
	\left[\dot G(T) G(t\!-\!t_0)+G(t_f\!-\!t)\right] &\left[\dot G(T) G(t'\!-\!t_0)+G(t_f\!-\!t')\right] \nonumber\\
	&\qquad + m^2 G(T)^2 G(t\!-\!t_0) G(t'\!-\!t_0) = G(T)^2 \dot G(t\!-\!t')\;.
\end{align}
So all of the terms proportional to $e^{-\e m(t+t'-2t_0)}$ will, in fact, cancel. Writing $\Ta(t$$-$$t')e^{-\e m(t-t')}$ $+$ $\Ta(t'$$-$$t)e^{\e m(t-t')} = e^{-\e m|t-t'|}$ and expressing the $J_+(t)J_-(t')$ term as the sum of $J_+(t)J_-(t')$ and $J_+(t') J_-(t)$ terms, I arrive at the result:
\begin{align}
	&\Phi(J_+,J_-) = \frac{i}{4m}\doubleint \left[ e^{-\e m|t-t'|} \dot G(t\!-\!t')\;J_-(t) J_-(t') \right. \nonumber\\
	& \left.+im\, e^{\e m(t-t')} G(t\!-\!t')\,\Ta(t'\!-\!t)\, J_+(t) J_-(t') + im\, e^{\e m(t'-t)} G(t'\!-\!t)\,\Ta(t\!-\!t')\, J_-(t) J_+(t') \right]\;. \label{eq:ie result}
\end{align}
No $J_+J_+$ terms, sublime. 
\\\\
To garnish the omelette, I will return to the original basis. With $J_\pm = J\pm J'$, the products of sources become:
\begin{align}
	& J_-(t) J_-(t') = [J(t)\!-\!J'(t)][J(t')\!-\!J'(t')] = J(t)J(t') + J'(t)J'(t') - J(t)J'(t')-J'(t)J(t')\;, \nonumber\\
	& J_+(t) J_-(t') = [J(t)\!+\!J'(t)][J(t')\!-\!J'(t')] = J(t)J(t')-J'(t)J'(t')-J(t)J'(t')+J'(t)J(t')\;, \nonumber\\
	& J_-(t) J_+(t') = [J(t)\!-\!J'(t)][J(t')\!+\!J'(t')] = J(t)J(t')-J'(t)J'(t') + J(t)J'(t')-J'(t)J(t')\;.
\end{align}
The $J_+J_-$ terms become:
\begin{align}
	&e^{\e m(t-t')}G(t\!-\!t')\Ta(t'\!-\!t)J_+(t)J_-(t') + e^{\e m(t'-t)}G(t'\!-\!t)\Ta(t\!-\!t') J_-(t) J_+(t') \nonumber\\
	&\nonumber\\
	&= \left[ e^{\e m(t-t')}G(t\!-\!t')\Ta(t'\!-\!t) + e^{\e m(t'-t)}G(t'\!-\!t)\Ta(t\!-\!t')\right][J(t)J(t')-J'(t)J'(t')] \nonumber\\
	&+\left[-e^{\e m(t-t')}G(t\!-\!t')\Ta(t'\!-\!t) + e^{\e m(t'-t)}G(t'\!-\!t)\Ta(t\!-\!t') \right] [J(t)J'(t')-J'(t)J(t')]\;. 
\end{align}
Note that the coefficient of $JJ-J'J'$ is symmetric, not antisymmetric:
\begin{align}
	C(t,t') &\equiv e^{\e m(t-t')}G(t\!-\!t')\Ta(t'\!-\!t) + e^{\e m(t'-t)}G(t'\!-\!t)\Ta(t\!-\!t')
\end{align}
\begin{align}
	C(t',t) &= e^{\e m(t'-t)} G(t'\!-\!t)\Ta(t\!-\!t') + e^{\e m(t-t')} G(t\!-\!t') \Ta(t'\!-\!t) \nonumber\\
	&= e^{\e m(t-t')}G(t\!-\!t')\Ta(t'\!-\!t) + e^{\e m(t'-t)} G(t'\!-\!t)\Ta(t\!-\!t') \nonumber\\
	&= C(t,t')\;.
\end{align}
Therefore, I find:
\begin{align}
	&e^{-\e m|t-t'|}\dot G(t\!-\!t') [J(t)J(t')+J'(t)J'(t')] + im C(t,t')[J(t)J(t')-J'(t)J'(t')] \nonumber\\
	&\nonumber\\
	&= \left\{ e^{-\e m(t-t')} \left[\dot G(t\!-\!t')\! -\! im\, G(t\!-\!t')\right] \Ta(t\!-\!t') + e^{\e m(t-t')}\left[ \dot G(t\!-\!t') \!+\! im\,G(t\!-\!t')\right]\Ta(t'\!-\!t)\right\}J(t)J(t') \nonumber\\
	&+\left\{ e^{-\e m(t-t')} \left[\dot G(t\!-\!t')\! +\! im\, G(t\!-\!t')\right] \Ta(t\!-\!t') + e^{\e m(t-t')}\left[ \dot G(t\!-\!t') \!-\! im\,G(t\!-\!t')\right]\Ta(t'\!-\!t)\right\} J'(t)J'(t') \nonumber\\
	&\nonumber\\
	&= \left[ e^{-\e m(t-t')} e^{-im(t-t')} \Ta(t\!-\!t') + e^{\,\e m(t-t')} e^{\,i m(t-t')} \Ta(t'\!-\!t)\right] J(t)J(t') \nonumber\\
	&+ \left[e^{-\e m(t-t')} e^{\,im(t-t')} \Ta(t\!-\!t') + e^{\,\e m(t-t')} e^{-im(t-t')} \Ta(t'\!-\!t)\right] J'(t)J'(t') \nonumber\\
	&\nonumber\\
	&= e^{-i(1-i\e)m |t-t'|} J(t)J(t') + e^{\,i(1+i\e)m|t-t'|}J'(t)J'(t')\;.\label{eq:improved feynman and dyson}
\end{align}
There you have the $i\e$ prescription as ordinarily understood: The frequency $m$ gets a small, negative, imaginary part in the forward direction, and a small, positive, imaginary part in the backward direction. Compare to Eqs.~(\ref{eq:G_F(t) for ground state}) and~(\ref{eq:G_D(t) for ground state}). 
\\\\
The new terms derived from Lindblad evolution are:
\begin{align}
	&e^{-\e m|t-t'|} \dot G(t\!-\!t') [-J(t)J'(t')-J'(t)J(t')] + im e^{\e m(t-t')} G(t\!-\!t')\Ta(t'\!-\!t) [-J(t)J'(t')+J'(t)J(t')] \nonumber\\
	&+im e^{\e m(t'-t)} G(t'\!-\!t)\Ta(t\!-\!t') [J(t)J'(t')-J'(t)J(t')] = \nonumber\\
	&\nonumber\\
	&-[e^{-\e m|t-t'|} \dot G(t\!-\!t') + im e^{\e m(t-t')} G(t\!-\!t') \Ta(t'\!-\!t) -im e^{\e m(t'-t)}G(t'\!-\!t)\Ta(t\!-\!t')]J(t)J'(t') \nonumber\\
	&-[e^{-\e m|t-t'|} \dot G(t\!-\!t') - im e^{\e m(t-t')} G(t\!-\!t')\Ta(t'\!-\!t) + im e^{\e m(t'-t)} G(t'\!-\!t)\Ta(t\!-\!t')] J'(t)J(t') = \nonumber\\
	&\nonumber\\
	&-\left\{ e^{-\e m(t-t')}\left[ \dot G(t\!-\!t')+im G(t\!-\!t')\right]\Ta(t\!-\!t') + e^{-\e m(t'-t)} \left[ \dot G(t\!-\!t') + im G(t\!-\!t')\right]\Ta(t'\!-\!t)\right\}J(t)J'(t') \nonumber\\
	&-\left\{ e^{-\e m(t-t')}\left[ \dot G(t\!-\!t') - im G(t\!-\!t')\right]\Ta(t\!-\!t') + e^{-\e m(t'-t)}\left[ \dot G(t\!-\!t') - im G(t\!-\!t') \right]\Ta(t'\!-\!t)\right\}J'(t)J(t')
	\nonumber\\
	&\nonumber\\
	&= -\left\{ e^{-\e m(t-t')} e^{\,im(t-t')}\Ta(t\!-\!t') + e^{\,\e m (t-t')} e^{\,im(t-t')}\Ta(t'\!-\!t)\right\} J(t)J'(t') \nonumber\\
	&-\left\{ e^{-\e m(t-t')} e^{-im(t-t')} \Ta(t\!-\!t') + e^{\,\e m(t-t')} e^{-im(t-t')} \Ta(t'\!-\!t)\right\} J'(t)J(t') \nonumber\\
	&\nonumber\\
	&= -\left[ e^{\,i(1+i\e)m(t-t')}\Ta(t\!-\!t') + e^{\,i(1-i\e)m(t-t')}\Ta(t'\!-\!t)\right] J(t)J'(t')\nonumber\\
	&-\left[ e^{-i(1-i\e)m(t-t')}\Ta(t\!-\!t') + e^{-i(1+i\e)m(t-t')}\Ta(t'\!-\!t) \right] J'(t)J(t') \nonumber\\
	&\nonumber\\
	&= -e^{\,i[1+\sign(t-t')i\e]m(t-t')} J(t)J'(t')-e^{-i[1-\sign(t-t')i\e]m(t-t')}J'(t)J(t')\;. \label{eq:improved wightman}
\end{align}
Compare to Eqs.~(\ref{eq:G_<(t) for ground state}) and~(\ref{eq:G_>(t) for ground state}). 
\\\\
Comparing Eqs.~(\ref{eq:improved feynman and dyson}) and~(\ref{eq:improved wightman}) with the general form of the influence phase from Eq.~(\ref{eq:ground state influence phase}), I infer the Feynman, Dyson, and Wightman functions produced by the Lagrangian from Eq.~(\ref{eq:new lagrangian for iepsilon}):
\begin{align}
	&G_F(t) = \frac{i}{2m}e^{-i(1-i\e)m|t|}\;,\;\; G_D(t) = -\frac{i}{2m}e^{\,i(1+i\e)m|t|}\;, \nonumber\\
	&G_<(t) = \frac{i}{2m} e^{\,i[1+\sign(t)i\e]m t}\;,\;\; G_>(t) = \frac{i}{2m}e^{-i[1-\sign(t)i\e]m t}\;.
\end{align}
That, ladies and gentlemen, is the $i\e$ prescription done right. 
\pagebreak
\section{Discussion}\label{sec:end}
%%%%
I have reviewed the Schwinger-Keldysh path integral and its generalization for OTOCs in presumably enough detail. I will conclude this treatise, and my career, with some words.
\\\\
It is common to speak of the temporal ``Keldysh contour,'' reflecting the steps that led to Eq.~(\ref{eq:keldysh}). I do not find it convenient to think that way. Instead, toward uplifting the formalism to higher-dimensional quantum field theory, I prefer to express the generating function as
\begin{align}
&Z(J,J') = \nonumber\\
&\;\;\!\INT \!\!dq_f\INT \!\!dq_0\INT \!\!dq_0'\;\rho(q_0,q_0')\int_{q(t_0)\,=\,q_0}^{q(t_f)\,=\,q_f}\!\!\!\!\!\!\!\! \Ds q(\cdot)\int_{q'(t_0)\,=\,q_0'}^{q'(t_f)\,=\,q_f}\!\!\!\!\!\!\!\! \Ds q'(\cdot)\;e^{\,i\left\{S(q,q') + \singleint\left[J(t)q(t)-J'(t)q'(t)\right]\right\}}\;,
\end{align}
with action
\begin{align}
	S(q,q') &= \singleint \left[ \half \dot q(t)^2 \!-\!\half m^2 q(t)^2 \!-\! \half \dot q'(t)^2 \!+\! \half m^2 q'(t)^2 + ... \right]\;,
\end{align}
and leave it at that. Similarly for the Larkin-Ovchinnikov path integral and whatever I defined in Eq.~(\ref{eq:who knows}). I see only one contour: A straight line from $t_0$ to $t_f$. Whether for thermal equilibrium it ``could make practical calculations much easier'' \cite{su1988} to reformulate the initial condition as a detour in imaginary time is beside the point. The advantage of the Schwinger-Keldysh formalism is to have a \textit{real-time} generating function for expectation values in arbitrary states.
\\\\
The novelties are instead two flavors of field (or four, or three, etc.), one with a wrong-sign kinetic term, and a constraint between them at $t_f$ \cite{calzetta1994}. This is just the well-trodden\footnote{But not well-developed. I do not recall discussions of charge conjugation, parity, and time reversal, or of other staples of relativistic quantum field theory in this context; and apparently a systematic account of renormalization has only just begun \cite{open_RG_scalar, open_RG_yukawa}.} $2\!\times\! 2$ matrix perspective, but I find it remarkable that quantum field theorists find it ``cleaner, both conceptually and notationally'' \cite{mottola_large-N} to \textit{return} to operator expressions like Eqs.~(\ref{eq:operator form of feynman path integral}), (\ref{eq:operator form of schwinger-keldysh}), and (\ref{eq:operator form of larkin-ovchinnikov}) instead of using them for a thing or two then throwing them away. 
\\\\
Undoubtedly this stems from the ``typical strategy in practice'' of using path integrals merely to ``derive equations of motion for a given set of correlation functions'' \cite{buchhold2016}---a statement those authors intended as a matter of fact, but which I read as an indictment. Feynman's paper dates to 1948, and yet fear, uncertainty, and doubt persist. Path integrals are not on ``very shaky mathematical grounds'' \cite{wilson_galitski}: They form the bedrock of modern physics \cite{wilson1983, polchinski_RG}. 
%\\\\
\vspace{50pt}
\pagebreak
\\
As for why I chose to study this formalism in the first place: Black-hole evaporation. 
%
%\\
%\pagebreak
\\\\
Hawking's prediction that isolated black holes will radiate and disappear was a conceptual breakthrough \cite{hawking1975}, but the formalism he used amounts to implicitly integrating out, instead of tracing out, a dynamical thermal environment and inferring detailed properties of the wavefunction of the radiation \cite{wald_evaporation}. That is not a sound basis for effective field theory. But given the apparently unassailable assumptions behind semiclassical gravity coupled to quantum fields, the question is why not. If the observation scale is macroscopic compared to the anticipated regime of quantum gravity, then why exactly can the system not be decoupled from the bath? How exactly does effective field theory predict its own demise? That is the information problem.\footnote{The question of what happens when an observer falls into a black hole is a different problem.} That is why it is considered a paradox, and not just one of many difficult dynamical problems in theoretical physics. 
\\\\
A useful comparison is to a modified\footnote{This is more or less the actual history, but with the crucial benefit of knowing that quantum field theory is the correct mathematical framework for fundamental particles. I am not old enough to have lived that history---my understanding is that it cannot be overstated just how difficult it was to test not only a model but also the edifice on which that model was built \cite{qft_rehab, qft_gross}. People had certainly heard of the renormalization group, but they did not trust it. As an aside within an aside: I think that nowadays many people trust the renormalization group too blindly, or at least too literally. That effective field theory predicts the values of superrenormalizable couplings to be of order the cutoff \cite{susskind_hierarchy_problem, polchinski_condmat_EFT} should not be interpreted as an inconsistency; instead, it should be interpreted as effective field theory saying that you have asked it a question it cannot answer. It tells you that the most straightforward high-energy completion is incorrect, nothing more. The hierarchy problem and the cosmological-constant problem are difficult dynamical problems but not paradoxes.} history of asymptotic freedom in QCD. Suppose that you know, whether by experiment or by encyclical, that the long-distance model for the strong force is the chiral Lagrangian, and that the short-distance model is the QCD Lagrangian. But suppose you know nothing about the renormalization group---never heard of Gell-Mann and Low, let alone of Wilson. How could those two models be mutually consistent? They cannot coexist, they do not imply each other. One has color, the other does not; one has massless particles, the other does not. Both describe the real world. Not just a hard problem, but a paradox. 
\\\\
The resolution was not to calculate the fermion condensate or nucleon masses in terms of the QCD coupling; as far as I know, that still cannot be done analytically. 't Hooft calculated the meson spectrum in 1+1 dimensions \cite{thooft2dQCD}, and Witten explained how baryons fit in \cite{witten2dQCD}, which demonstrated that \textit{in a toy model} the picture of long-distance pions and nucleons and of short-distance gluons and quarks was plausible.\footnote{See also the work on chiral-symmetry breaking by Coleman and Witten \cite{coleman_witten_chiralsym}.} But that was not what solved the paradox. What solved the paradox was the renormalization group: The explicit demonstration that the coupling of QCD shrinks as the distance scale decreases \cite{gross_wilczek_qcd, politzer_qcd}. That was the breakthrough, even though all it implies is that perturbation theory breaks down. It does not prove confinement, and it does not prove that there exists an energy gap at long distances. It simply shows that effective field theory predicts its own demise, paving the way for consistency between two unimpeachable models. 
%\\\\
%\vfill
\vspace{10pt}
\pagebreak
\\
That is what I want to see for black-hole evaporation. I want to see the analysis by Kiem, Verlinde, and Verlinde \cite{kiem_verlinde_verlinde} and by Polchinski \cite{polchinski_chaos} repeated in the Schwinger-Keldysh formalism, using the foundation I set up with the $i\e$ prescription. At minimum, I want to see an answer to ``why is the public paying us?'' \cite{polchinski_memoir}. Whether you could be super-duper smart and posit the correct pure state for the black hole plus radiation in JT gravity makes no difference, even if you enumerate all of the microstates that constitute the interior. That would not solve the information problem. The SYK model is great, and I owe my career to K, but it is not gravity.\footnote{I thank Eva Silverstein for sharing similar thoughts during my visit to Stanford in 2017.} It is like 't Hooft's toy model of QCD. It does not describe the real world.\footnote{I am well aware that near-extremal black holes have an $AdS_2$ factor \cite{extreme_kerr_AdS2}, and that realistic black holes spin quickly \cite{polarization_whorls}, but I will stake my reputation on predicting that mastering JT gravity will not be good enough to solve the real problem. If I am wrong, then hey, at least I never insisted that superpartners would be discovered at the LHC.} 
\\\\
Here are some observations I have made over the last few years, which may have some relevance given the recent work on replicas \cite{maldacena_replicas, stanford_replicas}. First, recall that for evaporation of a black hole formed from collapse, attempting to trace modes backward in time from future infinity\footnote{Mulling over that, I was led to the following formal observation, which I alluded to in Sec.~\ref{sec:generating functions}. As I have emphasized from square one, expectation values can be built from transition amplitudes, as in Eq.~(\ref{eq:expectation with n=1 v2}). Instead of pairing the density matrix with the \textit{initial} field configurations and summing over a common \textit{final} field configuration, I could do the reverse and thereby construct an ``out-out'' generating function for expectation values like Eq.~(\ref{eq:wightman correlator special case out-out}). Would that be useful?} would accelerate them to arbitrarily high energy, while tracing them forward in time from past infinity does not. Most people seem to consider that discrepancy harmless, but I had always interpreted it as evidence of a phase transition.\footnote{I made comments along those lines to Douglas Stanford and Nick Hunter-Jones at the Simons meeting in NYC in Dec.~2017, and to Beni Yoshida at Settlement Co.~in Waterloo sometime in 2018-2019.} I am pleased to see that a first-order phase transition at the Page time is now consensus \cite{kitaev_replicas}. 
\\\\
Second, between thinking about topological superconductors \cite{me_2d, bentov_zee_so18} and gravitational shockwaves \cite{bentov_swearngin, me_kerr-ads}, the first thing I thought of when seeing the commutation relation between the gravitationally backreacted ingoing and outgoing near-horizon fields is lattice dislocations.\footnote{That is an observation I shared with many people in the IQIM during 2015-2018, including Gil Refael and probably Justin Wilson. See in particular the order-disorder commutation relations in Marino et al.~\cite{path_integral_for_kinks}} That makes sense, because that is exactly what shockwave operators are: Translational defect lines. When you act with two opposing shockwaves in a specified order, it sure does look like it induces a puncture at what was formerly $U$ $=$ $V$ $=$ $0$ in the Kruskal diagram \cite{thooft1985}.\footnote{I have long wanted to characterize the topological nature of two opposing shockwaves by considering some kind of holonomy, but I have never figured out how. Alexei Kitaev gave me that idea, and I have spoken about it at length with Joe Swearngin and Alex Rasmussen. I think the pertinent connection is the translational gauge field of the Poincare group \cite{grignani1992}. Remember two things: Cartan gravity is better than Einstein gravity, and the translational gauge field is not the frame.} Are those the Lorentzian defects that proliferate to restore the black-hole metric plus the branching surface in the $n \to 1$ replica limit? That would make quite the shakshuka from my previous work.
%\\\\
\vspace{20pt}
\pagebreak
\\
But what do I know---I never did make it to Princeton or Harvard. These are just the idle musings of yet another soon-to-be-former physicist who has decided to renounce academia to make money \cite{derman, dash, zuckerman}. 
\\\\
The Schwinger-Keldysh formalism has also been in the back of my mind since senior year of college, when no less a deity than Polyakov entertained a meeting with a nobody like me and suggested I learn it. Thirteen years later, I finally did. While crafting this paper I was urged by various professionals to ``lower my standards'' and just write something; to ``pay the tax'' and give half-baked presentations to uninterested audiences; and to consider the arXiv less a place for ``finished products'' and more a place for ``thoughts'' or ``spreading knowledge.'' To any such suggestions to eviscerate my integrity, I have but two words:
\\\\
I dissent. 
%
%\vfill
%%
%\pagebreak
\\\\
\\\\
\begin{center}
	\textit{Acknowledgments}
\end{center}
%%%%
%%%%
I thank Michael Buchhold for acquiescing to my interrogation about the Schwinger-Keldysh formalism, and I thank Beatrice Bonga for putting up with a lot. I thank the IQIM at Caltech for letting me relive my glory days for the month of February, 2020. I thank in particular Xie Chen and Alexei Kitaev for always teaching me new things and for letting me vent. I thank Joe Swearngin and Justin Wilson for innumerable conversations, many of which contributed in one way or another to this project. I thank Michael Buchhold and Joe Swearngin a second time for feedback on a draft of this paper. I thank the CGWiki community for making coronavirus lockdown tolerable. Finally, I thank Leah and Leila at the Grand Surf Lounge for not ejecting me from the premises for explaining to some guy that the icosahedra he saw while struck by lightning were not a theory of wavefunctions. Research at Perimeter Institute is supported in part by the Government of Canada through the Department of Innovation, Science and Economic Development Canada and by the Province of Ontario through the Ministry of Economic Development, Job Creation and Trade.
%%%%
%%%%
\vfill
\pagebreak
\appendix
\section{Free-particle path integral}\label{sec:free particle}
In this appendix I will calculate the amplitude-generating function for the free particle,
\begin{align}
	Z(q_0,t_0;q_f,t_f|J) &\equiv \int_{q(t_0)\,=\,q_0}^{q(t_f)\,=\,q_f}\!\!\!\! \Ds q(\cdot)\;e^{\,i\int_{t_0}^{t_f}\!dt\,\left[\half \dot q(t)^2 \,+\, J(t)\, q(t)\right]}\;,
\end{align}
using the method of stationary phase. Then I will review how to evaluate the $J = 0$ version using time-slicing regularization. 
\subsection{Stationary-phase method}
I explained the method in Sec.~\ref{sec:oscillator}, but here is an abbreviated retelling. 
\\\\
%%%%
Split the field into classical solutions plus fluctuations: $q(t) \equiv \bar q(t) + Q(t)$, with $\bar q(t_0) = q_0$, $\bar q(t_f) = q_f$, and $Q(t_0) = Q(t_f) = 0$. The stationary-phase condition is
\begin{equation} \label{eq:free-particle equation of motion}
	\ddot{\bar q}(t) = J(t)\;.
\end{equation}
The kinetic term produces a total derivative:
%%%%
\begin{align}
	& \dot q^2 = (\dot {\bar q} + \dot Q)^2 = \dot{\bar q}^2 + 2\dot{\bar q}\dot Q + \dot Q^2 = \tfrac{d}{dt}(\bar q\dot{\bar q})-\bar q \ddot{\bar q} + \tfrac{d}{dt}(2\dot{\bar q} Q)-2\ddot{\bar q} Q+\dot Q^2\;.
\end{align}
Therefore,
\begin{align}
	&\int_{t_0}^{t_f}\!\!dt\left[\half \dot q(t)^2 + J(t) q(t)\right] \nonumber\\
	&\qquad = \left.\half \bar q \dot{\bar q}\right|_{t\,=\,t_0}^{t_f} + \left. \dot{\bar q} Q\right|_{t\,=\,t_0}^{t_f} + \int_{t_0}^{t_f}\!\! dt\left\{ \bar q(t)\left[-\half\ddot{\bar q}(t) + J(t)\right] + Q(t)\left[-\ddot{\bar q}(t) + J(t) \right] + \half \dot Q(t)^2\right\} \nonumber\\
	&\qquad = \left.\bar q \dot{\bar q} \right|_{t\,=\,t_0}^{t_f} + \int_{t_0}^{t_f}\!\! dt\, \half \bar q(t)J(t) + \int_{t_0}^{t_f}\!\! dt\,\half \dot Q^2\;.
\end{align}
The generating function is
\begin{equation}
	Z(q_0,t_0; q_f,t_f|J) = e^{\,i\half\left[\left.\bar q \dot{\bar q}\right|_{t\,=\,t_0}^{t_f}+\int_{t_0}^{t_f} \!dt\,\bar q(t) J(t) \right]} \int_{Q(t_0)\,=\,0}^{Q(t_f)\,=\,0}\!\! \Ds Q(\cdot)\;e^{\,i\int_{t_0}^{t_f}\! dt\;\half \dot Q(t)^2}\;.
\end{equation}
\subsection{Classical solution}
I will offer a slightly different take on the method from Sec.~\ref{sec:classical solution}. 
\subsubsection{Homogeneous and particular solutions}
The homogeneous solution of Eq.~(\ref{eq:free-particle equation of motion}) has the form 
\begin{equation}
	q_h(t) = (t\!-\!t_0)A + B\;,
\end{equation}
with some constants $A$ and $B$. The causal particular solution has the form
\begin{equation}
	q_p(t) = \int_{t_0}^{t_f}\!dt'\;\Ta(t\!-\!t')\,G_0(t\!-\!t')\,J(t')\;,
\end{equation}
with conditions on the function $G_0(t)$ to be determined by imposing $\ddot q_p(t) = J(t)$: 
\begin{align}
	&\dot q_p(t) = \int_{t_0}^{t_f}\! dt'\left[ \del(t\!-\!t') G_0(t\!-\!t') + \Ta(t\!-\!t') \dot G_0(t\!-\!t')\right]J(t') \nonumber\\
	&\qquad = G_0(0) J(t) + \int_{t_0}^{t_f}\! dt'\,\Ta(t\!-\!t')\,\dot G_0(t\!-\!t')J(t') \nonumber\\
	&\qquad \implies G_0(0) = 0 \\
	&\ddot q_p(t) = \int_{t_0}^{t_f}\! dt'\left[ \del(t\!-\!t') \dot G_0(t\!-\!t') + \Ta(t\!-\!t') \ddot G_0(t\!-\!t')\right] J(t') \nonumber\\
	&\qquad = \dot G_0(0) J(t) + \int_{t_0}^{t_f}\! dt'\,\Ta(t\!-\!t')\, \ddot G_0(t\!-\!t') J(t') \nonumber\\
	&\qquad \implies \dot G_0(0) = 1\;,\;\; \ddot G_0(t) = 0\;.
\end{align}
That is just a way to derive the conditions on $G(t)$ instead of announcing them from general principles. Either way, the free-particle Green's function is
\begin{equation}
	G_0(t) = t\;.
\end{equation}
I will use that to express the homogeneous solution as
\begin{equation}
	q_h(t) = G(t\!-\!t_0)A + \dot G(t\!-\!t_0)B\;,
\end{equation}
in line with the general expression from Eq.~(\ref{eq:homogeneous solution}). Combining the homogeneous solution with the causal particular solution, I obtain
\begin{equation}
	\bar q(t) = G_0(t\!-\!t_0)A + \dot G(t\!-\!t_0)B + \int_{t_0}^{t_f}\! dt'\,\Ta(t\!-\!t')\,G_0(t\!-\!t')\,J(t')\;.
\end{equation}
\subsubsection{Boundary conditions}
Since $\Ta(t_0-t') = 0$ within the physical interval, imposing $\bar q(t_0) = q_0$ gives 
\begin{equation}
	B = q_0\;.
\end{equation}
Since $\Ta(t_f-t') = 1$ within that interval, imposing $\bar q(t_f) = q_f$ gives (let $T \equiv t_f-t_0$):
\begin{align}
	&q_f = G_0(T) A + \dot G_0(T) q_0 + \int_{t_0}^{t_f}\!dt'\,G_0(t_f\!-\!t')\,J(t') \nonumber\\
	&\implies A = \frac{1}{G_0(T)}\left[q_f-\dot G_0(T)q_0 - \int_{t_0}^{t_f}\!dt'\,G_0(t_f\!-\!t')\,J(t')\right]\;.
\end{align}
\subsubsection{Complete solution}
The classical solution is then:
\begin{align}
	&\bar q(t) = \frac{G_0(t\!-\!t_0)}{G_0(T)}\left[q_f\!-\!\dot G_0(T)q_0 \!-\! \int_{t_0}^{t_f}\!\!\!dt'\,G_0(t_f\!-\!t')\,J(t')\right]\!+\! \dot G_0(t\!-\!t_0) \,q_0 \!+\! \int_{t_0}^{t_f}\!\!\! dt'\,\Ta(t\!-\!t')\,G_0(t\!-\!t')\,J(t') \nonumber\\
	&= \frac{1}{G_0(T)}\left\{G_0(t\!-\!t_0)q_f + \left[G_0(T) \dot G_0(t\!-\!t_0) - G_0(t\!-\!t_0)\dot G_0(T)\right]\!q_0 \right. \nonumber\\
	&\left.\qquad + \int_{t_0}^{t_f}\!\! dt'\left[\Ta(t\!-\!t')G_0(T)G_0(t\!-\!t')-G_0(t\!-\!t_0)G_0(t_f\!-\!t') \right]J(t')\right\} \nonumber\\
	&=\frac{1}{G_0(T)}\left\{ G_0(t\!-\!t_0)q_f + G_0(t_f\!-\!t)q_0 + \int_{t_0}^{t_f}\!\! dt'\left[\Ta(t\!-\!t')G_0(T)G_0(t\!-\!t')-G_0(t\!-\!t_0)G_0(t_f\!-\!t') \right]J(t')\right\}.
\end{align}
Notice that $G_0(T)\dot G_0(t$$-$$t_0)$$-$$G_0(t$$-$$t_0)\dot G_0(T) = T-(t$$-$$t_0) = t_f$$-$$t = G(t_f$$-$$t)$; trivial in this case, but a trigonometric identity for the oscillator, with $G_0(t) = t$ replaced by $G(t) = \frac{1}{m}\sin(m t)$.
\subsection{Action on classical solution}
To evaluate the action on the classical solution, I will need $\dot{\bar q}(t)$:
\begin{align}
	&\dot{\bar q}(t) = \frac{1}{G_0(T)}\Big\{ \dot G_0(t\!-\!t_0)q_f\!-\!\dot G_0(t_f\!-\!t)q_0 \Big. \nonumber\\
	&\qquad \qquad \qquad \Big. + \int_{t_0}^{t_f}\!\!\!\! dt'\left[\Ta(t\!-\!t')G_0(T)\dot G_0(t\!-\!t')-\dot G_0(t\!-\!t_0)G_0(t_f\!-\!t')\right]J(t') \Big\}\;.
\end{align}
At $t = t_0$:
\begin{align}
	&\dot{\bar q}(t_0) = \frac{1}{G_0(T)}\left\{ q_f-\dot G_0(T)q_0 - \int_{t_0}^{t_f}\!\! dt'\,G_0(t_f\!-\!t') J(t') \right\}\;.
\end{align}
At $t = t_f$:
\begin{align}
	&\dot{\bar q}(t_f) = \frac{1}{G_0(T)}\left\{ \dot G_0(T)q_f - q_0 + \int_{t_0}^{t_f}\!\! dt'\left[G_0(T)\dot G_0(t_f\!-\!t') - \dot G_0(T)G_0(t_f\!-\!t')\right]J(t') \right\}\;.
\end{align}
Therefore:
\begin{align}
	&G_0(T)\left.\bar q \dot{\bar q}\right|_{t\,=\,t_0}^{t_f} = q_f \dot{\bar q}(t_f)-q_0\dot{\bar q}(t_0) \nonumber\\
	&\nonumber\\
	&= q_f\left\{ \dot G_0(T)q_f - q_0 + \int_{t_0}^{t_f}\!\! dt'\left[G_0(T)\dot G_0(t_f\!-\!t') - \dot G_0(T)G_0(t_f\!-\!t')\right]J(t') \right\} \nonumber\\
	&-q_0\left\{ q_f-\dot G_0(T)q_0 - \int_{t_0}^{t_f}\!\! dt'\,G_0(t_f\!-\!t') J(t') \right\}
	\nonumber\\
	&\nonumber\\
	&= \dot G_0(T)(q_f^2+q_0^2)-2q_fq_0 + \int_{t_0}^{t_f}\!\! dt'\left\{ \left[G_0(T)\dot G_0(t_f\!-\!t') - \dot G_0(T)G_0(t_f\!-\!t')\right]q_f + G_0(t_f\!-\!t') q_0 \right\}J(t') \nonumber\\
	&\nonumber\\
	&= \dot G_0(T)(q_f^2+q_0^2)-2q_fq_0 + \int_{t_0}^{t_f}\!\! dt'\left\{ G_0(t'\!-\!t_0) q_f + G(t_f\!-\!t')q_0 \right\} J(t')\;.
\end{align}
I will need to combine that with
\begin{align}
	G_0(T) \int_{t_0}^{t_f}\!\! dt\,\bar q(t)\,J(t) &= \int_{t_0}^{t_f}\!\! dt\left[G_0(t\!-\!t_0)q_f + G_0(t_f\!-\!t)q_0\right] J(t) \nonumber\\
	&+\doubleint\left[\Ta(t\!-\!t')G_0(T)G_0(t\!-\!t') - G_0(t\!-\!t_0)G_0(t_f\!-\!t')\right]J(t)J(t').
\end{align}
The terms linear in $J$ add. Since $J(t)J(t') = J(t')J(t)$, the term quadratic in $J$ should be put into a manifestly symmetric form:
\begin{align}
	&\Ta(t\!-\!t')G_0(T)G_0(t\!-\!t') - G_0(t\!-\!t_0)G_0(t_f\!-\!t') = \Ta(t\!-\!t') \left[G_0(T)G_0(t\!-\!t')-G_0(t\!-\!t_0)G_0(t_f\!-\!t')\right] \nonumber\\
	&\qquad \qquad \qquad \qquad \qquad \qquad \qquad \qquad \qquad +\Ta(t'\!-\!t)\left[-G_0(t\!-\!t_0)G_0(t_f\!-\!t')\right] \nonumber\\
	&\nonumber\\
	&= -\Ta(t\!-\!t') G_0(t_f\!-\!t)G_0(t'\!-\!t_0) - \Ta(t'\!-\!t) G_0(t\!-\!t_0)G_0(t_f\!-\!t')\;.
\end{align}
With that, I obtain the following form of the generating function:
\begin{equation}
	Z(q_0,t_0;q_f,t_f|J) = C(T)\,e^{\,i S_{\text{cl}}(q_0,t_0;q_f,t_f|J)}\;,
\end{equation}
with action
\begin{align}
	&S_{\text{cl}}(q_0,t_0;q_f,t_f|J) = \frac{1}{2G_0(T)}\left\{\dot G_0(T)(q_f^2+q_0^2)-2q_fq_0 + 2\int_{t_0}^{t_f}\!\! dt\left[G_0(t\!-\!t_0)q_f + G_0(t_f\!-\!t)q_0 \right]J(t)\right. \nonumber\\
	&-\left.\doubleint \left[\Ta(t\!-\!t')G_0(t_f\!-\!t)G_0(t'\!-\!t_0) + \Ta(t'\!-\!t) G_0(t\!-\!t_0) G_0(t_f\!-\!t') \right]J(t)J(t') \right\}\;,
\end{align}
and overall factor
\begin{equation}
	C(T) \equiv \int_{Q(t_0)\,=\,0}^{Q(t_f)\,=\,0}\!\!\!\! \Ds Q(\cdot)\;e^{\,i\int_{t_0}^{t_f}\! dt\;\half \dot Q(t)^2}\;.
\end{equation}
\subsection{The overall factor}
The overall factor $C(T)$ does not depend on the external source $J$, so I will calculate it from the source-free path integral:
\begin{align}
	Z(q_0,t_0;q_f,t_f|0) = C(T)\,e^{\,i S_{\text{cl}}(q_0,t_0;q_f,t_f|0)}\;,
\end{align}
with
\begin{equation}
	S_{\text{cl}}(q_0,t_0;q_f,t_f|0) = \frac{\dot G_0(T)(q_f^2+q_0^2)-2q_fq_0}{2G_0(T)}\;.
\end{equation}
Recall that $Z(q_0,t_0;q_f,t_f|0) = \bra q_f|e^{-i\hat H T}|q_0\ket$. That amplitude satisfies a composition law:
\begin{align}
	Z(q_0,t_0;q_2,t_2|0) & = \bra q_2|e^{-i(t_2-t_0)\hat H}|q_0\ket = \bra q_2|e^{-i(t_2-t_1)\hat H} e^{-i(t_1-t_0)\hat H}|q_0\ket \nonumber\\
	&= \INT dq_1 \bra q_2|e^{-i(t_2-t_1)\hat H}|q_1\ket \bra q_1|e^{-i(t_1-t_0)\hat H}|q_0\ket \nonumber\\
	&\INT dq_1\;Z(q_1,t_1;q_2,t_2|0)\,Z(q_0,t_0;q_1,t_1|0)\;.
\end{align}
Because that relation is nonlinear, imposing it will fix the overall normalization of the path integral. To calculate the integral over the intermediate field value, I will need the sum of two actions evaluated on classical solutions:
\begin{align}
	&S_{\text{cl}}(q_0,t_0;q_1,t_1|0) + S_{\text{cl}}(q_1,t_1;q_2,t_2|0) = \frac{\dot G_0(t_1\!-\!t_0)(q_1^2+q_0^2)-2q_1q_0}{2G_0(t_1\!-\!t_0)} + \frac{\dot G_0(t_2\!-\!t_1)(q_2^2+q_1^2)-2q_2q_1}{2G_0(t_2\!-\!t_1)} \nonumber\\
	&\nonumber\\
	&= \frac{ G_0(t_2\!-\!t_1)\left[\dot G_0(t_1\!-\!t_0)(q_1^2\!+\!q_0^2)\!-\!2q_1q_0 \right] + G_0(t_1\!-\!t_0)\left[\dot G_0(t_2\!-\!t_1)(q_2^2\!+\!q_1^2)\!-\!2q_2q_1\right]}{2G_0(t_1\!-\!t_0)G_0(t_2\!-\!t_1)} \nonumber\\
	&\nonumber\\
	&= \frac{G_0(t_2\!-\!t_1)\dot G_0(t_1\!-\!t_0)q_0^2 + G_0(t_1\!-\!t_0)\dot G_0(t_2\!-\!t_1)q_2^2}{2G_0(t_1\!-\!t_0)G_0(t_2\!-\!t_1)} \nonumber\\
	&+ \frac{\left[G_0(t_2\!-\!t_1)\dot G_0(t_1\!-\!t_0) + G_0(t_1\!-\!t_0)\dot G_0(t_2\!-\!t_1)\right]q_1^2 - 2\left[G_0(t_2\!-\!t_1)q_0 + G_0(t_1\!-\!t_0) q_2 \right] q_1}{2G_0(t_1\!-\!t_0)G_0(t_2\!-\!t_1)} \nonumber\\
	&\nonumber\\
	&= \frac{G_0(t_2\!-\!t_1)\dot G_0(t_1\!-\!t_0)q_0^2 + G_0(t_1\!-\!t_0)\dot G_0(t_2\!-\!t_1)q_2^2}{2G_0(t_1\!-\!t_0)G_0(t_2\!-\!t_1)} \nonumber\\
	&+ \frac{G_0(t_2\!-\!t_0)q_1^2 - 2\left[G_0(t_2\!-\!t_1)q_0 + G_0(t_1\!-\!t_0) q_2 \right] q_1}{2G_0(t_1\!-\!t_0)G_0(t_2\!-\!t_1)}\;.
\end{align}
To prepare for integration, complete the square in the numerator of the second term:
\begin{align}
	&G_0(t_2\!-\!t_0)q_1^2 \!-\! 2\left[G_0(t_2\!-\!t_1)q_0 \!+\! G_0(t_1\!-\!t_0) q_2 \right] q_1 = G_0(t_2\!-\!t_0)\!\left[ q_1^2 \!-\! 2\left( \frac{G_0(t_2\!-\!t_1)q_0 + G_0(t_1\!-\!t_0)q_2}{G_0(t_2\!-\!t_0)} \right)q_1\right] \nonumber\\
	&= G_0(t_2\!-\!t_0)\left[ \left(q_1-\frac{G_0(t_2\!-\!t_1)q_0 + G_0(t_1\!-\!t_0)q_2}{G_0(t_2\!-\!t_0)}\right)^2-\left(\frac{G_0(t_2\!-\!t_1)q_0 + G_0(t_1\!-\!t_0)q_2}{G_0(t_2\!-\!t_0)}\right)^2 \right]
\end{align}
Therefore:
\begin{align}
	&2G_0(t_1\!-\!t_0)G_0(t_2\!-\!t_1)\left[ S_{\text{cl}}(q_0,t_0;q_1,t_1|0) + S_{\text{cl}}(q_1,t_1;q_2,t_2|0) \right] \nonumber\\
	&= G_0(t_2\!-\!t_0)\left(q_1-\frac{G_0(t_2\!-\!t_1)q_0 + G_0(t_1\!-\!t_0)q_2}{G_0(t_2\!-\!t_0)}\right)^2 \nonumber\\
	&+G_0(t_2\!-\!t_1)\dot G_0(t_1\!-\!t_0)q_0^2 + G_0(t_1\!-\!t_0)\dot G_0(t_2\!-\!t_1)q_2^2-\frac{\left[G_0(t_2\!-\!t_1)q_0 + G_0(t_1\!-\!t_0)q_2\right]^2}{G_0(t_2\!-\!t_0)} \nonumber\\
	&\nonumber\\
	&= G_0(t_2\!-\!t_0)\left(q_1-\frac{G_0(t_2\!-\!t_1)q_0 + G_0(t_1\!-\!t_0)q_2}{G_0(t_2\!-\!t_0)}\right)^2 \nonumber\\
	&+\frac{G_0(t_2\!-\!t_0)\left[G_0(t_2\!-\!t_1)\dot G_0(t_1\!-\!t_0)q_0^2 + G_0(t_1\!-\!t_0)\dot G_0(t_2\!-\!t_1)q_2^2 \right]}{G_0(t_2\!-\!t_0)}\nonumber\\
	&-\frac{G_0(t_2\!-\!t_1)^2 q_0^2 + G_0(t_1\!-\!t_0)^2 q_2^2 + 2G_0(t_2\!-\!t_1)G_0(t_1\!-\!t_0) q_2 q_0}{G_0(t_2\!-\!t_0)} \nonumber\\
	&\nonumber\\
	&= G_0(t_2\!-\!t_0)\left(q_1-\frac{G_0(t_2\!-\!t_1)q_0 + G_0(t_1\!-\!t_0)q_2}{G_0(t_2\!-\!t_0)}\right)^2 -\frac{2G_0(t_2\!-\!t_1)G_0(t_1\!-\!t_0) q_2q_0}{G_0(t_2\!-\!t_0)}\nonumber\\
	&+ \frac{G_0(t_2\!-\!t_1)\left[G_0(t_2\!-\!t_0)\dot G_0(t_1\!-\!t_0) \!-\! G_0(t_2\!-\!t_1)\right]\!q_0^2 + G_0(t_1\!-\!t_0)\left[G_0(t_2\!-\!t_0)\dot G_0(t_2\!-\!t_1) \!-\! G_0(t_1\!-\!t_0)\right]\!q_2^2}{G_0(t_2\!-\!t_0)}.
\end{align}
And so,
\begin{align}
	&S_{\text{cl}}(q_0,t_0;q_1,t_1|0) \!+\! S_{\text{cl}}(q_1,t_1;q_2,t_2|0) = \frac{G_0(t_2\!-\!t_0)}{2G_0(t_2\!-\!t_1)G_0(t_1\!-\!t_0)}\left(\!q_1\!-\frac{G_0(t_2\!-\!t_1)q_0 \!+\! G_0(t_1\!-\!t_0)q_2}{G_0(t_2\!-\!t_0)}\right)^2 \nonumber\\
	&+\frac{G_0(t_2\!-\!t_0)\dot G_0(t_1\!-\!t_0)-G_0(t_2\!-\!t_1)}{2G_0(t_2\!-\!t_0)G_0(t_1\!-\!t_0)}q_0^2 + \frac{G_0(t_2\!-\!t_0)\dot G_0(t_2\!-\!t_1) - G_0(t_1\!-\!t_0)}{2G_0(t_2\!-\!t_1)G_0(t_2\!-\!t_0)}q_2^2-\frac{2q_2q_0}{2G_0(t_2\!-\!t_0)}\;.
\end{align}
Since $G_0(t_2\!-\!t_0)\dot G_0(t_1\!-\!t_0) - G_0(t_2\!-\!t_1) = G_0(t_1\!-\!t_0) = G_0(t_1\!-\!t_0) \dot G_0(t_2\!-\!t_0)$ [I included that factor of $1 = \dot G_0(t_2\!-\!t_0)$ for reasons that will be come clear shortly], and since $G_0(t_2\!-\!t_0)\dot G_0(t_2\!-\!t_1) - G_0(t_1\!-\!t_0) = G_0(t_2\!-\!t_1) = G_0(t_2\!-\!t_1) \dot G_0(t_2\!-\!t_0)$, I obtain
\begin{align}
	S_{\text{cl}}(q_0,t_0;q_1,t_1|0) \!+\! S_{\text{cl}}(q_1,t_1;q_2,t_2|0) &= \frac{G_0(t_2\!-\!t_0)}{2G_0(t_2\!-\!t_1)G_0(t_1\!-\!t_0)}\left(\!q_1\!-\frac{G_0(t_2\!-\!t_1)q_0 \!+\! G_0(t_1\!-\!t_0)q_2}{G_0(t_2\!-\!t_0)}\right)^2 \nonumber\\
	&+\underbrace{\frac{\dot G_0(t_2\!-\!t_0)(q_0^2+q_2^2)-2q_2q_0}{2G_0(t_2\!-\!t_0)}}_{\begin{matrix}\,=\,S_{\text{cl}}(q_0,t_0;q_2,t_2|0) \end{matrix}}\;.
\end{align}
Therefore, the integral over the intermediate field value produces:
\begin{align}
	&\INT dq_1\;Z(q_0,t_0;q_1,t_1|0)\,Z(q_1,t_1;q_2,t_2|0) \nonumber\\
	&\qquad = C(t_1\!-\!t_0)C(t_2\!-\!t_1) e^{\,i S_{\text{cl}}(q_0,t_0;q_2,t_2|0)} \INT dq_1\;e^{\,\frac{i G_0(t_2\!-\!t_0)}{2G_0(t_2\!-\!t_1)G_0(t_1\!-\!t_0)} \left(q_1-\text{const}\right)^2} \nonumber\\
	&\qquad = C(t_1\!-\!t_0)C(t_2\!-\!t_1) e^{\,i S_{\text{cl}}(q_0,t_0;q_2,t_2|0)}\sqrt{\frac{2\pi i\, G_0(t_2\!-\!t_1)G_0(t_1\!-\!t_0)}{G_0(t_2\!-\!t_0)}} \nonumber\\
	&\qquad \equiv Z(q_0,t_0;q_2,t_2|0) = C(t_2\!-\!t_0)\,e^{\,iS_{\text{cl}}(q_0,t_0;q_2,t_2|0)} \nonumber\\
	&\nonumber\\
	&\implies C(t_1\!-\!t_0)C(t_2\!-\!t_1)\sqrt{2\pi i\,G_0(t_2\!-\!t_1)G_0(t_1\!-\!t_0)} = C(t_2\!-\!t_0) \sqrt{G_0(t_2\!-\!t_0)}\;.
\end{align}
The solution is
\begin{equation}
	C(t) = \frac{1}{\sqrt{2\pi i\, G_0(t)}}\;.
\end{equation}
\subsection{Result}
The generating function for the free particle is
\begin{equation}\label{eq:generating function for free particle with sources}
	Z(q_0,t_0;q_f,t_f|J) = \frac{1}{\sqrt{2\pi i\, G_0(T)}}\,e^{\,i S_{\text{cl}}(q_0,t_0;q_f,t_f|J)}\;,\;G_0(t) = t\;,\;\; T \equiv t_f-t_0
\end{equation}
with action
\begin{align}\label{eq:effective action for free particle with sources}
	&S_{\text{cl}}(q_0,t_0;q_f,t_f|J) = \frac{1}{2G_0(T)}\left\{\dot G_0(T)(q_f^2+q_0^2)-2q_fq_0 + 2\int_{t_0}^{t_f}\!\! dt\left[G_0(t\!-\!t_0)q_f + G_0(t_f\!-\!t)q_0 \right]J(t)\right. \nonumber\\
	&-\left.\doubleint \left[\Ta(t\!-\!t')G_0(t_f\!-\!t)G_0(t'\!-\!t_0) + \Ta(t'\!-\!t) G_0(t\!-\!t_0) G_0(t_f\!-\!t') \right]J(t)J(t') \right\}\;.
\end{align}
Remark 1: For $J = 0$, I recover the standard result
\begin{equation}
	Z(q_0,t_0;q_f,t_f|0) = \frac{1}{\sqrt{2\pi i\,T}} e^{\,\frac{i}{2T}(q_f-q_0)^2}\;.
\end{equation}
Remark 2: For $J \neq 0$, the coefficient of the $JJ$ term is
\begin{align}
	K(t,t') &\equiv -\frac{1}{2T}\left[\Ta(t\!-\!t')(t_f\!-\!t)(t'\!-\!t_0) + \Ta(t'\!-\!t)(t\!-\!t_0)(t_f\!-\!t')\right]\;.
\end{align}
%\\\\
Remark 3: The generating function and effective action for the harmonic oscillator take exactly the same forms as in Eqs.~(\ref{eq:generating function for free particle with sources}) and ~(\ref{eq:effective action for free particle with sources}), with $G_0(t) = t$ replaced by $G(t) = \frac{1}{m}\sin(m t)$.
\subsection{Time-slicing method}\label{sec:free particle from time slicing}
Now I will evaluate Eq.~(\ref{eq:generating function for free particle with sources}) for $J = 0$ (i.e., the 0-point amplitude) by time-slicing regularization. The defining expression is:
\begin{equation} \label{eq:regularized free particle}
	Z_0(q_N,t_N) = \int_{q(t_0)\,=\,q_0}^{q(t_N)\,=\,q_N} \!\! \!\! \Ds q(\cdot)\;e^{\,i\int_{t_0}^{t_N}\! dt\;\half \dot q(t)^2} \equiv C_N \INT dq_{N-1}\,...\INT dq_1\;e^{\,i\e \sum_{j\,=\,0}^{N-1} \half\left(\frac{q_{j+1} - q_j}{\e}\right)^2}\;.
\end{equation}
Here I am using the notation $q_N \equiv q_f$ and $t_N \equiv t_f$, and explicitly denoting the dependence of $Z_0$ only on those variables. The overall constant $C_N$ depends on $N$.
%\\\\
\subsubsection{Composition law}
Eq.~(\ref{eq:regularized free particle}) describes the composition law for the transition amplitude: The particle propagates from $q_0$ to $q_1$, then from $q_1$ to $q_2$, and so on.\footnote{This is the oscillatory version of the Chapman-Kolmogorov relation for Markov processes.} Evolving forward yet another step, from $(q_N,t_N)$ to $(q_{N+1},t_{N+1})$, would give
\begin{align}
	Z_0(q_{N+1},t_{N+1}) &= C_{N+1} \INT dq_N \INT dq_{N-1}\,...\INT dq_1\;e^{\,i\e \sum_{j\,=\,0}^N \half\left(\frac{q_{j+1} - q_j}{\e}\right)^2} \nonumber\\
	&= \frac{C_{N+1}}{C_N} \INT dq_N\;e^{\,i\e \half\left(\frac{q_{N+1}-q_N}{\e}\right)^2} Z_0(q_N,t_N)\;.
\end{align}
Change integration variables from $q_N$ to $q \equiv q_{N+1}-q_N$:
\begin{equation}
	Z_0(q_{N+1},t_{N+1}) = \frac{C_{N+1}}{C_N} \INT dq\;e^{\,\frac{i}{2\e} q^2} Z_0(q_{N+1}\!-\!q,t_N)\;.
\end{equation}
The pertinent limit is always $\e \to 0$, in which case the factor $e^{\frac{i}{2\e}q^2}$ oscillates wildly. So only the parts near $q = 0$ contribute appreciably to the integral, and $Z_0(q_{N+1}$$-$$q,t_N)$ can be expanded in a series:
\begin{align}
	&Z_0(q_{N+1},t_{N+1}) \approx \frac{C_{N+1}}{C_N}\INT dq\;e^{\,\frac{i}{2\e}q^2} \left[Z_0(q_{N+1},t_N) - q \frac{\pa}{\pa q_{N+1}} Z_0(q_{N+1},t_N) \right. \nonumber\\
	&\qquad\qquad\qquad \left. +\half q^2 \frac{\pa^2}{\pa q_{N+1}^2}Z_0(q_{N+1},t_N)\right] \nonumber
\end{align}
\begin{align}
	&= \frac{C_{N+1}}{C_N}\left[ Z_0(q_{N+1},t_N)\INT dq\;e^{\frac{i}{2\e}q^2} - \tfrac{\pa}{\pa q_{N+1}} Z_0(q_{N+1},t_N) \INT dq\;q\;e^{\,\frac{i}{2\e} q^2} \right. \nonumber\\
	&\qquad\qquad\qquad \left.+ \half \tfrac{\pa^2}{\pa q_{N+1}^2}Z_0(q_{N+1},t_N) \INT dq\;q^2\;e^{\,\frac{i}{2\e}q^2}\right] \nonumber\\
	&= \frac{C_{N+1}}{C_N}\left[ Z_0(q_{N+1},t_N) \sqrt{2\pi i\e} + 0 + \half \frac{\pa^2}{\pa q_{N+1}^2}Z_0(q_{N+1},t_N) (-2i)\left.\frac{\pa}{\pa \al} \sqrt{\frac{2\pi i}{\al}}\right|_{\al\,=\,\e^{-1}} \right] \nonumber\\
	&= \frac{C_{N+1}}{C_N}\sqrt{2\pi i\e}\left[ Z_0(q_{N+1},t_N) + \half i \e \frac{\pa^2}{\pa q_{N+1}^2}Z_0(q_{N+1},t_N)\right]\;.\label{eq:infinitesimal composition rule}
\end{align}
Meanwhile, since $t_{N+1} \equiv t_N + \e$, I can compare the above to a series expansion in time:
\begin{equation}
	Z_0(q_{N+1},t_{N+1}) = Z_0(q_{N+1},t_N+\e) \approx  Z_0(q_{N+1},t_N) + \e \frac{\pa}{\pa t_N}Z_0(q_{N+1},t_N)\;.
\end{equation}
This fixes the overall factor on the right-hand side of Eq.~(\ref{eq:infinitesimal composition rule}) to be 1, in which case
\begin{equation}\label{eq:C}
	\frac{C_{N+1}}{C_N} = \frac{1}{\sqrt{2\pi i \e}}\;.
\end{equation}
The recursive boundary condition for this is $C_0 \equiv 1$. (The amplitude for the particle to start at $q_0,t_0$ and end at $q_0,t_0$ is by definition 1.) Therefore, $C_1 = \frac{1}{\sqrt{2\pi i\e}} C_0 = \frac{1}{\sqrt{2\pi i\e}}$, $C_2 = \frac{1}{\sqrt{2\pi i\e}} C_1 = \frac{1}{(\sqrt{2\pi i\e})^2}$, and in general 
\begin{equation}
	C_N = \frac{1}{(\sqrt{2\pi i\e})^N}\;.
\end{equation}
That fixes the overall factor. Feynman's fundamental insight, moreover, was that matching the $O(\e)$ parts reproduces the Schrodinger equation,
\begin{equation}
	i\frac{\pa}{\pa t_N} Z_0(q_{N+1},t_N) = -\half \frac{\pa^2}{\pa q_{N+1}^2} Z_0(q_{N+1},t_N)\;.
\end{equation}
%\\\\
\subsubsection{Iterated integrals}
Eq.~(\ref{eq:C}) implies that Eq.~(\ref{eq:regularized free particle}) becomes
\begin{equation}\label{eq:normalized free-particle path integral}
	Z_0(q_N,t_N) = \frac{1}{(\sqrt{2\pi i \e})^{N}} \INT dq_{N-1} ... \INT dq_1\;e^{\,i\e \sum_{j\,=\,0}^{N-1} \half\left(\frac{q_{j+1}-q_j}{\e}\right)^2}\;.
\end{equation}
The task is to calculate those iterated integrals. First define
\begin{equation}
	I_1 \equiv \INT dq_1\;e^{\,\frac{i}{2\e}\left[(q_2-q_1)^2+(q_1-q_0)^2\right]} = \sqrt{\pi i\e}\,e^{\,\frac{i}{4\e}(q_2-q_0)^2}\;.
\end{equation}
Then define
\begin{align}
	I_2 &\equiv \INT dq_2\;e^{\,\frac{i}{2}\e(q_3-q_2)^2} I_1 = \sqrt{\pi i\e}\INT dq_2\;e^{\,\frac{i}{4\e}\left[2(q_3-q_2)^2+(q_2-q_0)^2\right]} \nonumber\\
	&=\sqrt{\pi i\e}\,e^{\,\frac{i}{6\e}(q_3-q_0)^2}\INT dq_2\;e^{\,\frac{3i}{4\e}\left[q_2-\frac{1}{3}(2q_3+q_0)\right]^2} \nonumber\\
	&= \sqrt{\frac{1}{2}(2\pi i\e)}\;\sqrt{\frac{2}{3}(2\pi i\e)}\;e^{\,\frac{i}{3(2\e)}(q_3-q_0)^2}\;.
\end{align}
That produces a pattern even I can spot, leading to
\begin{equation}
	I_{N-1} = \sqrt{\frac{(2\pi i\e)^{N-1}}{N}}\,e^{\,\frac{i}{N(2\e)}(q_N-q_0)^2}\;.
\end{equation}
So the free-particle path integral in Eq.~(\ref{eq:normalized free-particle path integral}) is (recall that $T = N\e$)
\begin{align}
	Z_0(q_N,t_N) &= \frac{1}{(\sqrt{2\pi i\e})^{N}}I_{N-1} = \frac{1}{\sqrt{2\pi i T}}\,e^{\,\frac{i}{2T}(q_N-q_0)^2}\;. \label{eq:overall free-particle factor for harmonic oscillator}
\end{align}
\subsubsection{Free-particle factor for harmonic oscillator}
Setting $q_N = q_0 = 0$ in Eq.~(\ref{eq:overall free-particle factor for harmonic oscillator}) produces the missing ingredient from Eq.~(\ref{eq:path integral up to a factor}):
\begin{equation}
	Z_0 = Z_0(0,0) = \frac{1}{\sqrt{2\pi i T}}\;.
\end{equation}
With that, the overall factor for the oscillator path integral matches Eq.~(\ref{eq:normalization for harmonic oscillator}) for $t = T$:
\begin{equation}
	Z(0,0|0) =  Z_0 \sqrt{\frac{T}{G(T)}} = \frac{1}{\sqrt{2\pi i G(T)}}\;.
\end{equation} 
%
%\pagebreak
%
\section{Integrating out vs. tracing out}\label{sec:tracing}
I asked two physicists the following question: ``What is the difference between integrating out a field and tracing out a field?" One replied, ``Why should those two have anything to do with each other?" and the other replied, ``Aren't those the same thing?"\footnote{Justin Wilson and A. Zee---I will let you guess who gave which reply. I thank both for interesting discussions about this.}
\\\\
I, for one, was about halfway between the two, maximally confused. Here is what I think. Let $q(t)$ and $x(t)$ be independent fields, and let their combined action have the following general form:
\begin{equation}
	S_{\text{tot}}(q,x) = S_{\text{sys}}(q) + S_{\text{bath}}(x) + S_{\text{int}}(q,x)\;.
\end{equation}
For example, $S_\text{sys}(q) = \half \dot q^2 - \half m^2 q^2$, $S_\text{bath}(x) = \half \dot x^2 - \half M^2 x^2$, and $S_\text{int}(q,x) = \g q x$ (for some coupling $\g$); but in this section I will not work through an explicit calculation.\footnote{Balasubramanian et al.~\cite{Balasubramanian2012} seem to have raised the right questions but then devolved into mumblings about the thermal partition function instead of applying the Schwinger-Keldysh formalism. I have not tried to work through their examples.} As suggested by the notation, I will formally think of $q(t)$ as the ``system'' field whose correlation functions I want to calculate, and I will think of $x(t)$ as the ``bath'' field, even though in this case it is just an oscillator with discrete energy levels. 
%\\\\
\subsection{Integrating out}
The amplitude-generating function for $q(t)$ is\footnote{In this section I will be pedantic about denoting which variables are fields and which are numbers, hence the proliferation of parentheses.}
\begin{align}\label{eq:amplitude-generating function for system}
	Z(q_0,x_0;q_f,x_f|J(\cdot)) &\equiv \int_{q(t_0)\,=\,q_0}^{q(t_f)\,=\,q_f}\!\!\!\!\!\!\Ds q(\cdot)\int_{x(t_0)\,=\,x_0}^{x(t_f)\,=\,x_f}\!\!\!\!\!\!\Ds x(\cdot)\;e^{\,i\left[S_{\text{tot}}(q(\cdot),\,x(\cdot)) + \int_{t_0}^{t_f}\!dt\;J(t)\,q(t)\right]}\;.
\end{align}
To \textit{integrate out} the bath means to calculate, possibly in some approximation, the ``induced'' action $S_{\text{ind}}(x_0,x_f|q(\cdot))$ defined by:
\begin{equation}\label{eq:induced action}
	e^{\,iS_{\text{ind}}(x_0,\,x_f|q(\cdot))} \equiv \int_{x(t_0)\,=\,x_0}^{x(t_f)\,=\,x_f}\!\!\!\!\!\! \Ds x(\cdot)\;e^{\,i\left[S_{\text{bath}}(x(\cdot)) + S_{\text{int}}(q(\cdot),\,x(\cdot)) \right]}\;.
\end{equation}
The amplitude-generating function for $q(t)$ would then have the form
\begin{align}
	Z(q_0,x_0;q_f,x_f|J(\cdot)) &= \int_{q(t_0)\,=\,q_0}^{q(t_f)\,=\,q_f}\!\!\!\!\!\!\Ds q(\cdot)\;e^{\,i\left[S_{\text{eff}}(q(\cdot)|x_0,x_f) + \int_{t_0}^{t_f}\! dt\; J(t)\,q(t)\right]}\;,
\end{align}
with ``effective'' action 
\begin{align}
	S_{\text{eff}}(q(\cdot)|x_0,x_f) = S_{\text{sys}}(q(\cdot)) + S_{\text{ind}}(x_0,x_f|q(\cdot))\;.
\end{align}
Note that the effective action depends on the boundary conditions of the bath.\footnote{This is now reminding me of scattering from monopoles \cite{polchinski_monopole} and impurities \cite{maldacena_ludwig}.}
\subsection{Tracing out}
Given the amplitude-generating function in Eq.~(\ref{eq:amplitude-generating function for system}), the expectation-value-generating function for $q(t)$ in an ensemble described by density matrix $\rho$ is
\begin{align}
	Z(J(\cdot),J'(\cdot)) &\equiv \INT \!\! dq_f\,dx_f\,dq_0\, dx_0\, dq_0'\, dx_0'\;\rho(q_0,x_0;q_0',x_0')\;Z(q_0,x_0;q_f,x_f|J(\cdot))\,Z(q_0',x_0';q_f,x_f|J'(\cdot))^*.
\end{align}
To \textit{trace out} the bath means to calculate, again possibly in some approximation, an ``effective'' density matrix 
\begin{align}\label{eq:effective density matrix}
	&\rho_{\text{eff}}(q_0,q_0';q(\cdot),q'(\cdot)) \equiv \INT \!\! dx_f\,dx_0\,dx_0'\;\rho\left(q_0,x_0;q_0',x_0'\right) \times \nonumber\\
	&\qquad \int_{x(t_0)\,=\,x_0}^{x(t_f)\,=\,x_f}\!\!\!\!\!\! \Ds x(\cdot)\;e^{\,i\left[S_{\text{bath}}(x(\cdot)) + S_{\text{int}}(q(\cdot),\,x(\cdot)) \right]}\int_{x'(t_0)\,=\,x_0'}^{x'(t_f)\,=\,x_f}\!\!\!\!\!\! \Ds x'(\cdot)\;e^{-i\left[S_{\text{bath}}(x'(\cdot)) + S_{\text{int}}(q'(\cdot),\,x'(\cdot)) \right]}\;.
\end{align}
The expectation-value-generating function for $q(t)$ would then have the form
\begin{align}
	Z(J(\cdot),J'(\cdot)) &= \INT \!\!dq_f\,dq_0\,dq_0'\int_{q(t_0)\,=\,q_0}^{q(t_f)\,=\,q_f}\!\!\!\!\!\!\Ds q(\cdot) \int_{q'(t_0)\,=\,q_0'}^{q'(t_f)\,=\,q_f}\!\!\!\!\!\!\Ds q'(\cdot)\;\rho_{\text{eff}}(q_0,q_0';q(\cdot),q'(\cdot))\;\times \nonumber\\
	&e^{\,i\left\{S_{\text{sys}}(q(\cdot)) + \singleint J(t)\,q(t) - \left[S_{\text{sys}}(q'(\cdot)) +\singleint J'(t)\,q'(t)\right]\right\}}\;.
\end{align}
Note that, in general, the effective density matrix depends on the entire history of the field. 
\subsection{Remarks}
Since I will not continue with an example, I should probably stop here and let the equations speak for themselves. But this is an appendix, after all. 
\\\\
In a sense, both physicists were right. Integrating out means performing the path integral over the bulk fluctuations, while tracing out means integrating over the boundary conditions (against a density matrix)---two different operations that, a priori, have nothing to do with each other. And yet, calculating the effective density matrix in Eq.~(\ref{eq:effective density matrix}) requires as input the induced action from Eq.~(\ref{eq:induced action}). 
\\\\
Said that way, it seems obvious: The Feynman path integral is the evolution operator, which is used to construct the Schrodinger-picture wavefunction, and the effective Schrodinger-picture density matrix requires the effective wavefunction. 
%
%\pagebreak
%
\section{List of results}\label{sec:list}
In this appendix I will collect some key formulas and results. 
\subsection{Feynman path integrals}
The amplitude-generating function (Feynman path integral) is
\begin{equation}\label{eq:definition of amplitude-generating function}
	Z(q_0,q_f|J) \equiv \int_{q(t_0) \,=\, q_0}^{q(t_f)\,=\,q_f}\!\!\!\! \Ds q(\cdot)\;e^{\,i\int_{t_0}^{t_f}\! dt\left[\la(q(t)) + J(t)q(t)\right]}\;.
\end{equation}
For translationally invariant systems, $Z(q_0,q_f|J)$ will depend on $t_0$ and $t_f$ only in the combination 
\begin{equation}
	T \equiv t_f\!-\!t_0\;.
\end{equation}
\subsubsection{Free particle}
The free particle has Lagrangian $\la(q) = \half \dot q^2$, and its generating function is
\begin{equation}
	Z(q_0,q_f|J) = \frac{1}{\sqrt{2\pi i G_0(T)}}\;e^{\,iS_{\text{cl}}(q_0,q_f|J)}\;,\;\; G_0(t) = t\;,
\end{equation}
with action 
\begin{align}
	&S_{\text{cl}}(q_0,q_f|J) = \frac{1}{2G_0(T)}\left\{\dot G_0(T)\left(q_f^2+q_0^2\right) - 2q_fq_0 + 2\singleint\left[G_0(t\minus t_0)q_f + G_0(t_f\minus t)q_0\right]J(t) \right. \nonumber\\
	&\qquad \left. -\doubleint \left[\Ta(t\minus t') G_0(t_f\minus t)G_0(t'\minus t_0) + \Ta(t'\minus t)G_0(t\minus t_0)G_0(t_f\minus t') \right] J(t)J(t')\right\}\;.
\end{align}
\subsubsection{Harmonic oscillator}
The harmonic oscillator has Lagrangian $\la(q) = \half \dot q^2 - \half m^2 q^2$, and its generating function is
\begin{equation}\label{eq:Z(q_0,q_f|J) for oscillator}
	Z(q_0,q_f|J) = \frac{1}{\sqrt{2\pi i G(T)}}\;e^{\,iS_{\text{cl}}(q_0,q_f|J)}\;,\;\; G(t) = \frac{1}{m}\sin(mt)\;,
\end{equation}
with action 
\begin{align}\label{eq:S_cl(q_0,q_f|J) for oscillator}
	&S_{\text{cl}}(q_0,q_f|J) = \frac{1}{2G(T)}\left\{\dot G(T)\left(q_f^2+q_0^2\right) - 2q_fq_0 + 2\singleint\left[G(t\minus t_0)q_f + G(t_f\minus t)q_0\right]J(t) \right. \nonumber\\
	&\qquad \left. -\doubleint \left[\Ta(t\minus t') G(t_f\minus t)G(t'\minus t_0) + \Ta(t'\minus t)G(t\minus t_0)G(t_f\minus t') \right] J(t)J(t')\right\}\;.
\end{align}
\subsection{Schwinger-Keldysh path integrals}
The expectation-value-generating function (Schwinger-Keldysh path integral) is
\begin{equation}\label{eq:definition of expectation-value-generating function}
	Z(J,J') \equiv \INT dq_f \INT dq_0 \INT dq_0'\;\rho(q_0,q_0')\;Z(q_0,q_f|J)\,Z(q_0',q_f|J')^*\;.
\end{equation}
In general it depends on the choice of density matrix $\rho$ and on the initial time $t_0$; for translationally invariant systems, $t_0$ will drop out. The results for $Z(J,J')$ will be expressed in terms of the influence phase
\begin{equation}
	\Phi(J,J') \equiv -i\ln Z(J,J')\;.
\end{equation}
\subsubsection{Quadratic actions}
The influence phase for quadratic actions has the general form
\begin{align}\label{eq:general form of influence phase for quadratic actions}
	\Phi(J,J') &= \half\doubleint \left[G_F(t,t') J(t) J(t') - G_D(t,t') J'(t)J'(t') \right. \nonumber\\
	&\left. \qquad \qquad \qquad -G_<(t,t') J(t)J'(t') - G_>(t,t') J'(t)J(t') \right]\;.
\end{align}
The property $\Phi(J,J) = 0$ implies
\begin{equation}
	G_F(t,t') - G_D(t,t') - G_<(t,t') - G_>(t,t') = 0\;.
\end{equation}
The property $\Phi(J,J')^* = -\Phi(J',J)$ implies 
\begin{equation}
	G_F(t,t')^* = G_D(t,t')\;,\;\; G_<(t,t')^* = -G_<(t,t')\;.
\end{equation}
Invariance of $\Phi(J,J')$ under $t \leftrightarrow t'$ implies 
\begin{equation}
	G_F(t',t) = G_F(t,t')\;,\;\; G_D(t',t) = G_D(t,t')\;,\;\; G_<(t',t) = G_>(t,t')\;.
\end{equation}
For translationally invariant systems, $G_F(t,t') = G_F(t-t')$, $G_D(t,t') = G_D(t-t')$, $G_<(t,t') = G_<(t-t')$, and $G_>(t,t') = G_>(t-t')$. 
\subsubsection{Harmonic oscillator in ground state}
The ground-state wavefunction of the harmonic oscillator is $\psi_0(q) = \left(\frac{m}{\pi}\right)^{1/4} e^{-\half m q^2}$. Putting the harmonic oscillator in its ground state means setting 
\begin{equation}
	\rho(q,q') = \psi_0(q)\psi_0(q')^* = \sqrt{\tfrac{m}{\pi}}\;e^{-\half m(q^2+q'^2)}\;.
\end{equation}
The resulting influence phase has the form in Eq.~(\ref{eq:general form of influence phase for quadratic actions}) with
\begin{equation}\label{eq:lesser propagator for harmonic oscillator in ground state}
	G_<(t) = \frac{i}{2m}e^{\,i m t}\;,
\end{equation}
$G_>(t) = -G_<(t)^* = \frac{i}{2m}e^{-imt}$, $G_F(t) = \Ta(t)G_>(t) + \Ta(-t)G_<(t) = \frac{i}{2m}e^{-im|t|}$, and $G_D(t) = G_F(t)^* = -\frac{i}{2m}e^{\,im|t|}$.
\subsubsection{Harmonic oscillator in thermal state}
The thermal density matrix in the field basis is
\begin{equation}
	\rho(q,q') = \sqrt{\tfrac{\dot G(-i\beta)-1}{i\pi G(-i\beta)}} \;e^{\frac{i}{2G(-i\beta)}\left[\dot G(-i\beta)(q^2+q'^2)-2qq'\right]}\;,
\end{equation}
with $G(t) = \frac{1}{m}\sin(m t)$. The resulting influence phase is
\begin{equation}\label{eq:result for thermal influence phase}
	\Phi(J,J') = \Phi_0(J,J') + \half\doubleint \Delta(t\!-\!t')\left[J(t)\minus J'(t)\right]\left[J(t')\minus J'(t')\right]\;,
\end{equation}
with 
\begin{equation}
	\Delta(t) = \frac{1}{e^{\beta m} - 1 }\frac{i}{m} \cos(m t)\;,
\end{equation}
where $\Phi_0(J,J')$ is the influence phase for the ground state. The result in Eq.~(\ref{eq:result for thermal influence phase}) also has the form in Eq.~(\ref{eq:general form of influence phase for quadratic actions}), with 
\begin{equation}
	G_<(t) = G_<^0(t) + \Delta(t)\;\;,
\end{equation}
where $G_<^0(t) = \frac{i}{2m}e^{\,imt}$ from Eq.~(\ref{eq:lesser propagator for harmonic oscillator in ground state}). 
\subsubsection{Quenched oscillator in ground state}
The quenched oscillator has Lagrangian $\la(q) = \half \dot q^2 - \half m(t)^2q^2$, with $m(t>t_0) = m$ and $m(t < t_0) = m_0$. Putting the system into its ground state amounts to using $Z(q_0,q_f|J)$ for the oscillator with frequency $m$ [Eqs.~(\ref{eq:Z(q_0,q_f|J) for oscillator}) and~(\ref{eq:S_cl(q_0,q_f|J) for oscillator})] but calculating $Z(J,J')$ with density matrix
\begin{equation}
	\rho(q,q') = \sqrt{\tfrac{m_0}{\pi}}\;e^{-\half m_0(q^2+q'^2)}\;.
\end{equation}
The resulting influence phase has the form in Eq.~(\ref{eq:general form of influence phase for quadratic actions}) with
\begin{equation}
	G_<(t,t') = \frac{i}{2m_0}\,\z(t\minus t_0)\, \z(t'\minus t_0)^*\;,\;\; \z(t) = \dot G(t) + m_0 G(t)\;,\;\; G(t) = \frac{1}{m}\sin(mt)\;.
\end{equation}
\subsubsection{Harmonic oscillator in first excited state}
The first excited state of the harmonic oscillator has wavefunction $\psi_1(q) = \sqrt{2m}\, q\, \psi_0(q)$, leading to a density matrix \begin{equation}
	\rho(q,q') = 2m\,qq'\,\psi_0(q)\psi_0(q')^* = 2m\sqrt{\tfrac{m}{\pi}}\;e^{-\half m(q^2+q'^2) + \ln(qq')}\;.
\end{equation}
Since $\ln \rho$ is not quadratic, the influence phase will not have the form in Eq.~(\ref{eq:general form of influence phase for quadratic actions}). The result is
\begin{equation}
	\Phi(J,J') = \Phi_0(J,J') -i\ln\left\{1-\frac{1}{2m}\doubleint \cos[m(t\minus t')]\left[J(t)\minus J'(t)\right]\left[J'(t)\minus J'(t')\right]\right\}\;.
\end{equation}
\subsection{Larkin-Ovchinnikov path integrals}
The OTOC-generating function (Larkin-Ovchinnikov path integral) is
\begin{align}\label{eq:definition of OTOC-generating function}
	Z(J,J',J',J'') &\equiv \INT dq_0 \INT dq_0' \INT dq_0''\INT dq_f \INT dq_f'\;\rho(q_0,q_0'')\; \times \nonumber\\
	& Z(q_0,q_f|J)\,Z(q_0',q_f|J')^*\,Z(q_0',q_f'|J'')\,Z(q_0'',q_f'|J''')^*\;.
\end{align}
The generalized influence phase
\begin{equation}
	\Phi(J,J',J'',J''') \equiv -i\ln Z(J,J',J'',J''')
\end{equation}
for quadratic actions is
\begin{align}
	&\Phi(J,J',J'',J''') = \half \doubleint\!\!\left\{ \phantom{\tfrac{1}{2}}\right. \nonumber\\
	&\qquad \;\;\;\;\, G_F(t\!-\!t')\left[J(t)J(t')+J''(t)J''(t')\right]-G_D(t\!-\!t')\left[J'(t)J'(t')+J'''(t)J'''(t')\right]\nonumber\\
	&\qquad -G_<(t\!-\!t')\left[J(t)J'(t')+J''(t)J'''(t')\right] -G_>(t\!-\!t')\left[J'(t)J(t')+J'''(t)J''(t')\right]\nonumber\\
	&\qquad +G_<(t\!-\!t')\left[J(t)J''(t')+J'(t)J'''(t') - J(t)J'''(t')-J'(t)J''(t')\right]  \nonumber\\
	&\qquad +G_>(t\!-\!t')\left[J''(t)J(t')+J'''(t)J'(t') - J'''(t)J(t')-J''(t)J'(t')\right] \nonumber\\
	&\left. \phantom{\tfrac{1}{2}}\right\}\;.
\end{align}
\subsection{Lindblad-improved $i\e$ prescription}
The Lagrangian for a harmonic oscillator with $i\e$ prescription is
\begin{align}
	\la(q,q') &= \half(1\plus i\e)\dot q^2-\half(1\minus i\e)m^2 q^2 -\left[\half(1\minus i\e)\dot q'^2 - \half(1\plus i\e)m^2 q'^2\right] \nonumber\\
	&-i\e\left[\dot q\dot q' + m^2 q q' + im(\dot q q'\minus \dot q' q)\right]\;.
\end{align}
The expectation-value-generating function is similar to Eq.~(\ref{eq:definition of expectation-value-generating function}), except that the integrals over forward and backward fields do not factorize:
\begin{align}
	&Z(J,J') = \nonumber\\
	&\INT \!\!dq_f \INT \!\! dq_0 \INT \!\!dq_0'\;\rho(q_0,q_0')\int_{q(t_0)\,=\,q_0}^{q(t_f)\,=\,q_f}\!\!\!\!\!\!\Ds q(\cdot) \int_{q'(t_0)\,=\,q_0'}^{q'(t_f)\,=\,q_f} \!\!\!\!\!\!\Ds q'(\cdot)\;e^{\,i\singleint\left[\la(q(t),\,q'(t)) + J(t)q(t)-J'(t)q'(t)\right]}\;.
\end{align}
The influence phase has the form in Eq.~(\ref{eq:general form of influence phase for quadratic actions}) with lesser Green's function
\begin{equation}
	G_<(t) = \frac{i}{2m}\,e^{\,i\left[1+\sign(t)i\e\right] m t}\;.
\end{equation}
The greater Green's function is $G_>(t) = -G_<(t)^* = \frac{i}{2m}\,e^{-i\left[1-\sign(t)i\e\right]mt}$. The Feynman and Dyson functions are $G_F(t) = \Ta(t)G_>(t)+\Ta(-t)G_<(t) = \frac{i}{2m}\,e^{-i\left(1-i\e\right)m|t|}$ and $G_D(t) = G_F(t)^*= -\frac{i}{2m}\,e^{\,i\left(1+i\e\right) m|t|}$.
\bibliographystyle{unsrt}
\bibliography{References_for_Keldysh.bib}
%%%%
\end{document}